\documentclass[apj,twocolumn]{openjournal}
\usepackage{amsmath}
\usepackage{graphicx}
\usepackage{booktabs}

\usepackage{supertabular}
\let\supertabularfirsthead\tablefirsthead
\let\supertabularhead\tablehead
\let\supertabulartail\tabletail
\let\supertabularlasttail\tablelasttail
\usepackage{multirow}
\usepackage{color}
\usepackage{soul}
\usepackage{placeins}

\usepackage[dvipsnames]{xcolor} 

\usepackage[breaklinks,colorlinks,citecolor=blue,urlcolor=blue]{hyperref}
\usepackage{orcidlink}

\defcitealias{Shen2018}{S18}

\renewcommand{\arraystretch}{1.2}  
\usepackage{listings}
\usepackage{color}
\definecolor{dkgreen}{rgb}{0,0.6,0}
\definecolor{gray}{rgb}{0.5,0.5,0.5}
\definecolor{mauve}{rgb}{0.58,0,0.82}
\definecolor{golden}{rgb}{0.86,0.65,0.01}
\begin{document}


\title{Spectroscopic follow-up of compact object binary candidates from {\it Gaia} DR3: White dwarfs, neutron stars, black holes, and the parallax zeropoint}

\author{\vspace{-22pt}Kareem El-Badry\,\orcidlink{0000-0002-6871-1752}$^{1,3}$}
\author{David W. Latham\,\orcidlink{0000-0001-9911-7388}$^{2}$}
\author{Hans-Walter Rix\,\orcidlink{0000-0003-4996-9069}$^{3}$}
\author{Allyson Bieryla\,\orcidlink{0000-0001-6637-5401}$^{2}$}
\author{Lars A. Buchhave\,\orcidlink{0000-0003-1605-5666}$^{4}$}
\author{Sahar Shahaf\,\orcidlink{0000-0001-9298-8068}$^{3}$}
\author{Tsevi Mazeh\,\orcidlink{0000-0002-3569-3391}$^{5}$}
\author{Johanna M\"uller-Horn\,\orcidlink{0000-0001-9590-3170}$^{3,6}$}
\author{Pranav Nagarajan\,\orcidlink{0000-0002-1386-0603}$^{1}$}
\author{Natsuko Yamaguchi\,\orcidlink{0000-0001-6970-1014}$^{1}$}
\author{Casey Y. Lam\,\orcidlink{0000-0002-6406-1924}$^{7}$}
\author{Dominick Rowan\,\orcidlink{0000-0003-2431-981X}$^{8}$}
\author{Joshua D. Simon\,\orcidlink{0000-0002-4733-4994}$^{7}$}
\author{Jessica R. Lu\,\orcidlink{0000-0001-9611-0009}$^{8}$}
\author{Howard Isaacson\,\orcidlink{0000-0002-0531-1073}$^{8}$}

\affiliation{$^1$Department of Astronomy, California Institute of Technology, 1200 E. California Blvd., Pasadena, CA 91125, USA}
\affiliation{$^2$Center for Astrophysics $|$ Harvard \& Smithsonian, 60 Garden Street, Cambridge, MA 02138, USA}
\affiliation{$^3$Max-Planck-Institut f\"ur Astronomie, K\"onigstuhl 17, 69117 Heidelberg, Germany}
\affiliation{$^4$DTU Space, National Space Institute, Technical University of Denmark, Elektrovej 327, DK-2800 Kgs. Lyngby, Denmark}
\affiliation{$^5$School of Physics and Astronomy, Tel Aviv University, Tel Aviv 69978, Israel}
\affiliation{$^6$Fakult\"at f\"ur Physik und Astronomie, Universit\"at Heidelberg, Im Neuenheimer Feld 226, 69120 Heidelberg, Germany}
\affiliation{$^7$The Observatories of the Carnegie Institution for Science, 813 Santa Barbara Street, Pasadena, CA 91101, USA}
\affiliation{$^8$Department of Astronomy, University of California, Berkeley, Berkeley, CA 94720, USA}
\email{Corresponding author: kelbadry@caltech.edu}

\begin{abstract}
Astrometry and radial velocities (RVs) from {\it Gaia} DR3 yielded orbits for hundreds of thousands of binary systems, including several samples proposed to contain black holes (BHs), neutron stars (NSs), and white dwarfs (WDs). We present results of a systematic spectroscopic follow-up program targeting systems with high inferred companion masses. Beginning with a sample of 227 sources, we used a combination of archival data and many-epoch spectroscopic follow-up to characterize more than 200. We obtained 1292 high-quality RVs over a period of four years using the TRES and FEROS spectrographs, achieving a typical precision of $0.05\,{\rm km\,s^{-1}}$ and at least 10 RVs for 60 sources. We use these data to test the {\it Gaia} orbital solutions and tighten constraints on orbital parameters and component masses. Joint fitting of astrometry and RVs allows us to directly constrain flux ratios, verifying that undetected companions are genuinely dark. We find that $\sim 60\%$ of the astrometric candidates have reliable orbital solutions and indeed host compact objects, including the two known Gaia BHs, 27 NS candidates, and a dozen massive WDs. We show that tight WD+WD binaries may masquerade as NSs within this sample. The spectroscopic candidates have lower purity: $\sim 50\%$ have spurious spectroscopic solutions, and a majority of the rest are post-mass-transfer binaries or hierarchical triples. Joint astrometry+RV fits of binaries with dark companions yield direct, parallax-independent distance measurements. Using 40 such systems, we measure the {\it Gaia} DR3 parallax zeropoint for astrometric orbital solutions. We find $Z=-0.0362\pm0.0053$ mas, perfectly consistent with the single-star zeropoint for sources of similar color and magnitude. These results will guide the selection of cleaner candidate samples from {\it Gaia} DR4, where a longer observing baseline will enable discovery of many more compact object binaries. Ground-based RV follow-up will remain important for confirming individual systems, particularly those with extreme parameters.

\end{abstract}

\maketitle

\section{Introduction}
\label{sec:intro}

The {\it Gaia} mission \citep{GaiaCollaboration2016, GaiaCollaboration2018, GaiaCollaboration2023} is revolutionizing the study of binary and multiple star systems. {\it Gaia} scanned the sky continuously for almost 11 years, from 2014 to 2025, collecting many-epoch astrometric and photometric measurements for about 2 billion stars, as well as many-epoch spectroscopy and radial velocities (RVs) for tens of millions of them. These data can be used to study binary populations and characterize rare objects across an enormous range of physical scales \citep[e.g.][]{Campbell2015, Kervella2019, El-Badry2021, GaiaCollaboration2024, Stefansson2025}.

Only the first $\sim$1000 days of {\it Gaia} observations have been published. The mission's third data release in 2022 (``DR3'') contained the first astrometric and spectroscopic orbital solutions for binaries \citep{GaiaCollaboration2023_binaries} based on observations taken in $2014-2017$. DR3 contained orbital solutions for $\sim 170,000$ astrometric binaries \citep{Halbwachs2023} and $\sim 190,000$ spectroscopic binaries \citep{Gosset2025}, most of which have orbital periods shorter than 1000 days. The fourth data release is scheduled for December 2026 and will roughly double the DR3 observing baseline and number of epochs, likely increasing the number of orbital solutions by almost an order of magnitude \citep[e.g.][]{El-Badry2024_genmod, Lam2025}.

While interesting classes of binaries can be selected based on {\it Gaia} data alone \citep[e.g.][]{ElBadryRix2022, Fu2022, Andrews2022, Gomel2023, Shahaf2023, Jayasinghe2023, Tanikawa2023}, spectroscopic follow-up is often useful to vet and refine orbital solutions from {\it Gaia}. Follow-up can identify sources with spurious orbital solutions \citep[e.g.][]{Simon2026}, check for the presence of a luminous secondary \citep[e.g.][]{Seeburger2026}, directly measure the flux ratio for systems with astrometric solutions \citep[e.g.][]{Yamaguchi2024b}, improve stellar parameter estimates for the luminous stars \citep[e.g.][]{MuellerHorn2025}, and yield significantly tighter constraints on orbital parameters than are achievable with {\it Gaia} data alone \citep[e.g.][]{Nagarajan2024, Yamaguchi2024}.

After {\it Gaia} DR3, we initiated a multi-epoch spectroscopic follow-up program targeting compact object binary candidates (i.e., systems containing a luminous star and a black hole (BH), neutron star (NS), or white dwarf (WD)) with {\it Gaia} astrometric or spectroscopic orbital solutions. Results for some individual systems and sub-populations characterized throughout the program have been published in previous work \citep[e.g.][]{ElBadry2023BH1, ElBadry2023BH2, Yamaguchi2024, Nagarajan2024, ElBadry2024NS1, El-Badry2024_ns, MuellerHorn2025, Simon2026}. Here, we detail the selection of the initial sample, describe all of our follow-up observations, and present orbits for all the systems that can be solved with our data. Through a combination of our follow-up and analysis of external data, we characterize more than 90\% of our initial sample of 227 compact object binary candidates. The systematic nature of our follow-up allows us to characterize the population of compact object binaries accessible with {\it Gaia}, the purity of {\it Gaia}-selected samples of high-mass function binaries, and the various classes of astrophysical false positives. 

The remainder of this paper is organized as follows.
Section~\ref{sec:sample} describes how our follow-up sample was selected from
DR3 catalogs of astrometric and spectroscopic binary orbits and describes our initial vetting with archival data.
Section~\ref{sec:followup-observations} summarizes our spectroscopic follow-up and RV measurements.  Section~\ref{sec:sed-fitting} is focused on SED fits used to estimate luminous star masses, and
Section~\ref{sec:orbital-solutions} presents orbital solutions inferred from fitting of our RVs and/or the {\it Gaia} solutions. In Section~\ref{sec:discussion}, we discuss the nature of false positives and different classes of compact objects revealed by our search. We summarize our primary results and conclude in Section~\ref{sec:conclusions}. The Appendices present a source-by-source summary of our follow-up. All the spectra, RVs, and orbital solutions resulting from the program are made publicly available together with this paper. 

\section{Sample selection}
\label{sec:sample}
We selected compact object binaries from the \texttt{gaiadr3.nss\_two\_body\_orbit} catalog with full astrometric solutions (\texttt{nss\_solution\_type = Orbital} or \texttt{AstroSpectroSB1}) or full single-lined spectroscopic solutions (\texttt{nss\_solution\_type = SB1} or \texttt{SB1C}). For both classes of solutions, we estimated foreground extinction values using 3D dust maps available in 2022. We used the \citet{Green2019} map for sources in the north ($\delta > -28$ deg) and the \citet{Lallement2019} map for sources farther south. 

We estimated extinction-corrected colors and absolute magnitudes in the {\it Gaia} bands for all sources assuming
$E(G_{\rm BP}-G_{\rm RP})=1.33E(B-V)$ and $A_G=2.66E(B-V)$, which is appropriate for typical sources in the sample with $T_{\rm eff} \approx 6000$\,K. We then separated sources on or near the main sequence from evolved sources using simple cuts in the extinction-corrected color-magnitude diagram. We classified sources as lying on the main sequence if they satisfied 
\begin{equation}
    \label{eq:ms_cut}
    M_{G,0} > 4.5
    \quad{\rm or}\quad
    M_{G,0} > -9.37 + 13.42(G_{\rm BP}-G_{\rm RP})_0,
\end{equation}
where $(G_{\rm BP}-G_{\rm RP})_0$ and $M_{G,0}$ represent extinction-corrected color and absolute magnitude. These cuts are illustrated in Figure~\ref{fig:sample-summary}.

For sources classified as falling on the main sequence, we estimated the primary mass $\widetilde{M}_1$ using the empirical $M_{G,0}$--mass relation compiled by \citet[][their Figure 3]{Janssens2022}. 
These photometric masses are necessarily crude because they do not attempt to model the detailed evolutionary state of each source or light contributions from possible luminous companions, but they are suitable for selection of initial candidates. We denote these estimates $\widetilde{M}_1$, and constraints on companion mass based on them as $\widetilde{M}_2$, to emphasize these limitations. We infer more reliable mass estimates from SED fitting -- including estimates for evolved stars -- in Section~\ref{sec:sed-fitting}.

\subsection{Astrometric candidates}
\label{sec:astrometric_candidates}

We selected astrometric candidates with \texttt{Orbital} and \texttt{AstroSpectroSB1} solutions indicative of a massive dark companion. We calculated the angular photocenter semimajor axis, $\mathring{a}_0$, and its error, $\sigma_{\mathring{a}_0}$, from the best-fit Thiele-Innes parameters using the \texttt{nsstools} package \citep{Halbwachs2023}. We then calculated the astrometric mass ratio function \citep[AMRF;][]{Shahaf2019},
\begin{equation}
    \label{eq:AMRF}
    \mathcal{A}
    =
    \left(\frac{\mathring{a}_0}{\varpi}\right)
    \left(\frac{\widetilde{M}_1}{M_\odot}\right)^{-1/3}
    \left(\frac{P_{\rm orb}}{\rm yr}\right)^{-2/3},
\end{equation}
where $P_{\rm orb}$ is the orbital period and $\varpi$ is the parallax.  If the companion contributes negligible light in the $G$ band, this quantity is related to the mass ratio $q=M_2/M_1$ by
\begin{equation}
    \label{eq:AMRF_vs_q}
    \mathcal{A} = \frac{q}{(1+q)^{2/3}}.
\end{equation}
We inverted Equation~\ref{eq:AMRF_vs_q} numerically and then calculated an initial estimate of the companion mass, $\widetilde{M}_2 = q\widetilde{M}_1$. For a given $\widetilde{M}_1$, this represents the minimum companion mass consistent with the astrometric photocenter orbit: a luminous companion would imply a larger companion mass. 

We constructed a sample of astrometric candidates for follow-up by combining several subsamples: 

\begin{enumerate}
\item A ``primary'' sample of NS or BH candidates satisfying
\begin{equation}
    \label{eq:primary}
    \begin{aligned}
        {\rm MS = true},\\
        \widetilde{M}_2 > 1.4\,M_\odot,\\
        \widetilde{M}_2/\widetilde{M}_1 > 1.2,\\
        F_2 < 10,\\
        P_{\rm orb}<1200\,{\rm d},\\
        G < 15,
    \end{aligned}
\end{equation}
where ${\rm MS}={\rm true}$ refers to sources that pass the main-sequence cut in Equation~\ref{eq:ms_cut}, and
$F_2$ is the \texttt{goodness\_of\_fit} statistic, for which a larger value corresponds to a worse fit.\footnote{The $F_2$ threshold below which solutions are ``good'' likely depends on apparent magnitude: somewhat higher values of $F_2$ are expected for good solutions at $G<13$ than at fainter sources \citep[e.g.][]{ElBadry2023BH2}. Nevertheless, we find in Section~\ref{sec:astrometric-joint-fits} that most solutions with $F_2 > 6$ are poor.} Our cut of $G < 15$ makes high-resolution spectroscopy possible with 2m-class telescopes; 92\% of sources with astrometric orbital solutions satisfy it. This selection produced 47 sources.  

\item We also considered the BH candidates listed in Table~E1 of \citet{ElBadry2023BH1}.  Compared to Equation~\ref{eq:primary}, this selection targeted higher companion masses and did not use cuts on \texttt{goodness\_of\_fit} or require sources to fall on the main sequence. We include in our sample the five sources with $G<15$, of which two were already included in the primary sample. 

\item We also included the BH and NS candidates with $G<15$ that were selected by \citet{Andrews2022}. Their selection is similar in spirit to our primary sample but differs from it in several details: they neglected extinction corrections, assumed a primary mass of $\widetilde{M}_1= 1\,M_\odot$ when constructing the sample, and imposed additional cuts on companion mass uncertainty and $F_2$. This selection yields 16 sources, of which 12 were already included in our samples described above. 

\item Finally, we included a sample of sources suspected to host dark companions with masses below $1.4\,M_\odot$, which could be massive WDs or low-mass NSs. These were selected as 
\begin{equation}
    \begin{aligned}
        {\rm MS = true},\\
        1.05 \leq \widetilde{M}_2/M_\odot \leq 1.40,\\
        \sigma_{\widetilde{M}_2}\leq0.105\,M_\odot,\\
        P_{\rm orb}\leq900\,{\rm d},\\
        G < 15.
    \end{aligned}
\end{equation}
These cuts yielded 22 sources, none of which were included in the three subsamples listed above.
\end{enumerate}

The union of these selections contains 76 unique sources, which are listed and classified in Table~\ref{tab:initial-astrometric-candidates}. We characterized 70 of these via a combination of our own follow-up and analysis of archival spectra and light curves. In 16 cases, follow-up or archival RVs showed the {\it Gaia} solution to be spurious before we obtained a many-epoch orbit. Light curves revealed one secure short-period eclipsing binary, implying that the source is a hierarchical triple, and one additional possible eclipsing system. We ultimately obtained many-epoch follow-up and measured orbital solutions for 51 sources.

\subsection{Spectroscopic candidates}
\label{sec:sb1_cands}

We also constructed a sample of BH and NS candidates with single-lined spectroscopic binary solutions in DR3 \citep{Gosset2025}. We calculated the binary mass function for each source,
\begin{equation}
    \label{eq:fm}
    f_m =
    \frac{P_{\rm orb}K_1^3}{2\pi G}
    \left(1-e^2\right)^{3/2},
\end{equation}
where $K_1$ is the RV semi-amplitude and $e$ is the orbital eccentricity. Assuming the primary has mass $\widetilde{M}_1$, we numerically solve for $\widetilde{M}_{2,\min}$, the companion mass implied by the orbit for an edge-on inclination. We then considered as candidates all sources satisfying \texttt{significance > 10}, where \texttt{significance} is defined as $K_1/\sigma_{K_1}$, as well as either of two cuts:
\begin{equation}
    f_m > 3\,M_\odot,
\end{equation}
or
\begin{equation}
    \widetilde{M}_{2,\min} > 1.4\,M_\odot
    \quad{\rm and}\quad \widetilde{M}_{2,\min} > \widetilde{M}_1.
\end{equation}
Since $\widetilde{M}_1$ is only defined for sources on the main sequence (Equation~\ref{eq:ms_cut}), the second selection applies only to such sources.

These criteria select 151 sources. Of these, 136 satisfy the main-sequence minimum-companion-mass selection and 30 satisfy the $f_m > 3\,M_\odot$ selection, with 15 sources satisfying both. The full list of 151 sources and our characterization of them are given in Table~\ref{tab:initial-sb1-candidates}. We summarize the primary classes of false-positives we encountered below. 

Fitting the SEDs of all candidates (Section~\ref{sec:sed-fitting}) revealed many sources for which a single-star model provided a poor fit, but a two-component model containing a hot star and cool star yielded a much better fit. Such sources are labeled ``two-temperature SED'' in Table~\ref{tab:initial-sb1-candidates}. Many of these sources also show high-amplitude ellipsoidal variability indicative of near Roche lobe-filling donors. As shown by previous work \citep[e.g.][]{ElBadryRix2022, Rowan2024, Seeburger2026}, these sources are predominantly Algol-type binaries with ongoing or recently-terminated mass transfer. They are preferentially selected by our cuts because their donors have lower masses than implied by the single-star mass-radius relation. Such sources are labeled ``Algol'' in Table~\ref{tab:initial-sb1-candidates}. 

Our follow-up revealed many of the sources with the largest mass functions to have strong Balmer emission lines due to ongoing or recent mass transfer. Most of these are mass-transfer binaries with evolved donors \citep[e.g.][]{El-Badry2022_hd15124, MuellerHorn2025}; i.e., more massive and/or more evolved Algol-type binaries. We obtained full orbits for two such systems, {\it Gaia} DR3 251157906379754496 and 3331748140308820352.

We rejected several sources after reconnaissance spectra showed them to contain two or more luminous components of similar luminosity. Such sources are labeled ``SB2'' in Table~\ref{tab:initial-sb1-candidates}. We suspect that they entered the sample because their photometric masses were overestimated or because the {\it Gaia} orbits are spurious. We label all systems with detected eclipses as ``EB'' in Table~\ref{tab:initial-sb1-candidates}. Those for which the eclipse period is much shorter than the {\it Gaia} SB1 period are hierarchical triples, with the SB1 solution tracing the outer orbit. We generally avoided follow-up of such sources but obtained full outer orbits in cases where we only discovered the eclipses late in the program.

Two sources in the SB1 sample, {\it Gaia} DR3 263578264603666560 and 5728328827639713792, also have astrometric solutions with similar periods, but they did not enter the astrometric sample. As we show in Section~\ref{sec:sb1_ast}, we believe this is because these sources have luminous companions that dilute their photocenter.

Our SED fitting (Section~\ref{sec:sed-fitting}) suggested a majority of sources in the SB1 sample to be somewhat evolved, meaning that they have radii significantly larger than predicted by models of the same temperature near the zero-age main sequence. As it is difficult to rule out a luminous companion or inner binary in such cases, we generally prioritized sources closer to the ZAMS. Since the {\it Gaia} orbital solutions make a prediction for each source's RV as a function of time, it is easy to identify sources with spurious solutions. We generally deprioritized sources for which we measured RVs inconsistent with the {\it Gaia} solution for further follow-up, but we did obtain complete orbits for nine such sources.

We vetted all 151 sources through a combination of spectroscopic follow-up, analysis of light curves and spectral energy distributions (SEDs), and literature searches. The dominant contaminants in the SB1 sample are systems whose SEDs and/or light curves indicate multiple luminous stars. We identified 52 candidates as having two-temperature SEDs, including 20 Algol-type binaries with strong ellipsoidal variability. Light curves from TESS \citep{Ricker2015} and ASAS-SN \citep{Shappee2014,Kochanek2017} revealed 25 eclipsing binaries. For 14 of these, the eclipse period is much shorter than the {\it Gaia} SB1 period, identifying the system as a hierarchical triple; for the other 11, the eclipsing binary may itself produce the {\it Gaia} solution. Our follow-up also revealed 15 obviously double-lined binaries (SB2s). We rejected 23 systems with RVs clearly inconsistent with the {\it Gaia} SB1 orbit. We obtained multi-epoch follow-up for 24 candidates and measured independent RV-only orbits for 19. Our classifications are summarized in Table~\ref{tab:initial-sb1-candidates}.

\subsection{Summary of the candidate sample}
\label{sec:summary_cands}

\begin{figure*}[!t]
    \centering
    \includegraphics[width=\textwidth]{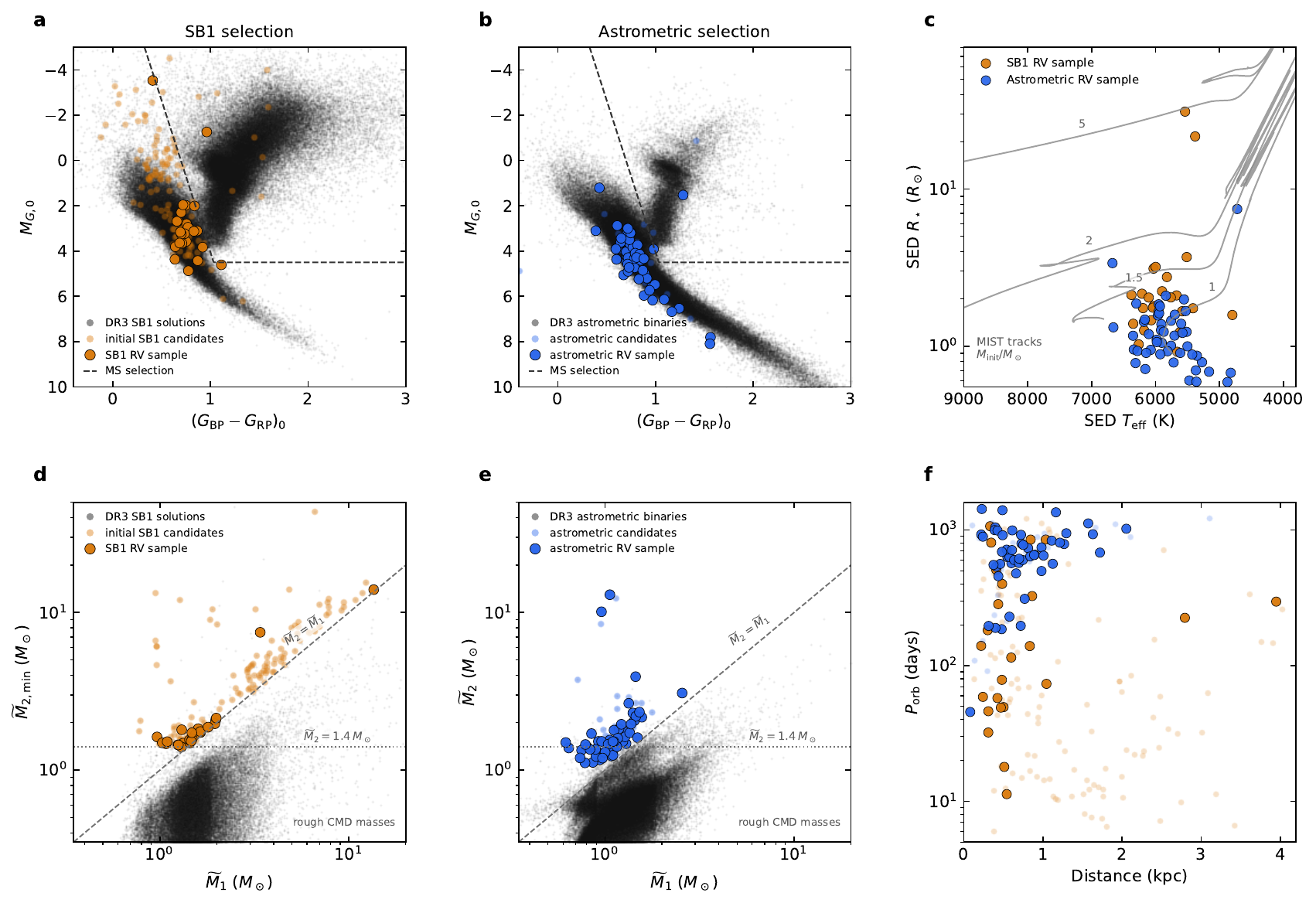}
    \caption{
    Basic properties of the SB1 (gold) and astrometric (blue) candidates. Panels (a) and (b) show extinction-corrected CMDs; small black points show the full DR3 astrometric orbit and SB1 catalogs from which our candidates are selected.  Dashed lines show the main-sequence selection used to assign initial photometric masses. Small colored points show initial candidates rejected due to evidence for a second luminous component or unreliable orbit; larger points show objects for which we obtained many-epoch RV follow-up. Panel (c) compares temperatures and radii inferred from SED fitting for objects in the many-epoch RV sample to MIST evolutionary tracks. Panels (d) and (e) show the approximate primary and companion masses used in the sample selection. Panel (f) shows orbital periods and distances. 
    \label{fig:sample-summary}}
\end{figure*}

Figure~\ref{fig:sample-summary} shows basic properties of the astrometric and SB1 samples, separating sources in the initial samples (small transparent points) from those for which we obtained many-epoch RV follow-up (larger solid points).

The upper panels show CMDs and SED-inferred temperatures and radii (Section~\ref{sec:sed-fitting}). Most of the systems in both the astrometric and spectroscopic RV follow-up samples are solar-type stars: nearly all sources with apparently massive or evolved primaries were rejected early on due to evidence for multiple luminous components or a spurious orbit. The SB1 targets are on average more luminous than the astrometric targets. Many of the astrometric targets, particularly the lowest-luminosity sources, fall on the blue edge of the main sequence, likely due to light contributions from WD companions. 

The lower panels show photometrically-inferred primary and companion masses. For the full catalogs, we only show these estimates for sources satisfying the main-sequence cut. For candidates that are giants, we show mass estimates based on comparison of the sources' CMD positions to single-star MIST evolutionary models. The dashed lines show our selection cuts for both samples. The SB1 sample contains a significant number of sources with $\widetilde{M}_{2,\min} > 3\,M_\odot$, which we initially considered as BH candidates. However, all of these turned out to be luminous binaries or have spurious orbital solutions. The astrometric sample contains two BH candidates with $\widetilde{M}_{2} \approx 10\,M_\odot$, which have been confirmed previously \citep{ElBadry2023BH1, ElBadry2023BH2}. The other candidates with $\widetilde{M}_{2} > 2\,M_\odot$ are refuted or have their masses revised downward by our analysis below. 

The orbital periods of the SB1 candidates are on average shorter than those of the astrometric candidates, reflecting the fact that astrometric wobble increases with increasing orbital period while RV variability amplitude increases at shorter periods. Both samples are largely confined to $P_{\rm orb} \lesssim 1000$ d, the DR3 observing baseline, and $d \lesssim 2$\,kpc. 

\subsubsection{Comparison with \citet{Simon2026}}
\citet{Simon2026} recently presented spectroscopic follow-up of a complementary sample selected to probe the most extreme BH and NS candidates in DR3, emphasizing binaries with inferred companion masses above $2\,M_\odot$. Besides {\it Gaia} BH1, BH2, and NS1, their sample includes 20 sources with DR3 orbital solutions. Of these, 11 are also in our samples: seven in the astrometric sample and four in the SB1 sample. None are in the final samples of 51 astrometric and 24 SB1 systems with multi-epoch RV follow-up.

The nine sources included in the \citet{Simon2026} sample and absent from ours were mainly excluded by quality cuts. Five SB1 sources fail our requirement of $\texttt{significance}>10$. Two astrometric sources have $F_2 >10$, and one is fainter than $G=15$. The final astrometric solution fails our cut of $\widetilde{M}_2/\widetilde{M}_1 > 1.2$.

\section{Follow-up observations}
\label{sec:followup-observations}

We obtained optical high-resolution spectra of candidates from both samples with several instruments. The primary goal was measurement of multi-epoch RVs, but we also used the spectra to search for features from luminous companions and infer atmospheric parameters and projected rotation velocity. 

\subsection{TRES}
\label{sec:tres-observations}

We observed 56 targets with the Tillinghast Reflector Echelle Spectrograph (TRES; \citealt{Szentgyorgyi2007,Furesz2008}) on the 1.5\,m Tillinghast Reflector at the Fred Lawrence Whipple Observatory.  TRES is a fiber-fed echelle spectrograph with resolving power $R \approx 44{,}000$ and wavelength coverage from approximately 3900--9100\,\AA. The spectra were extracted as described in \citet{Buchhave2010}. The TRES observations used here span 2022 June 23 to 2026 March 25. In total, we report RVs from 815 TRES spectra. 

\subsection{FEROS}
\label{sec:feros-observations}

We also obtained spectra with the Fiber-fed Extended Range Optical Spectrograph (FEROS; \citealt{Kaufer1999}) on the MPG/ESO 2.2\,m telescope at La Silla Observatory (programs P109.A-9001, P110.A-9014, P111.A-9003, P112.A-6010, P113.26XB, P114.27SS, P115.28KE, and P116.29G9).  FEROS has resolving power $R\approx 50{,}000$ and covers approximately 3500--9200\,\AA.  The FEROS spectra were reduced with the CERES pipeline \citep{Brahm2017}.  The observations analyzed here span 2022 August 22 to 2026 April 3. In total, we report RVs from 477 spectra of 42 targets. 

\subsection{Other spectrographs}
\label{sec:other-observations}
Early in the follow-up program, we obtained spectra and measured RVs for candidates using several spectrographs, including Keck/DEIMOS \citep{Faber2003}, Keck/ESI \citep{Sheinis2002}, Magellan/MagE \citep{Marshall2008}, and LBT/PEPSI \citep{Strassmeier2015}. We retrieved archival spectra for several targets from surveys including LAMOST \citep{Zhao2012}, APOGEE \citep{Majewski2017}, RAVE \citep{Steinmetz2006,Steinmetz2020}, and GALAH \citep{DeSilva2015,Buder2021}. In most cases only one or two spectra are available for each instrument for a given target. We did not include RVs from these spectra in our fits, since each new instrument requires an additional offset parameter, but we used them to reject a few candidates with spurious {\it Gaia} solutions. These spectra are documented in Tables~\ref{tab:initial-astrometric-candidates} and~\ref{tab:initial-sb1-candidates}.

\subsection{Radial velocities}
\label{sec:RVs}

\begin{table*}[!t]
\centering
\caption{RV measurements. The complete table is available in machine-readable form.}
\label{tab:radial-velocities}
\begin{tabular}{lrrrl}
\toprule
Gaia DR3 source ID & BJD & RV & $\sigma_{\rm RV}$ & Instrument \\
 &  & \multicolumn{2}{c}{(km\,s$^{-1}$)} & \\
\midrule
1007185297091149824 & 2459916.814371 & -48.452 & 0.328 & TRES \\
1007185297091149824 & 2459924.854298 & -47.619 & 0.127 & TRES \\
1007185297091149824 & 2460013.735043 & -45.536 & 0.135 & TRES \\
1007185297091149824 & 2460254.875412 & -50.946 & 0.073 & TRES \\
1007185297091149824 & 2460286.912289 & -51.753 & 0.085 & TRES \\
1007185297091149824 & 2460326.859825 & -52.925 & 0.070 & TRES \\
1007185297091149824 & 2460345.754957 & -53.534 & 0.077 & TRES \\
1007185297091149824 & 2460411.693138 & -55.756 & 0.062 & TRES \\
\ldots & \ldots & \ldots & \ldots & \ldots \\
\bottomrule
\end{tabular}
\begin{minipage}{0.94\textwidth}
\vspace{2pt}\footnotesize
\textit{Note.} RVs are barycentric and are reported on the native zero point of each instrument. The quoted uncertainties are inferred from the scatter across echelle orders and do not include fitted jitter. 
\end{minipage}
\end{table*}

\begin{figure*}[!t]
    \centering
    \includegraphics[width=\textwidth]{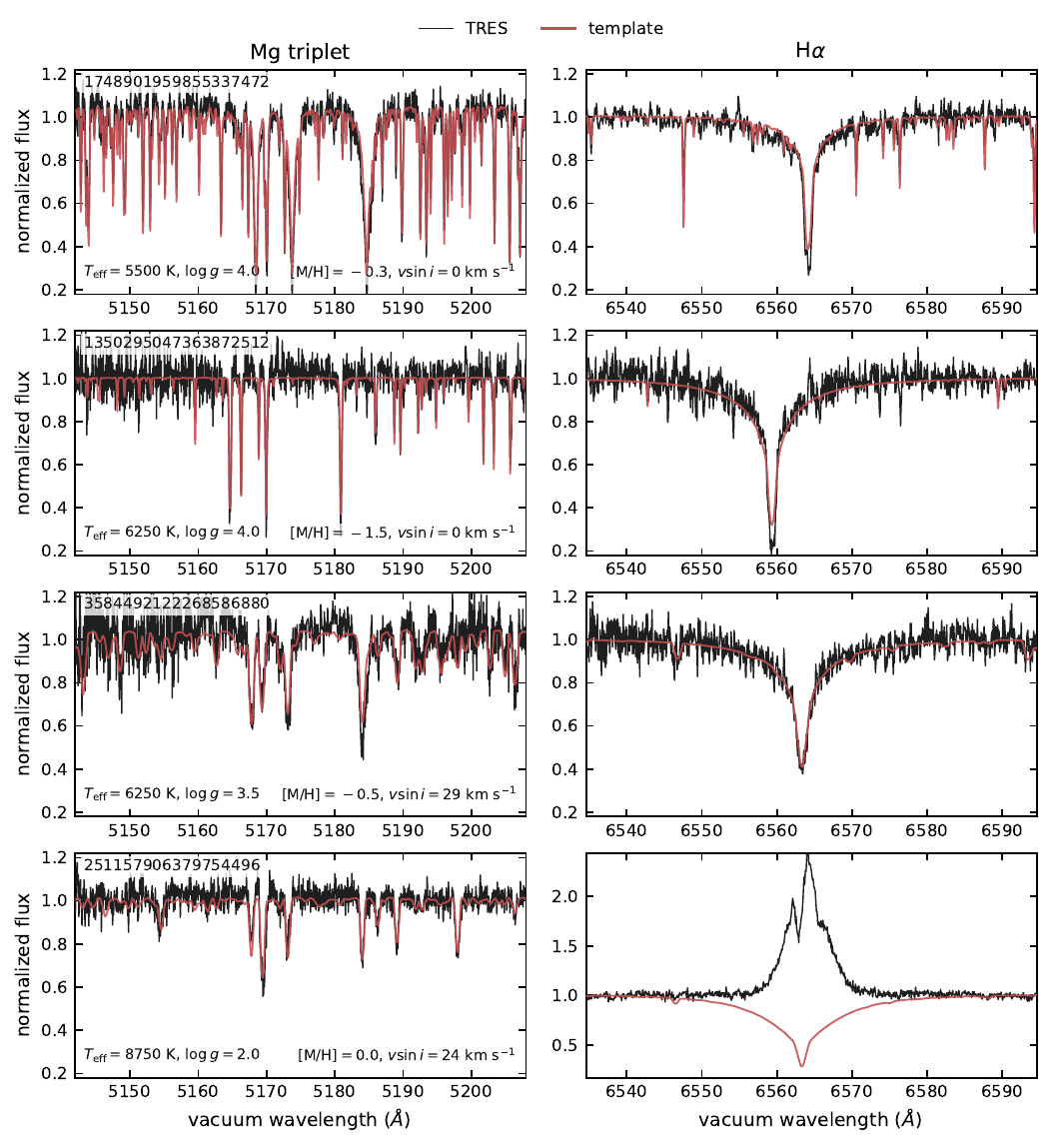}
    \caption{
    Normalized TRES spectra of four representative sources compared to the Kurucz model spectrum used for RV measurements. From top to bottom, we show a solar-type star, a low-metallicity MS star, a star with significant rotational broadening, and an evolved hot star in a post-mass transfer binary. In the last case, H$\alpha$ emission traces a disk around the companion.
    \label{fig:spectral-model-examples}}
\end{figure*}

We measured RVs by cross-correlating template spectra with the TRES and FEROS observations. We used templates from the BOSZ/Kurucz grid \citep{kurucz_model_1979,Kurucz1992,Bohlin2017}, varying effective temperature $T_{\rm eff}$, surface gravity $ \log\left[g/\left({\rm cm\,s^{-2}}\right)\right]$, and metallicity $[\rm Fe/H]$. We found the template spectrum that best matched each target by comparing the highest-SNR spectrum to all template spectra over a fine grid of RVs and projected rotation velocity, $v \sin i$, values. We modeled rotational broadening using \texttt{RotBroadInt} \citep{Carvalho2023}. 

 Figure~\ref{fig:spectral-model-examples} compares the spectra and best-fit models for four representative targets spanning the range of atmospheric parameters in the sample. A majority of the sources are best-fit by models for solar-type main-sequence stars with $4500 < T_{\rm eff}/{\rm K} < 6500$, $3 < \log\left[g/\left({\rm cm\,s^{-2}}\right)\right] < 5$, and negligible rotation, but a handful of sources are better fit by models with low metallicity or significant rotation. Several sources show emission lines in the Balmer series, likely tracing a disk around the companion.

After selecting a template for each source, we measure RVs for each echelle order of each spectrum via cross-correlation. We iteratively mask orders for which the measured RV deviates from the median across all orders by more than 2$\sigma$. We then report the median RV across surviving orders and calculate the uncertainty as the standard deviation across orders divided by the square root of the number of surviving orders. For the FEROS spectra, we fit the 15 orders covering wavelengths between 450 and 670 nm. For the TRES spectra, we fit 31 orders covering 420-670 nm. 

The individual RV measurements adopted in our orbital analysis are reported in Table~\ref{tab:radial-velocities}. The printed table shows the first eight rows; the complete machine-readable table contains all 1292 adopted measurements, together with the spectrum filenames. We report the velocities on the native zero point of each instrument and account for the offset between TRES and FEROS in the orbital fits below.

\begin{figure}[!t]
    \centering
    \includegraphics[width=\columnwidth]{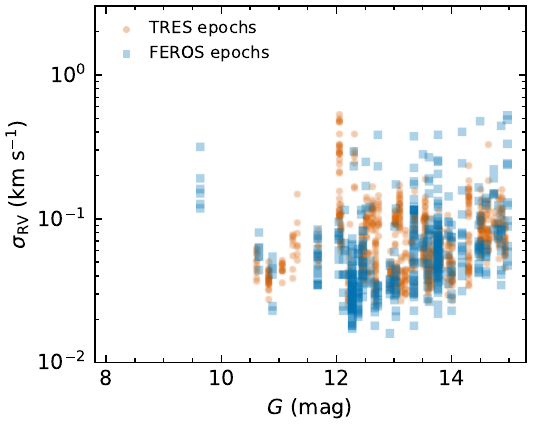}
    \caption{Measured RV uncertainties for all reported spectra as a function of $G$-band magnitude. The median uncertainty is $0.065\,{\rm km\,s^{-1}}$ for 815 TRES epochs and $0.049\,{\rm km\,s^{-1}}$ for 477 FEROS epochs.}
    \label{fig:rv-precision}
\end{figure}

Figure~\ref{fig:rv-precision} summarizes these epoch-level uncertainties as a function of $G$-band magnitude for both instruments. We achieve a median uncertainty of $\approx 0.06\,{\rm km\,s^{-1}}$ across all spectra, with similar performance for TRES and FEROS. Uncertainties increase gradually toward fainter magnitudes at $G>13$ but in most cases remain below $0.1\,{\rm km\,s^{-1}}$ down to $G = 15$. There is significant scatter at fixed magnitude due to sources with different spectral types and variable conditions across observations of a single source.

\section{SED fitting and stellar masses}
\label{sec:sed-fitting}

To estimate temperatures, radii, and masses of the sources in our astrometry and SB1 follow-up samples, we performed single-star fits of the sources' broadband spectral energy distributions (SEDs). We fit the Gaia XP synthetic photometry \citep{GaiaCollaboration2023SyntheticPhotometry} in
SDSS $ugriz$ bands, together with 2MASS $JHK_s$ \citep{Skrutskie2006} and WISE $W1/W2$ \citep{Wright2010}
photometry for each source. We also retrieved UV photometry from GALEX \citep{Martin2005, Bianchi2017} when it was available. To diagnose the presence of possible UV excess due to a hot companion, we did not include FUV data in the fits, and we only included NUV data in the fits for sources for which the optical--infrared SED fit yielded $T_{\rm eff}>5000$\,K. 

We placed a prior on the extinction for each source based on the 3D dust maps discussed in Section~\ref{sec:sample}, assuming a \citet{Cardelli1989} extinction law with $R_V=3.1$. We placed a prior on the parallax from the binary orbital solution for sources in the astrometric sample, and from the \texttt{gaia\_source} single-star solution for sources in the SB1 sample. As shown by \citet{El-Badry2025}, the single-star parallaxes for some of these sources are likely moderately biased by orbital motion, but we do not attempt to correct for this. 

We fixed the metallicity for each source to the value measured from {\it Gaia} XP low-resolution spectra, using the XGBoost method described by \citet{Andrae2023}. As described by \citet{El-Badry2024_ns}, the metallicities published by \citet{Andrae2023} are likely less accurate for sources with astrometric orbital solutions, because they are based on the parallaxes from the sources' single-star fits, which are biased \citep{El-Badry2025}. We therefore recomputed metallicities for sources in the astrometric sample using the \citet{Andrae2023} model but replacing the parallaxes with the parallax from the astrometric orbital solution.  
  
We fit each source for distance, $d$, effective temperature, $T_{\rm eff}$, radius, $R_\star$, and extinction, $A_V$, using BaSeL model spectra \citep{Lejeune1998} interpolated through \texttt{pystellibs}\footnote{\url{https://github.com/mfouesneau/pystellibs}} and computing synthetic photometry with \texttt{pyphot}\footnote{\url{https://github.com/mfouesneau/pyphot}}. We adopted a $T_{\rm eff}$ prior centered on the XGBoost value with a 150\,K standard deviation, and we placed a hard bound on $A_V$ at $\pm 1$ standard deviation of the value predicted by the 3D dust maps. We added a conservative 5\% flux-error floor to the formal flux uncertainties in all bands to account for systematic calibration uncertainties. 

The resulting stellar parameters and the spectral templates adopted for our RV measurements are reported in Table~\ref{tab:stellar-template-parameters}. The printed table shows the first eight rows; the complete machine-readable table contains all 75 sources.

\begin{table*}[!t]
\centering
\caption{Stellar parameters inferred from SED fitting and model templates adopted for the RV measurements. The complete table is available in machine-readable form.}
\label{tab:stellar-template-parameters}
\scriptsize
\setlength{\tabcolsep}{3.0pt}
\begin{tabular}{lrrrrrrrrr}
\toprule
Gaia DR3 source ID & $T_{\rm eff}$ & $R_\star$ & $[\mathrm{M/H}]$ & $A_V$ & $M_\star$ & $T_{\rm eff,temp}$ & $\log g_{\rm temp}$ & $[\mathrm{M/H}]_{\rm temp}$ & $v\sin i$ \\
 & (K) & ($R_\odot$) &  & (mag) & ($M_\odot$) & (K) &  &  & (km\,s$^{-1}$) \\
\midrule
220012968211559296 & $5360_{-76}^{+91}$ & $0.59_{-0.01}^{+0.01}$ & -0.07 & $0.33_{-0.09}^{+0.10}$ & $0.85_{-0.02}^{+0.02}$ & 5000 & 4.5 & -0.5 & 0.0 \\
251157906379754496 & $5598_{-76}^{+76}$ & $31.15_{-0.42}^{+0.42}$ & -1.07 & $2.02_{-0.05}^{+0.08}$ & $3.48_{-0.33}^{+0.56}$ & 8750 & 2.0 & 0.0 & 24.2 \\
263578264603666560 & $5775_{-76}^{+91}$ & $2.03_{-0.02}^{+0.02}$ & -0.14 & $1.56_{-0.08}^{+0.08}$ & $1.08_{-0.06}^{+0.04}$ & 6500 & 4.0 & -0.3 & 11.0 \\
465093354131112960 & $6342_{-138}^{+119}$ & $0.95_{-0.01}^{+0.01}$ & 0.09 & $0.85_{-0.10}^{+0.10}$ & $1.12_{-0.04}^{+0.03}$ & 6500 & 4.5 & -0.3 & 16.2 \\
497238847877377792 & $5527_{-76}^{+91}$ & $3.69_{-0.04}^{+0.04}$ & -0.01 & $0.32_{-0.08}^{+0.10}$ & $1.68_{-0.07}^{+0.07}$ & 4750 & 3.5 & -0.5 & 0.0 \\
556131091543510656 & $6218_{-182}^{+76}$ & $1.03_{-0.01}^{+0.01}$ & -0.50 & $0.36_{-0.13}^{+0.05}$ & $0.95_{-0.06}^{+0.06}$ & 6250 & 4.5 & -0.5 & 16.0 \\
747174436620510976 & $5517_{-45}^{+45}$ & $0.61_{-0.01}^{+0.01}$ & 0.09 & $0.06_{-0.04}^{+0.07}$ & $0.87_{-0.02}^{+0.02}$ & 5000 & 4.5 & -0.5 & 0.0 \\
809741149368202752 & $5303_{-30}^{+45}$ & $0.79_{-0.01}^{+0.01}$ & -0.07 & $0.03_{-0.02}^{+0.05}$ & $0.87_{-0.03}^{+0.03}$ & 5250 & 4.5 & -0.5 & 0.0 \\
\ldots & \ldots & \ldots & \ldots & \ldots & \ldots & \ldots & \ldots & \ldots & \ldots \\
\bottomrule
\end{tabular}
\begin{minipage}{0.96\textwidth}
\vspace{2pt}\footnotesize
\textit{Note.} SED parameters are posterior medians with 16th--84th percentile intervals. The final four columns give the template parameters and rotational broadening used for the RV measurements.
\end{minipage}
\end{table*}

We converted the inferred $T_{\rm eff}$, $R_\star$, and metallicity constraints into constraints on stellar mass and evolutionary state by comparison to MIST stellar-evolution models \citep{Dotter2016,Choi2016}, restricting the grid to ages younger than 13.8\,Gyr. We use the resulting mass constraints in our orbit fits to constrain the masses of the unseen companions (Section~\ref{sec:astrometric-orbit-solutions}). We also consider the inferred radii and temperatures of the stars when assessing whether a luminous companion could escape detection. 

SED fits for our full astrometric and SB1 follow-up samples are shown in Appendix~\ref{sec:appendix-seds}. In most cases, the observed SEDs are well reproduced by single-star models. One object in the astrometric sample, with DR3 source ID 2919995917769953408, shows an obvious UV excess due to a companion that is either a single ultramassive WD or a tight WD+WD binary. Several objects in the SB1 sample, including 3331748140308820352 and 556131091543510656, are poorly fit by a single-star model, likely indicative of a luminous companion. 

The inferred temperatures and radii of all sources in the astrometric and SB1 follow-up samples are shown in Figure~\ref{fig:sample-summary}. Sources in the SB1 sample have systematically larger radii at fixed temperature, indicating that they are more evolved. This likely reflects, at least in part, the fact that the SB1 sample is biased toward more luminous sources, since only sources with $G_{\rm RVS} < 12$ are eligible to receive SB1 solutions. The fact that the companions are more evolved, however, makes it difficult to rule out nondegenerate companions in these systems, primarily tight inner binaries within hierarchical triples.

\section{Orbital solutions}
\label{sec:orbital-solutions}

We now fit orbital solutions to the measured RVs. To assess whether the {\it Gaia} solutions are broadly consistent with our follow-up RVs, we first compare the RVs to predictions of the {\it Gaia} solution alone. In cases where the RVs and {\it Gaia} solution appear consistent, we fit them jointly. In cases where they are inconsistent -- i.e., where including the {\it Gaia} constraints requires significant inflation of the RV uncertainties -- we fit only our RVs. 

\subsection{Astrometric orbits}
\label{sec:astrometric-orbit-solutions}

\begin{figure*}[!tbp]
    \centering
    \includegraphics[width=\textwidth,height=0.84\textheight,keepaspectratio]{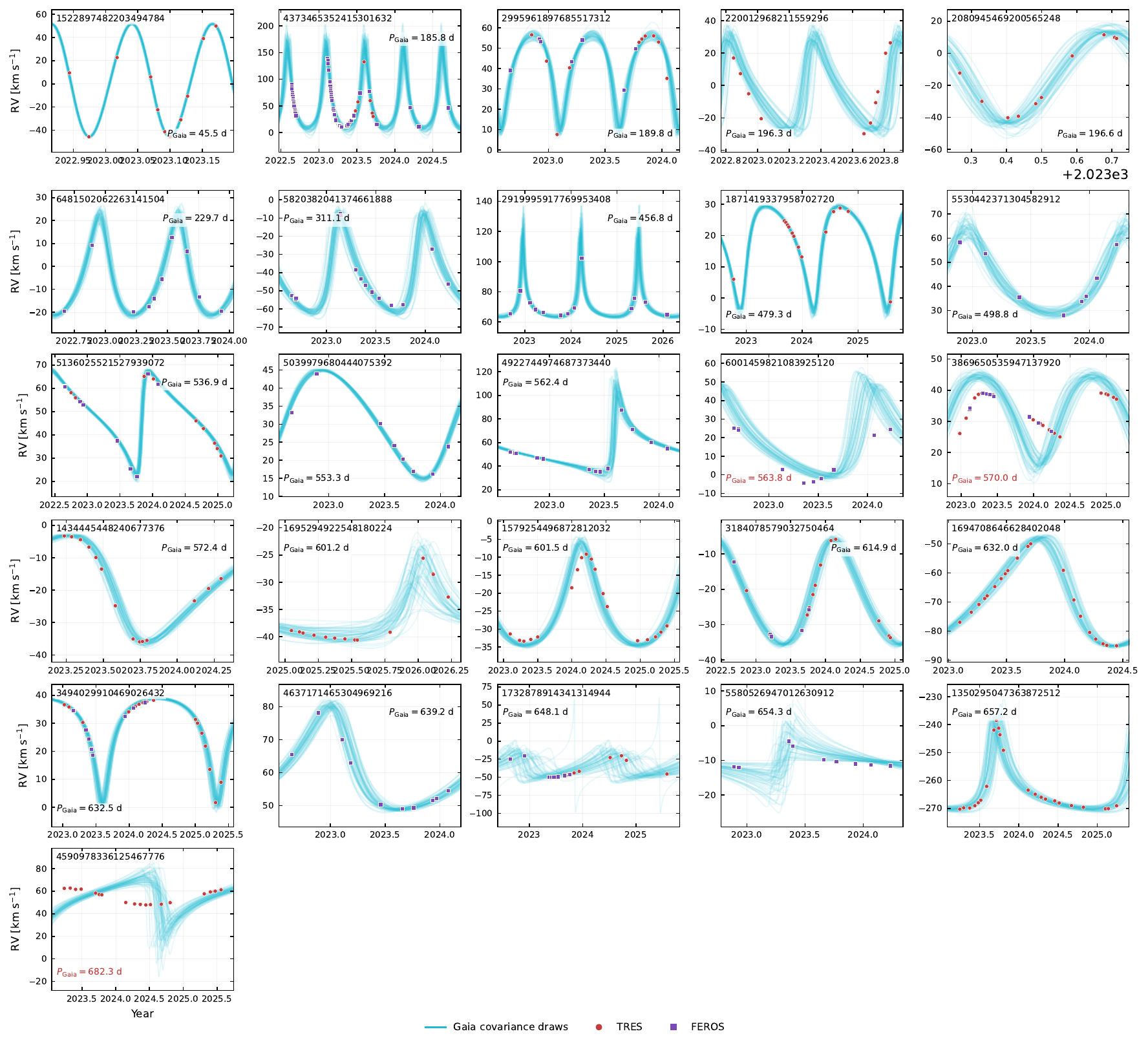}
    \caption{Observed and {\it Gaia}-predicted RVs for sources in the astrometric sample. Cyan curves show predicted RVs for draws from the {\it Gaia} astrometric covariance matrix. Systemic RVs are fixed to the values inferred from the joint astrometry-plus-RV fits (Figure~\ref{fig:astrometric-good-joint-rv-panels}). Red circles show TRES data and purple squares show FEROS. {\it Gaia} period labels are shown in red for sources whose astrometric solutions are inconsistent with our RVs. Panels are sorted by increasing {\it Gaia} orbital period. This figure is continued in Figure~\ref{fig:astrometric-gaia-rv-panels}b.
    }
    \label{fig:astrometric-gaia-rv-panels}
\end{figure*}

\begin{figure*}[!tbp]
    \centering
    \includegraphics[width=\textwidth,height=0.84\textheight,keepaspectratio]{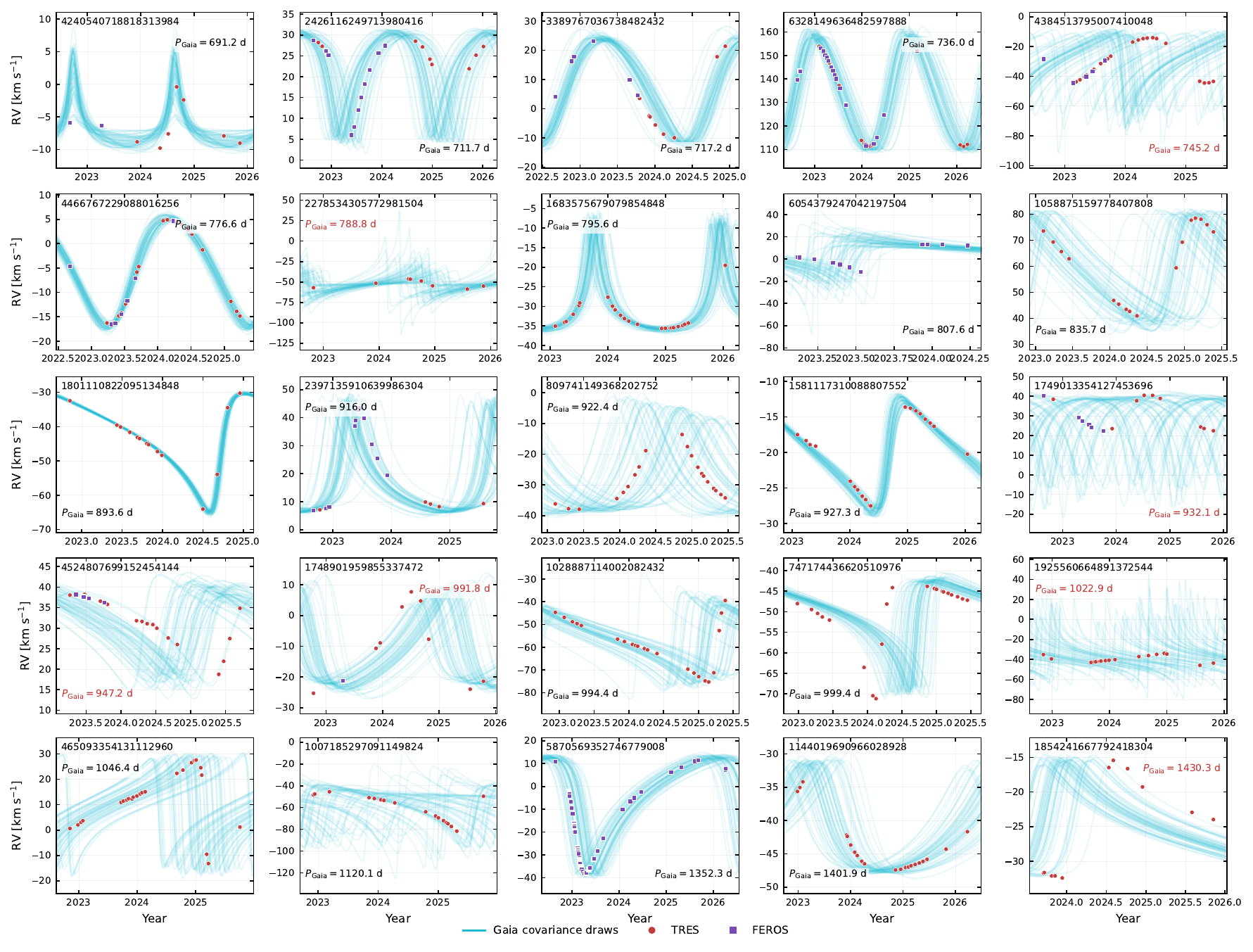}
    \vspace{0.5em}
    \noindent\footnotesize{\textbf{Figure~\ref{fig:astrometric-gaia-rv-panels}b.}
    Continuation of Figure~\ref{fig:astrometric-gaia-rv-panels}.}
\end{figure*}

Figure~\ref{fig:astrometric-gaia-rv-panels} compares the measured RVs to predictions of {\it Gaia}-only astrometric orbits for all sources in the astrometric follow-up sample. For each source, we draw samples from the covariance matrix of the {\it Gaia} astrometric solution and use them to predict the RVs of the luminous source as a function of time, assuming a dark companion. For purely astrometric solutions (i.e., \texttt{Orbital} rather than  \texttt{AstroSpectroSB1} solutions), the {\it Gaia} solutions do not constrain the systemic velocity or the sign of the inclination. In Figure~\ref{fig:astrometric-gaia-rv-panels}, we set the systemic velocity to the value inferred from the joint astrometry+RV fits described in Section~\ref{sec:astrometric-joint-fits}, and we choose the inclination sign that better matches the observed RVs. 

A majority of sources in Figure~\ref{fig:astrometric-gaia-rv-panels} show qualitatively good agreement between observed and predicted RVs. The observed RVs show variability with a period qualitatively similar to the {\it Gaia} period in all cases. In several sources with long periods (e.g. 809741149368202752, 465093354131112960), the {\it Gaia} predictions appear qualitatively consistent with the RVs but show significant scatter due to accumulated phase uncertainty. In other cases (e.g. 6001459821083925120, 3869650535947137920, and 4590978336125467776), the shape of the observed RV curves differs systematically from the {\it Gaia} predictions, indicating a problem with the {\it Gaia} solution.

\subsubsection{Joint astrometry + RV fits}
\label{sec:astrometric-joint-fits}

We next fit our follow-up RVs simultaneously with the {\it Gaia} astrometric orbital solution. Here we assume that the companion is dark, so the photocenter traces the luminous star; this assumption is relaxed in Section~\ref{sec:fluxratio-zpt}.  We sample the 16-parameter vector
\begin{equation}
\begin{split}
\boldsymbol{\theta} = (&\alpha,\delta,\varpi,\mu_{\alpha*},\mu_\delta,
P_{\rm orb},T_p,e,\omega,\Omega,i,\gamma,\\
&M_1,M_2,\sigma_{\rm jit},\Delta v_{\rm FEROS}).
\end{split}
\end{equation}
Here $\alpha$ and $\delta$ are the source's celestial coordinates, $\varpi$ is the parallax, $\mu_{\alpha*}$ and $\mu_\delta$ are proper motions, $P_{\rm orb}$ is the orbital period, $T_p$ is the epoch of periastron, $e$ is the eccentricity, $\omega$ is the luminous star's argument of periastron, $\Omega$ is the longitude of the ascending node, $i$ is the inclination, $\gamma$ is the systemic velocity, $M_1$ and $M_2$ are the masses of the luminous star and companion, $\sigma_{\rm jit}$ is an RV ``jitter'' term accounting for underestimated uncertainties or additional sources of RV noise, and $\Delta v_{\rm FEROS}$ is a FEROS-minus-TRES velocity offset. For systems observed with only one of TRES or FEROS, we eliminate the offset term. To reduce parameter degeneracies, we in practice sample $\ln P_{\rm orb}$, $\ln M_2$, $\ln\sigma_{\rm jit}$, the periastron phase $T_p/P_{\rm orb}$, and $\sqrt{e}\cos\omega$ and $\sqrt{e}\sin\omega$.

We use a Gaussian prior on $M_1$, centered on the value inferred from SED fitting (Section~\ref{sec:sed-fitting}), with standard deviation equal to the larger of the formal fitting uncertainty and $0.05\,M_\odot$. For each sampler step, we transform the sampled parameters into the parameters of the {\it Gaia} astrometric solution. For \texttt{Orbital} solutions, the {\it Gaia} parameter vector is
\begin{equation}
\boldsymbol{y}_{\rm Gaia} =
(\alpha,\delta,\varpi,\mu_{\alpha*},\mu_\delta,A,B,F,G,e,P_{\rm orb},T_p).
\end{equation}
For \texttt{AstroSpectroSB1} solutions, the Gaia solution also constrains the
line-of-sight Thiele-Innes elements and the systemic RV, so we use
\begin{equation}
\boldsymbol{y}_{\rm Gaia} =
(\alpha,\delta,\varpi,\mu_{\alpha*},\mu_\delta,A,B,F,G,C,H,e,P_{\rm orb},T_p,\gamma).
\end{equation}
Here $A$, $B$, $F$, $G$, $C$ and $H$ are Thiele-Innes elements, and $\gamma$ is the systemic velocity.  We first compute
the physical semimajor axis of the luminous star's orbit,
\begin{equation}
\label{eq:a1}
a_1 =
\left[\left(\frac{P_{\rm orb}}{1\,{\rm yr}}\right)^2(M_1+M_2)\right]^{1/3}
\frac{M_2}{M_1+M_2}\,{\rm AU}.
\end{equation}
For a dark companion, the angular semimajor axis of the photocenter orbit is $\mathring{a}_0=a_1\varpi$. The Thiele-Innes elements are then given by
\begin{align}
A &= \mathring{a}_0(\cos\omega\cos\Omega-\sin\omega\sin\Omega\cos i),\\
B &= \mathring{a}_0(\cos\omega\sin\Omega+\sin\omega\cos\Omega\cos i),\\
F &= -\mathring{a}_0(\sin\omega\cos\Omega+\cos\omega\sin\Omega\cos i),\\
G &= -\mathring{a}_0(\sin\omega\sin\Omega-\cos\omega\cos\Omega\cos i),\\
C &= a_1\sin\omega\sin i,\\
H &= a_1\cos\omega\sin i.
\end{align}
Here $A, B, F,$ and $G$ have angular units and are constrained by astrometry, while $C$ and $H$ have physical units and are constrained by RVs.  The {\it Gaia} contribution to the likelihood is then the multivariate Gaussian defined by the covariance matrix,
\begin{equation}
\ln\mathcal{L}_{\rm Gaia}
= -\frac{1}{2}
(\boldsymbol{y}_{\rm Gaia}-\boldsymbol{\mu}_{\rm Gaia})^T
\mathbf{C}_{\rm Gaia}^{-1}
(\boldsymbol{y}_{\rm Gaia}-\boldsymbol{\mu}_{\rm Gaia}),
\end{equation}
where $\boldsymbol{\mu}_{\rm Gaia}$ and $\mathbf{C}_{\rm Gaia}$ are the best-fit {\it Gaia} parameters and covariance matrix.

The predicted RV semi-amplitude is
\begin{equation}
K_1 = \frac{2\pi a_1\sin i}{P_{\rm orb}\sqrt{1-e^2}}.
\end{equation}
For an RV measurement $v_i$ with formal uncertainty $\sigma_i$, the variance is
\begin{equation}
s_i^2 = \sigma_i^2 + \sigma_{\rm jit}^2,
\end{equation}
and the RV likelihood is
\begin{equation}
\ln\mathcal{L}_{\rm RV}
= -\frac{1}{2}\sum_i
\left[
\frac{(v_i-v_{\rm model}(t_i))^2}{s_i^2}
+ \ln(2\pi s_i^2)
\right].
\end{equation}
Here $v_{\rm model}(t_i)$ is the predicted RV of the luminous star at time $t_i$. The full log likelihood is a sum of the {\it Gaia} and RV terms,
\begin{equation}
\ln\mathcal{L}
= \ln\mathcal{L}_{\rm Gaia}+\ln\mathcal{L}_{\rm RV}.
\end{equation}
We use broad, flat priors on all parameters except $M_1$, for which we adopt the Gaussian SED-based prior described above. We sample from the posterior using \texttt{emcee} \citep{Foreman-Mackey2013}, with 64 walkers, 1500 burn-in steps, and 3000 production steps. We initialize each sampler near the {\it Gaia} solution, optimize the posterior, and then begin drawing samples at the maximum-posterior solution.   

Sources for which the measured RVs are inconsistent with predictions of the {\it Gaia} astrometric solution will be best-fit by solutions with large $\sigma_{\rm jit}$. We classify sources as having astrometric orbits consistent with our RVs if they satisfy $\sigma_{\rm jit}/K_{\rm Gaia} < 0.025$; here $K_{\rm Gaia}$ is the RV semi-amplitude of the luminous star predicted by the {\it Gaia} solution. This empirical threshold was chosen after inspecting the joint-fit RV curves to distinguish systems requiring modest excess scatter from those showing systematic discrepancies relative to their orbital amplitudes.  Systems with $\sigma_{\rm jit}/K_{\rm Gaia} > 0.025$ -- roughly speaking, systems with typical RV residuals that are more than 2.5\% of the {\it Gaia}-predicted RV semi-amplitude -- are classified as having unreliable astrometric orbits. 10 systems in our astrometric follow-up sample fail this cut; we fit them with pure SB1 orbits. 

\begin{figure*}[!tbp]
    \centering
    \includegraphics[width=\textwidth,height=0.84\textheight,keepaspectratio]{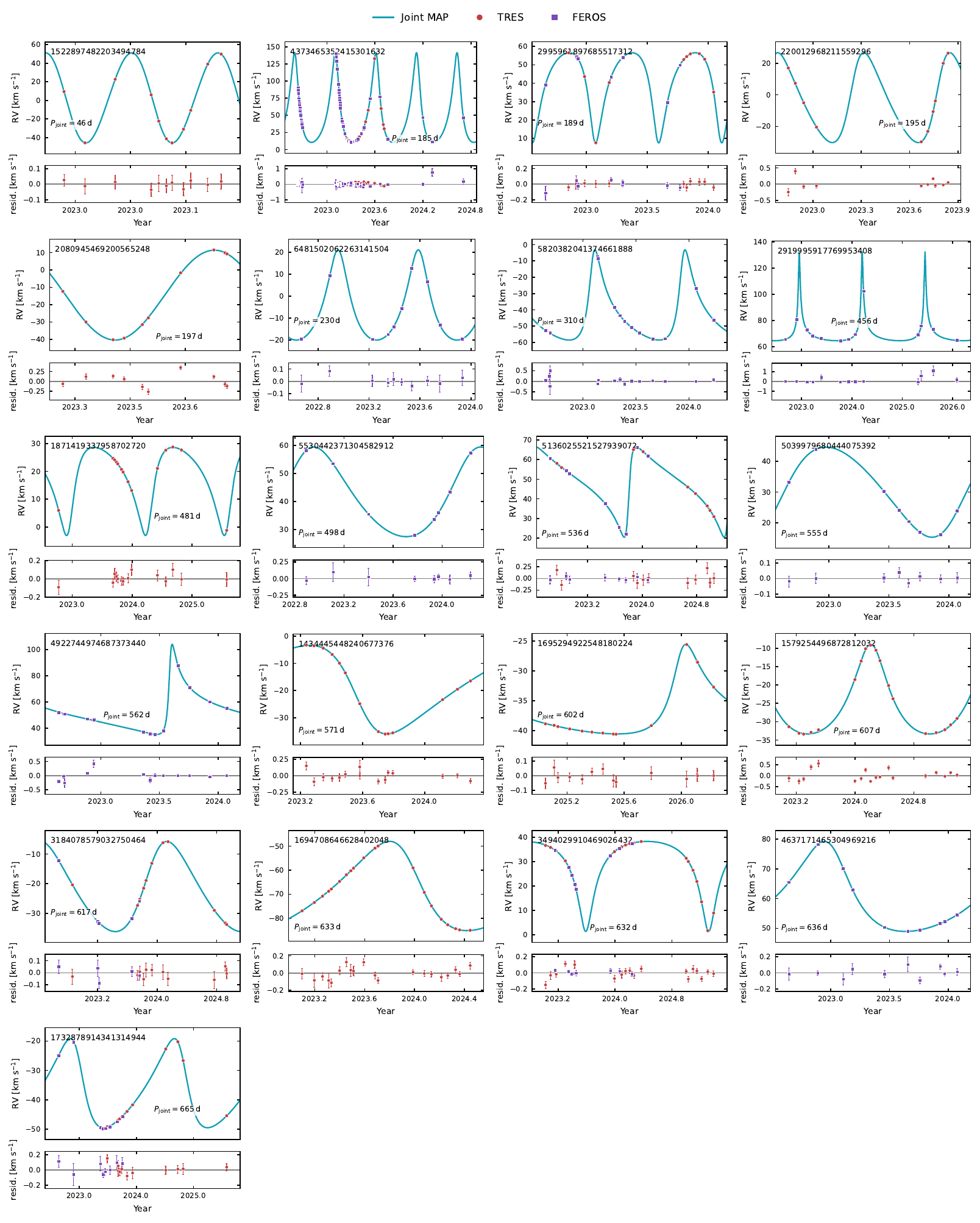}
    \caption{
    Joint astrometry+RV orbital fits for 41 systems in the astrometric sample for which RVs are consistent with the {\it Gaia} solution.  Cyan curves show the maximum-posterior joint model; lower sub-panels show RV residuals. The fitted FEROS-minus-TRES offsets are included when calculating the FEROS residuals but are not shown as separate RV curves. Most sources have RV residuals consistent with 0, indicating reliable uncertainties and a good combined fit. This
    figure is continued in Figure~\ref{fig:astrometric-good-joint-rv-panels}b.
    }
    \label{fig:astrometric-good-joint-rv-panels}
\end{figure*}

For the 41 systems with astrometric orbits consistent with our RVs, Figure~\ref{fig:astrometric-good-joint-rv-panels} shows the maximum-posterior joint model, our measured RVs, and residuals. The residuals are generally consistent with 0. Across all 793 RVs that appear in the figure, 494 (62.3\%) have residuals that are consistent with 0 within 1 sigma when only the formal RV uncertainties are considered, and 705 (88.9\%) are consistent within 2 sigma. If we instead use the quadrature sum of the formal RV uncertainty and the fitted jitter term, 563 (71.0\%) are consistent with 0 within 1 sigma, and 758 (95.6\%) are consistent within 2 sigma. 

Twelve of these systems have RVs from both TRES and FEROS, allowing us to fit the relative instrumental zero point. Defining the offset as FEROS minus TRES, the median fitted offset across these systems is $-0.12\,{\rm km\,s^{-1}}$, with a middle 68\% range from $-0.26$ to $-0.07\,{\rm km\,s^{-1}}$.

The inferred parameters of the 41 reliable joint solutions are reported in Table~\ref{tab:astrometric-joint-dark-orbits}. The printed table shows the first eight rows; the complete machine-readable table provides posterior summaries for all fitted and derived orbital parameters. This first set of solutions assumes a dark companion and does not correct the {\it Gaia} parallax zeropoint.

\begin{table*}[!t]
\centering
\caption{Joint astrometry+RV solutions for reliable astrometric orbits, assuming a dark companion and no parallax zeropoint correction. The complete table is available in machine-readable form.}
\label{tab:astrometric-joint-dark-orbits}
\scriptsize
\setlength{\tabcolsep}{3.2pt}
\begin{tabular}{lrrrrrrrr}
\toprule
Gaia DR3 source ID & $P_{\rm orb}$ & $e$ & $i$ & $\mathring{a}_0$ & $\varpi$ & $M_2$ & $\gamma$ & $\sigma_{\rm jit}$ \\
 & (d) &  & (deg) & (mas) & (mas) & ($M_\odot$) & (km\,s$^{-1}$) & (km\,s$^{-1}$) \\
\midrule
1522897482203494784 & $45.519_{-0.001}^{+0.001}$ & $0.050_{-0.000}^{+0.000}$ & $80.1_{-0.3}^{+0.3}$ & $2.559_{-0.005}^{+0.005}$ & $12.446_{-0.015}^{+0.015}$ & $1.32_{-0.04}^{+0.04}$ & $2.81_{-0.01}^{+0.01}$ & $0.002_{-0.001}^{+0.010}$ \\
4373465352415301632 & $185.39_{-0.01}^{+0.01}$ & $0.432_{-0.000}^{+0.000}$ & $53.3_{-0.3}^{+0.3}$ & $2.630_{-0.014}^{+0.014}$ & $2.098_{-0.016}^{+0.016}$ & $9.18_{-0.13}^{+0.13}$ & $48.90_{-0.03}^{+0.03}$ & $0.055_{-0.010}^{+0.011}$ \\
2995961897685517312 & $189.13_{-0.03}^{+0.03}$ & $0.388_{-0.001}^{+0.001}$ & $53.3_{-0.9}^{+0.9}$ & $1.228_{-0.014}^{+0.014}$ & $2.506_{-0.014}^{+0.014}$ & $1.35_{-0.04}^{+0.04}$ & $41.16_{-0.01}^{+0.01}$ & $0.003_{-0.002}^{+0.017}$ \\
220012968211559296 & $194.73_{-0.05}^{+0.05}$ & $0.235_{-0.003}^{+0.005}$ & $88.4_{-0.7}^{+0.7}$ & $1.592_{-0.015}^{+0.013}$ & $3.220_{-0.022}^{+0.021}$ & $1.22_{-0.04}^{+0.03}$ & $-4.67_{-0.08}^{+0.10}$ & $0.220_{-0.077}^{+0.158}$ \\
2080945469200565248 & $196.79_{-0.07}^{+0.07}$ & $0.002_{-0.001}^{+0.003}$ & $85.3_{-1.6}^{+1.6}$ & $0.663_{-0.005}^{+0.005}$ & $1.403_{-0.009}^{+0.010}$ & $1.43_{-0.03}^{+0.03}$ & $-14.49_{-0.09}^{+0.08}$ & $0.241_{-0.067}^{+0.118}$ \\
6481502062263141504 & $230.21_{-0.08}^{+0.07}$ & $0.307_{-0.002}^{+0.002}$ & $49.8_{-1.1}^{+1.2}$ & $0.949_{-0.012}^{+0.012}$ & $1.750_{-0.019}^{+0.019}$ & $1.33_{-0.04}^{+0.05}$ & $-5.76_{-0.02}^{+0.02}$ & $0.003_{-0.003}^{+0.032}$ \\
5820382041374661888 & $310.16_{-0.12}^{+0.11}$ & $0.534_{-0.002}^{+0.002}$ & $83.2_{-0.9}^{+0.9}$ & $0.901_{-0.009}^{+0.010}$ & $1.344_{-0.012}^{+0.013}$ & $1.39_{-0.03}^{+0.04}$ & $-42.94_{-0.12}^{+0.15}$ & $0.015_{-0.015}^{+0.060}$ \\
2919995917769953408 & $456.08_{-0.13}^{+0.13}$ & $0.816_{-0.007}^{+0.006}$ & $60.7_{-1.0}^{+1.1}$ & $2.143_{-0.037}^{+0.036}$ & $2.298_{-0.023}^{+0.022}$ & $1.31_{-0.06}^{+0.06}$ & $70.83_{-0.07}^{+0.07}$ & $0.003_{-0.003}^{+0.024}$ \\
\ldots & \ldots & \ldots & \ldots & \ldots & \ldots & \ldots & \ldots & \ldots \\
\bottomrule
\end{tabular}
\begin{minipage}{0.96\textwidth}
\vspace{2pt}\footnotesize
\textit{Note.} Values are posterior medians with 16th--84th percentile intervals. The machine-readable table also reports $T_p$, $\omega$, $\Omega$, $M_1$, $K_1$, and FEROS--TRES offset.
\end{minipage}
\end{table*}

\begin{figure*}[!tbp]
    \centering
    \includegraphics[width=\textwidth,height=0.84\textheight,keepaspectratio]{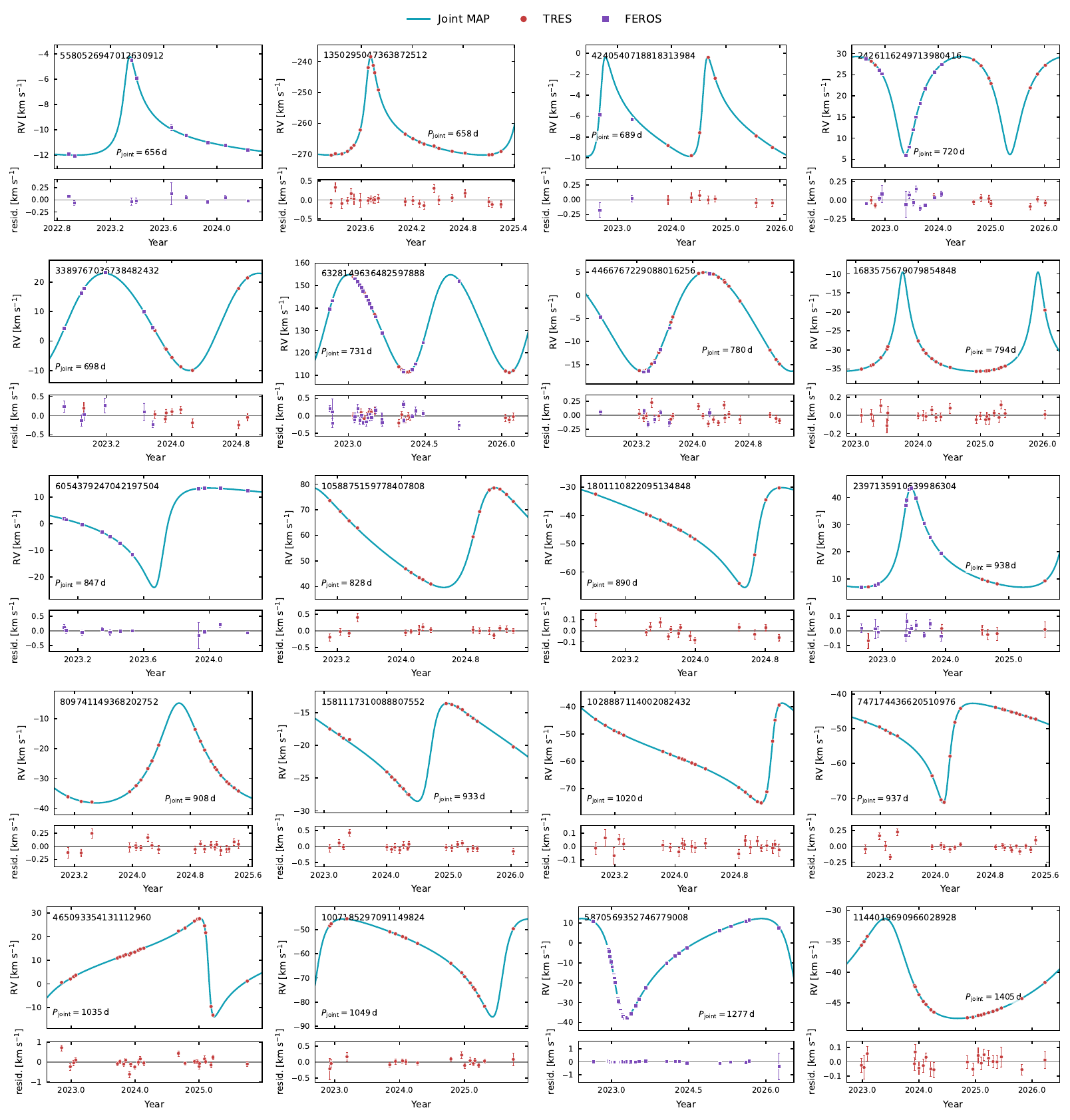}
    \vspace{0.5em}
\noindent\footnotesize{\textbf{Figure~\ref{fig:astrometric-good-joint-rv-panels}b.}
    Continuation of Figure~\ref{fig:astrometric-good-joint-rv-panels}.}
\end{figure*}

\begin{figure*}[!t]
    \centering
    \includegraphics[width=\textwidth,height=0.68\textheight,keepaspectratio]{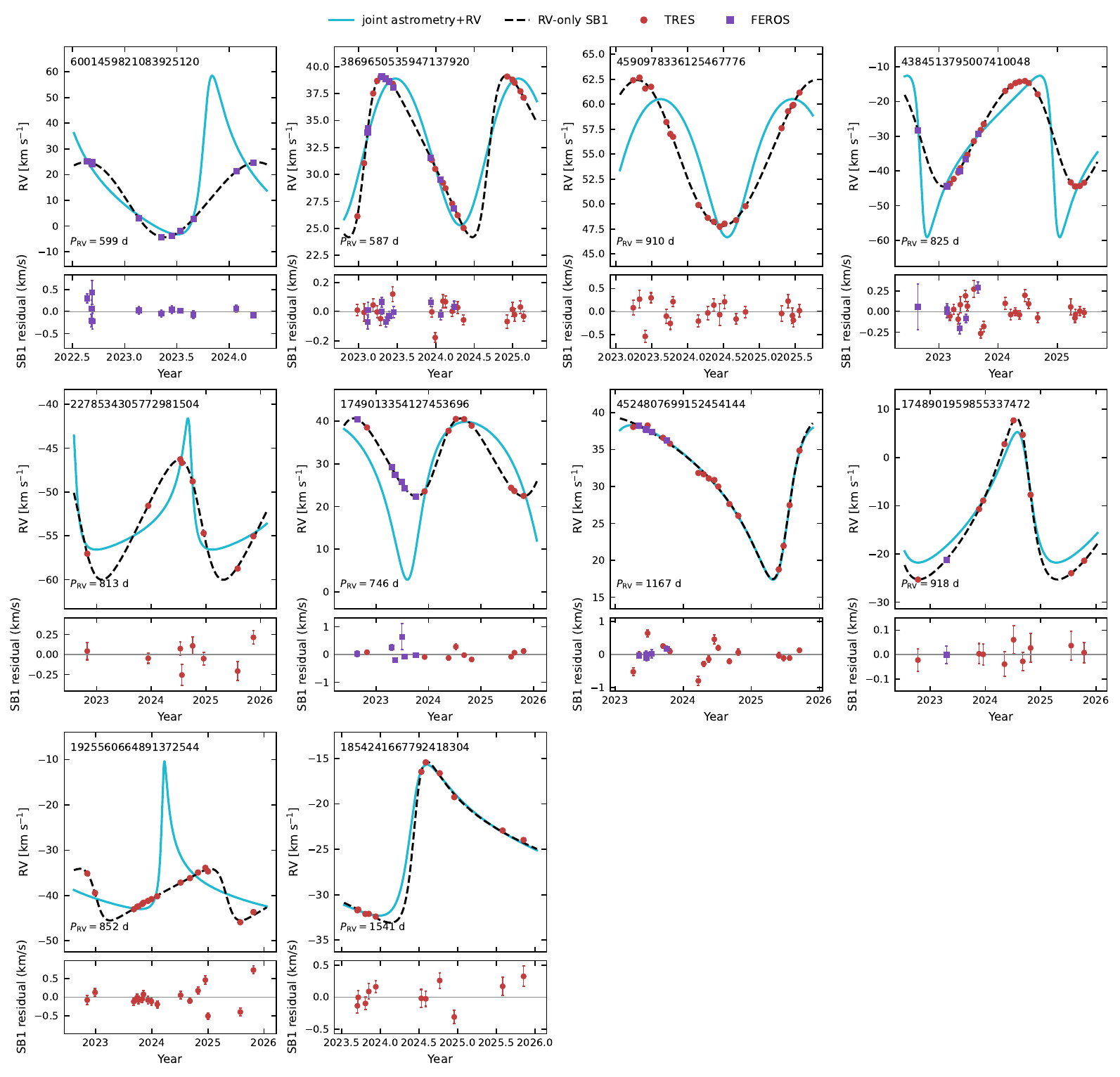}
    \caption{
    Joint astrometry+RV fits compared to RV-only fits for the 10 sources with RVs inconsistent with their astrometric solutions. Cyan curves show the best-fit joint astrometry+RVs model, while dashed black curves show the best-fit RV-only model.  Lower sub-panels show residuals relative to the SB1-only model. In most cases these are relatively small, implying that the SB1 model provides a reasonably good fit, but larger than for the sources in Figure~\ref{fig:astrometric-good-joint-rv-panels}. The SB1 solutions have orbital periods in the same ballpark as the {\it Gaia} solutions but significantly different orbital parameters. 
    }
    \label{fig:astrometric-flagged-joint-vs-sb1}
\end{figure*}

For the 10 systems where our joint fit yielded $\sigma_{\rm jit}/K_{\rm Gaia} > 0.025$, we also fit an unconstrained Keplerian SB1 model to the follow-up velocities alone, ignoring the astrometric solution. Figure~\ref{fig:astrometric-flagged-joint-vs-sb1} compares these spectroscopy-only fits to the joint fits.  In most of these cases, the RV-only orbit follows the data substantially better than the joint fit, indicating that the {\it Gaia} astrometric solution is not reliably describing the motion of the luminous star traced by the spectra.

One exception is source 4524807699152454144, for which the SB1 and joint fits are similar, and even the SB1 solution shows significant residuals. This suggests that the astrometric solution is likely reliable, but there is some additional RV scatter, the origin of which is uncertain. Another interesting case is 1748901959855337472, for which the SB1 and joint solutions predict RV curves of similar shape, but the SB1 solution better matches the observed RV amplitude. This suggests that the astrometric solution is reliable, but the flux ratio is nonzero (Section~\ref{sec:fluxratio-zpt}). The RV-only orbital solutions for these 10 systems are reported in Table~\ref{tab:astrometric-flagged-rv-only-orbits}. 

\begin{table*}[!t]
\centering
\caption{RV-only SB1 solutions for the systems in Figure~\ref{fig:astrometric-flagged-joint-vs-sb1}.}
\label{tab:astrometric-flagged-rv-only-orbits}
\footnotesize
\setlength{\tabcolsep}{3.0pt}
\resizebox{\textwidth}{!}{%
\begin{tabular}{lrrrrrrrrr}
\toprule
Gaia DR3 source ID & $P_{\rm orb}$ & $T_p-2457389$ & $e$ & $\omega$ & $K_1$ & $f_m$ & $\gamma$ & $\sigma_{\rm jit}$ & $\Delta v_{\rm FEROS}$ \\
 & (d) & (d) &  & (deg) & (km\,s$^{-1}$) & ($M_\odot$) & (km\,s$^{-1}$) & (km\,s$^{-1}$) & (km\,s$^{-1}$) \\
\midrule
3869650535947137920 & $586.8_{-1.4}^{+1.4}$ & $233.5_{-5.7}^{+5.4}$ & $0.366_{-0.003}^{+0.003}$ & $263.7_{-0.6}^{+0.6}$ & $7.48_{-0.03}^{+0.03}$ & $0.0205_{-0.0002}^{+0.0002}$ & $31.97_{-0.03}^{+0.03}$ & $0.049_{-0.012}^{+0.014}$ & $-0.10_{-0.03}^{+0.03}$ \\
6001459821083925120 & $600.6_{-6.3}^{+7.9}$ & $190.1_{-22.0}^{+17.6}$ & $0.070_{-0.004}^{+0.004}$ & $110.7_{-6.6}^{+7.4}$ & $14.59_{-0.05}^{+0.06}$ & $0.192_{-0.004}^{+0.005}$ & $10.60_{-0.13}^{+0.16}$ & $0.003_{-0.002}^{+0.103}$ & \nodata \\
1749013354127453696 & $747.7_{-2.2}^{+2.8}$ & $738.8_{-12.3}^{+5.4}$ & $0.136_{-0.008}^{+0.008}$ & $264.5_{-4.8}^{+3.6}$ & $9.17_{-0.07}^{+0.07}$ & $0.0582_{-0.0013}^{+0.0014}$ & $31.61_{-0.08}^{+0.09}$ & $0.172_{-0.049}^{+0.070}$ & $0.05_{-0.13}^{+0.15}$ \\
2278534305772981504 & $814.1_{-1.3}^{+3.1}$ & $808.6_{-11.1}^{+3.4}$ & $0.219_{-0.014}^{+0.011}$ & $86.8_{-3.2}^{+2.2}$ & $6.81_{-0.12}^{+0.09}$ & $0.0247_{-0.0013}^{+0.0010}$ & $-53.22_{-0.05}^{+0.07}$ & $0.008_{-0.006}^{+0.282}$ & \nodata \\
4384513795007410048 & $827.8_{-3.7}^{+1.6}$ & $0.7_{-3.1}^{+18.2}$ & $0.163_{-0.004}^{+0.004}$ & $123.0_{-1.4}^{+1.5}$ & $15.20_{-0.04}^{+0.05}$ & $0.289_{-0.003}^{+0.003}$ & $-27.99_{-0.06}^{+0.06}$ & $0.134_{-0.023}^{+0.030}$ & $-0.21_{-0.08}^{+0.08}$ \\
1925560664891372544 & $855.4_{-3.1}^{+5.7}$ & $844.5_{-12.5}^{+5.5}$ & $0.443_{-0.024}^{+0.025}$ & $92.2_{-3.4}^{+3.5}$ & $5.68_{-0.17}^{+0.19}$ & $0.0117_{-0.0009}^{+0.0010}$ & $-39.65_{-0.09}^{+0.10}$ & $0.327_{-0.067}^{+0.089}$ & \nodata \\
4590978336125467776 & $913.5_{-16.2}^{+19.9}$ & $52.4_{-30.7}^{+36.6}$ & $0.051_{-0.013}^{+0.015}$ & $54.5_{-16.7}^{+26.7}$ & $7.33_{-0.09}^{+0.09}$ & $0.0370_{-0.0018}^{+0.0021}$ & $54.90_{-0.16}^{+0.19}$ & $0.190_{-0.054}^{+0.067}$ & \nodata \\
1748901959855337472 & $917.8_{-2.8}^{+2.3}$ & $431.4_{-7.1}^{+8.4}$ & $0.397_{-0.002}^{+0.002}$ & $52.7_{-0.3}^{+0.3}$ & $16.72_{-0.03}^{+0.03}$ & $0.343_{-0.002}^{+0.002}$ & $-12.61_{-0.02}^{+0.02}$ & $0.002_{-0.001}^{+0.004}$ & $-0.13_{-0.08}^{+0.08}$ \\
4524807699152454144 & $1205.5_{-34.8}^{+64.1}$ & $1047.9_{-128.8}^{+70.0}$ & $0.491_{-0.010}^{+0.012}$ & $231.5_{-2.9}^{+2.6}$ & $11.18_{-0.25}^{+0.31}$ & $0.116_{-0.009}^{+0.014}$ & $32.07_{-0.30}^{+0.43}$ & $0.316_{-0.058}^{+0.076}$ & $0.06_{-0.20}^{+0.21}$ \\
1854241667792418304 & $1517.6_{-69.3}^{+19.4}$ & $48.8_{-37.1}^{+132.4}$ & $0.668_{-0.032}^{+0.022}$ & $286.4_{-3.4}^{+2.2}$ & $8.81_{-0.25}^{+0.29}$ & $0.0441_{-0.0011}^{+0.0015}$ & $-25.92_{-0.10}^{+0.21}$ & $0.139_{-0.138}^{+0.141}$ & \nodata \\
\bottomrule
\end{tabular}
}
\begin{minipage}{0.96\textwidth}
\vspace{2pt}\footnotesize
\textit{Note.} Values are posterior medians with 16th--84th percentile intervals. These fits use only the follow-up RVs. $\Delta v_{\rm FEROS}$ is the FEROS-minus-TRES offset. 
\end{minipage}
\end{table*}

\begin{figure*}[!t]
    \centering
    \includegraphics[width=0.92\textwidth]{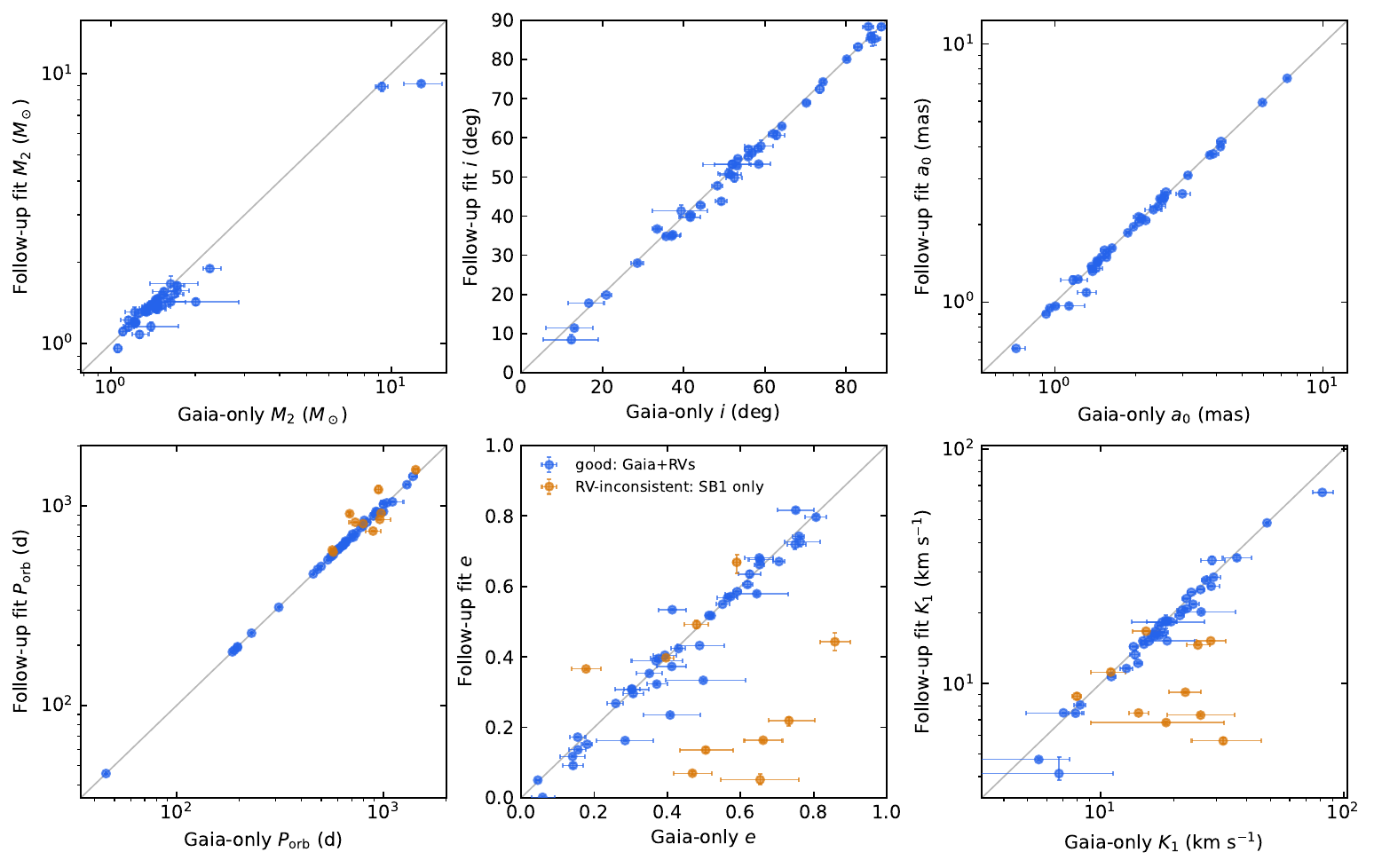}
    \caption{Comparison of parameters for systems in the astrometric sample inferred from the {\it Gaia} solutions alone compared to those constrained by our RV follow-up. Blue points show results of joint astrometry+RV fits for sources with consistent RVs and astrometry.  Orange points show RV-inconsistent systems, for which the follow-up values come from RV-only SB1 fits. For systems with consistent astrometry and RVs, the joint fits yield much tighter constraints than are achievable with {\it Gaia} data alone. The inferred companion masses are generally consistent with the {\it Gaia}-only constraints.    }
    \label{fig:astrometric-followup-vs-gaia-params}
\end{figure*}

In Figure~\ref{fig:astrometric-followup-vs-gaia-params}, we compare the orbital parameters inferred from the {\it Gaia} solutions alone to those measured with our follow-up RVs. For the 41 systems in Figure~\ref{fig:astrometric-good-joint-rv-panels} with ``good'' astrometric solutions, we show parameters from the joint astrometry+RV fits in blue. For the 10 systems in Figure~\ref{fig:astrometric-flagged-joint-vs-sb1} with inconsistent astrometry and RVs, we instead compare the {\it Gaia} solutions to the RV-only fits. For the astrometry+RV fits, the parameters inferred from the joint fit are generally consistent with the {\it Gaia}-only parameters, but the uncertainties are much smaller. For the RV-only fits, the parameters inferred from our follow-up are, unsurprisingly, inconsistent with the {\it Gaia} parameters. 

\begin{figure*}[!t]
    \centering
    \includegraphics[width=0.70\textwidth]{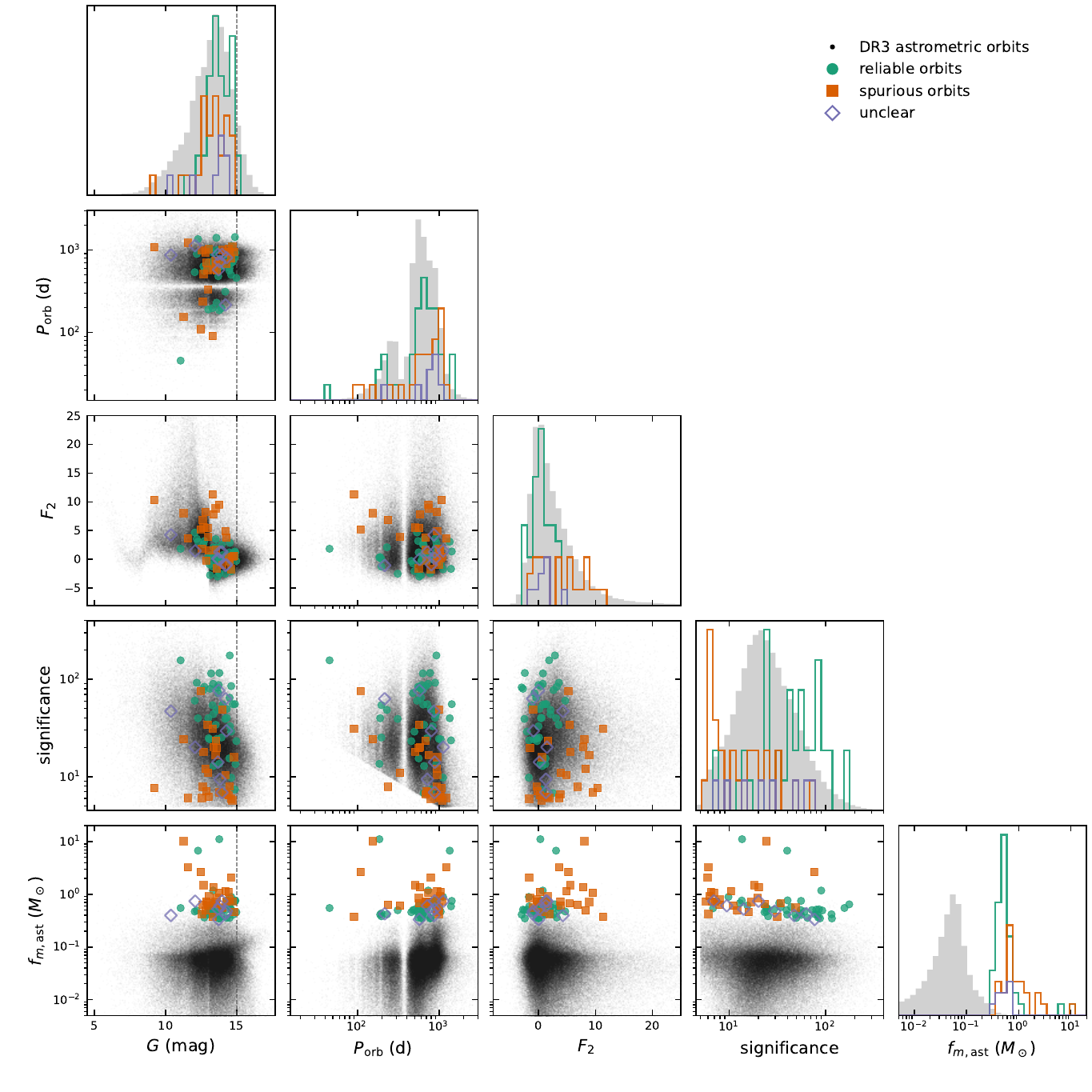}
    \caption{
    Comparison of the astrometric sample (Section~\ref{sec:astrometric_candidates}; colored symbols) to the full DR3 astrometric orbit catalog (black points). Green circles mark sources whose {\it Gaia} astrometric orbits appear reliable, orange squares mark sources whose astrometric orbits were invalidated by RV follow-up, and open purple diamonds show uncertain cases. Vertical dashed line marks the $G=15$ observability cut. Our sample targets objects with the largest astrometric mass functions, $f_{m,\,{\rm ast}}$. Sources with high \texttt{significance} and low $F_2$ are more likely to have reliable orbits, but there is no simple set of cuts that cleanly divides reliable and spurious orbits.}
    \label{fig:astrometric-catalog-status}
\end{figure*}

Figure~\ref{fig:astrometric-catalog-status} shows our astrometric sample in the context of the full DR3 astrometric orbit sample, separating sources with reliable  (green) and spurious (orange) orbits as determined by our follow-up. Sources in the astrometric sample with orbits of uncertain reliability -- primarily those for which we obtained few or no RVs -- are marked with open purple diamonds. 

A majority of sources with \texttt{significance > 20} have reliable orbits, while spurious sources preferentially have lower \texttt{significance}.  The typical \texttt{goodness\_of\_fit} statistic, $F_2$, is significantly larger for sources with $G< 13$ than for sources with $G>13$, likely due to the change in {\it Gaia} window class at $G=13$ \citep{Rowell2021, Lindegren2021}. Most sources in our sample with $F_2 > 6$ and $G < 13$ turned out to have spurious orbits, as did sources with $F_2 > 4$ and $G > 13$. The fraction of spurious orbits does not depend strongly on $G$ or $P_{\rm orb}$ within our sample. 

The figure shows that it is not possible to clearly separate reliable and spurious orbits based on {\it Gaia} quality metrics alone. However, a majority of spurious solutions can be discarded with a cut in the \texttt{significance} versus $F_2$ plane \citep[see also][]{Simon2026}.

Our follow-up suggests that about 60\% of all astrometric candidates have reliable orbits. Sources with spurious solutions were preferentially dropped from follow-up, and thus a smaller fraction of them have complete orbits. We emphasize that the fraction of spurious orbits in our astrometric sample is very likely higher than in the full DR3 astrometric orbit catalog: true solutions with high $f_m$ are rare, so a small number of spurious solutions will represent a disproportionate fraction of the total sample at high $f_m$.  

\subsubsection{Flux ratios}
\label{sec:fluxratio-zpt}
Thus far, our joint astrometry+RV fits have assumed that the companion is dark. This is generally a reasonable assumption because most sources in the sample are ``class III'' sources for which the astrometric triage algorithm \citep{Shahaf2019, Shahaf2023} requires the companion to be dark. However, it remains instructive to leave the flux ratio a free parameter in order to assess whether the RV and astrometric orbits are consistent.

If the companion contributes $G$-band light, the photocenter no longer traces the primary, but falls between the two components. In this case, the angular photocenter semimajor axis is
\begin{equation}
    \mathring{a}_0
    =
    \varpi a_1
    \left[
        1
        -
        \frac{f_G(1+q)}{q(1+f_G)}
    \right],
    \label{eq:photocenter-flux-ratio}
\end{equation}
where $a_1$ is the physical semimajor axis of the primary's orbit (Equation~\ref{eq:a1}), $q=M_2/M_1$ is the mass ratio, and $f_G=F_{2,G}/F_{1,G}$ is the $G$-band flux ratio, with $F_{1,G}$ and $F_{2,G}$ representing the flux from the primary and secondary in the $G$-band. We now repeat the joint fits for sources with consistent astrometry and RVs, varying $f_G$ as a free parameter. For a binary containing a dark companion and systematic-free astrometry, the inferred $f_G$ should be consistent with 0. 

\begin{figure*}[!t]
    \centering
    \includegraphics[width=0.90\textwidth]{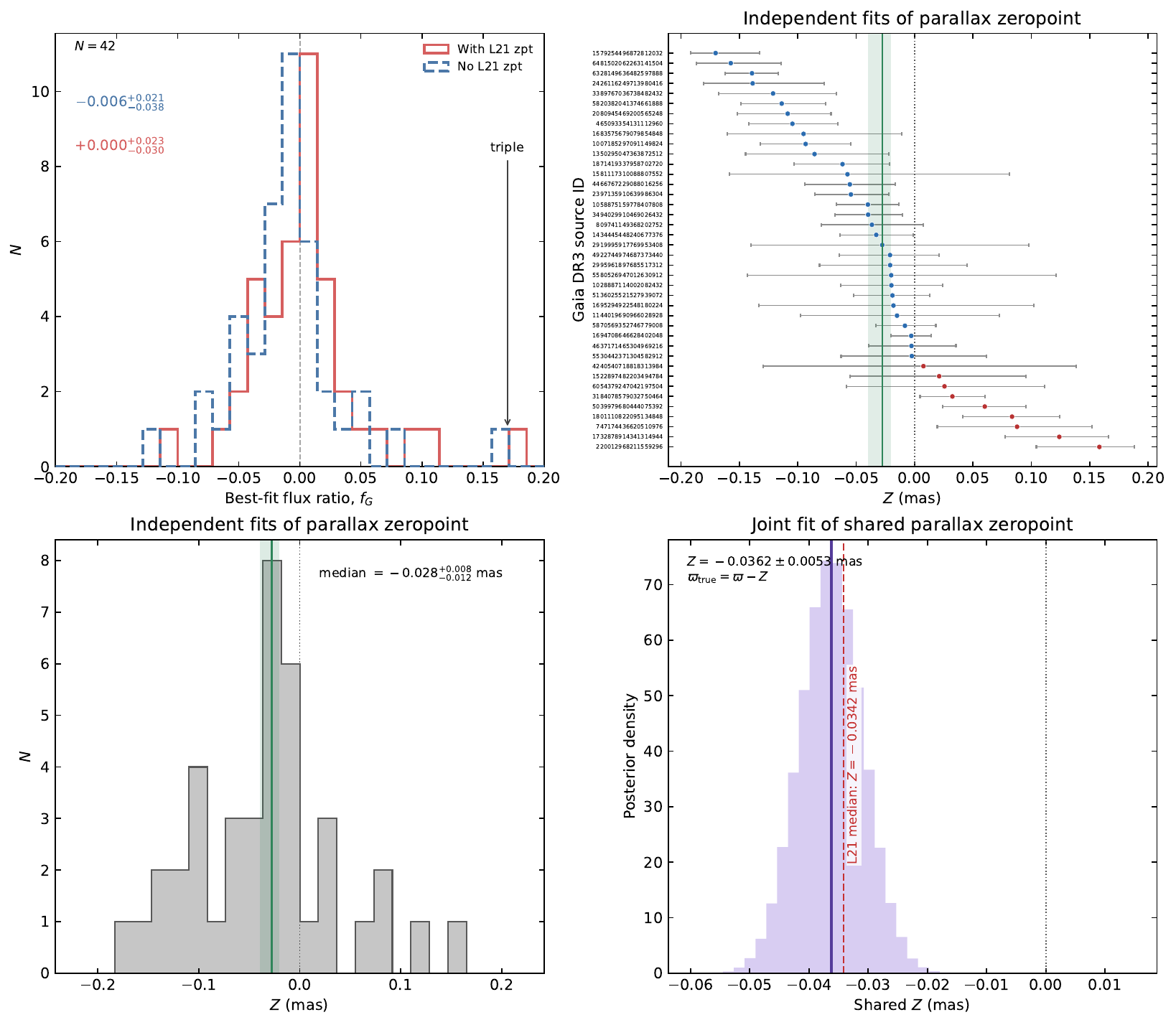}
    \caption{
    Constraints on the flux ratio and parallax zeropoint from joint astrometry+RV fits. Upper-left panel shows the best-fit $G$-band flux ratios inferred with the raw DR3 parallaxes (blue dashed) and after applying the zeropoint correction from \citet{Lindegren2021} (red). Without the parallax zeropoint correction, the distribution is peaked at an unphysical negative flux ratio. Applying the \citet{Lindegren2021} correction -- which is based on single-star solutions -- shifts the distribution's peak to zero. Upper-right and bottom left panels show the zeropoints $Z_i$ and their confidence intervals inferred independently for each of 40 binaries. Lower-right panel shows constraints on the zeropoint inferred from a joint fit of all 40 binaries. Red dashed line marks the median \citet{Lindegren2021} correction for these same sources, which is very similar to the value we infer.}
    \label{fig:fluxratio-zpt-summary}
\end{figure*}

The results are shown with a blue histogram in the left panel of Figure~\ref{fig:fluxratio-zpt-summary}. There is one source, {\it Gaia} DR3 1748901959855337472, with inferred $f_{\rm G} = 0.168 \pm 0.010$; as we show in Section~\ref{sec:triples}, this source is a hierarchical triple with a luminous inner binary, so a nonzero flux ratio is expected. This source has consistent astrometry and RVs once the flux ratio is allowed to vary, raising the number of good solutions to 42. The source  {\it Gaia} DR3  220012968211559296 has the next-highest inferred flux ratio, $f_{\rm G} = 0.055 \pm 0.019$, suggesting either a companion that contributes $\approx 6\%$ of the light or astrometry modestly affected by systematics. All remaining sources have flux ratios consistent with or smaller than 0. Several sources have best-fit flux ratios significantly below zero. As we show below, this is partially due to the parallax zeropoint, but the most extreme cases likely reflect astrometric systematics. The source with the smallest inferred $f_G$ is Gaia BH1, with $f_G = -0.13\pm 0.02$, perhaps reflecting the incomplete orbital phase coverage of the source's DR3 astrometry \citep{ElBadry2023BH1}. 

Negative flux ratios are unphysical but mathematically allowed: they imply that the RV variability amplitude is smaller than predicted by the astrometry. The distribution of best-fit $f_G$ values is centered on $-0.006$, and 30 of 42 sources have a best-fit $f_G < 0$. This suggests a small but systematic offset between the astrometry and RV orbits, which we show below to be naturally explained by the {\it Gaia} parallax zeropoint. 

We first try simply applying the nominal DR3 parallax zeropoint correction inferred by \citet{Lindegren2021}. We  replace the parallax in the {\it Gaia} likelihood with a corrected value, $\varpi_{\rm corr}$,
\begin{equation}
    \varpi_{\rm corr} = \varpi - Z(G,\nu_{\rm eff},\beta,N_{\rm par}).
    \label{eq:nominal-zpt-correction}
\end{equation}
Here $Z$ represents the parallax zeropoint, which is a function of apparent magnitude, color, ecliptic latitude, and the type of single-star solution in the {\it Gaia} archive (i.e., 5- vs. 6-parameter solutions). The likelihood then compares $\varpi_{\rm corr}$ with $\varpi_{\rm true}= 1/d$ for a binary of true distance $d$.

The red histogram in the upper left panel of Figure~\ref{fig:fluxratio-zpt-summary} shows the distribution of best-fit inferred $f_G$ obtained by repeating the joint fits after applying the zeropoint correction from \citet{Lindegren2021}. This distribution is now centered on zero, suggesting that correcting for the zeropoint improves the parallaxes for sources with orbital solutions, at least on average. Below, we {\it infer} the zeropoint for 12- and 15-parameter solutions directly.

\subsubsection{Measuring the parallax zeropoint}
Astrometry provides a measurement of the size of a binary orbit that scales linearly with distance. RVs, on the other hand, provide a distance-independent measurement of the orbit size, provided that the inclination is known. Since astrometry also constrains the inclination, astrometric orbits for binaries with a dark companion provide an opportunity to directly measure distance, independent of parallax. It is this fact that makes it possible to measure the distance to the Galactic center based on the orbits of stars orbiting the Galaxy's central black hole \citep[e.g.][]{GRAVITYCollaboration2019, Do2019}. 

Most previous work on sources with {\it Gaia} astrometric orbits -- including our own -- has not applied a parallax zeropoint correction \citep{Lindegren2021}, because it was not clear that this correction, which is measured based on single-star astrometric solutions, should still apply to binary solutions. Here, we infer a global parallax zeropoint correction for binaries with astrometric orbits. 

We repeated the dark-companion astrometry+RV fits for the 40 sources with well-constrained RV curves and no evidence for a luminous companion, excluding the hierarchical triple {\it Gaia} DR3 1748901959855337472 and Gaia BH1. 35 of these sources have \texttt{Orbital} solutions; five have \texttt{AstroSpectroSB1} solutions. We first tried fitting a separate zeropoint $Z_i$ for each binary, such that the true parallax of the $i$th binary is compared to $\varpi_i - Z_i$. The upper right panel of Figure~\ref{fig:fluxratio-zpt-summary} summarizes the results. The best-fit zeropoint is negative for 31 of the 40 sources, though 14 of these $Z_i$ are individually consistent with 0. Bootstrapping across all sources, we find a median inferred zeropoint of $Z=-0.028^{+0.008}_{-0.012}\,{\rm mas}$, which is shown with a green shaded region in the upper right panel of Figure~\ref{fig:fluxratio-zpt-summary}. A distribution of the best-fit zeropoints is shown in the bottom left panel. 

Finally, we fit all 40 binaries simultaneously with a single zeropoint that is shared across all sources. This entailed a nonlinear optimization problem in 641 dimensions, but it is tractable because the parameters of individual binaries are not strongly covariant. Optimization yields a best-fit joint zeropoint of $Z=-0.0362\pm0.0053\,{\rm mas}$, where the uncertainty is calculated from the Fisher matrix. This value is quite similar to the median $Z=-0.0342\,{\rm mas}$ for the same 40 sources calculated with the prescription from \citet{Lindegren2021}; the latter is shown by the red dashed line in the lower-right panel of Figure~\ref{fig:fluxratio-zpt-summary}. The posterior probability for this best-fit solution is higher by $\Delta\ln P=14.3$ than the value obtained when we set the zeropoint for each source according to the \citet{Lindegren2021} prescription. 
 
Several other works have investigated the DR3 parallax zeropoint using a variety of astrophysical standard candles \citep[e.g.][]{Zinn2021, Groenewegen2021, Huang2021, Ren2021, Bobylev2025}. All of these have found global zeropoints with amplitudes of $0.02-0.04$ mas, consistent with our results. However, all these works have focused on five- and six-parameter solutions for single stars. We conclude that the 12- and 15-parameter solutions for astrometric binaries in DR3 have a similar zeropoint to 5- and 6-parameter single-star solutions. \citet{Nagarajan2024_wbs} reached a similar conclusion by analyzing hierarchical triples containing an astrometric binary with a wide, co-moving companion; here, we infer the zeropoint directly from the astrometric binaries. 

The 40 sources analyzed here have a relatively narrow range of color and magnitude, with 68\% falling within $0.75 < G_{\rm BP} - G_{\rm RP} < 1.16$ and $13 < G < 14.58$. The corresponding ranges for all sources with \texttt{Orbital} and \texttt{AstroSpectroSB1} solutions in DR3 are $0.75 < G_{\rm BP} - G_{\rm RP} < 1.44$ and $11.53 < G< 14.53$. Our sample is thus slightly fainter and bluer than the full sample of sources with astrometric binary solutions, likely because compact object binaries can be detected to larger distances than luminous binaries and because our primary selection excluded red giants. After subtracting the  \citet{Lindegren2021} correction,  we find no significant correlation in the zeropoint residuals with either magnitude or color. We also find no significant difference between the zeropoints for 12- and 15-parameter solutions. Nevertheless, we cannot rule out trends in color or magnitude beyond the regime probed by our sample. \citet{Zinn2021} found a $\sim 15\mu$as bias in the \citet{Lindegren2021} correction for 5-parameter sources with $G < 10.8$. Our sample does not contain any sources this bright, so we cannot assess whether a similar bias exists for binary solutions.

\begin{figure}[!tbp]
    \centering
    \includegraphics[width=\columnwidth]{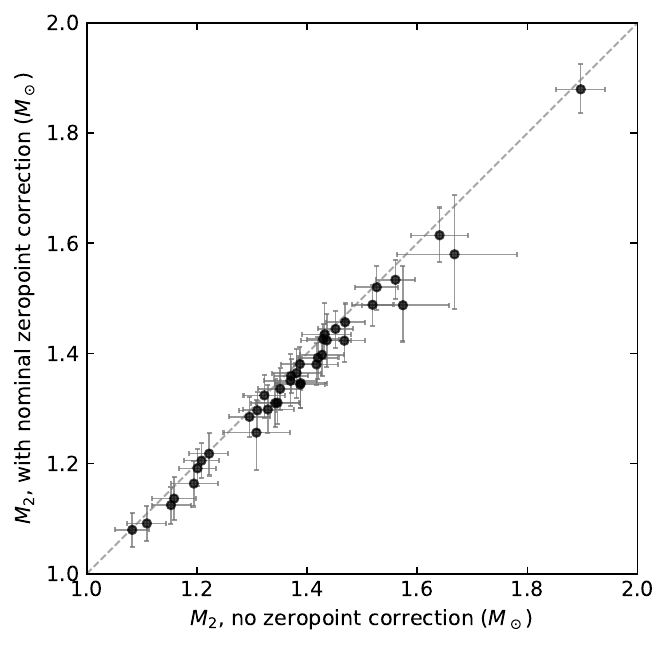}
    \caption{
    Effect of the \citet{Lindegren2021} parallax zeropoint correction on the companion masses inferred from joint astrometry+RV fits. Applying the zeropoint correction decreases the inferred masses slightly but systematically because it places the binaries at smaller distances. The median shift is $-0.018\,M_\odot$.}
    \label{fig:zpt-mass-comparison}
\end{figure}

Applying the parallax zeropoint correction slightly changes the inferred companion masses, since it brings the binaries closer and reduces the implied physical size of their astrometric orbits. Figure~\ref{fig:zpt-mass-comparison} compares $M_2$ constraints obtained with and without the \citet{Lindegren2021} parallax zeropoint correction. To highlight small changes resulting from the correction, we only show binaries with inferred $M_2$ between $1\,M_\odot$ and $2\,M_\odot$, excluding Gaia BH1 and BH2. Applying the correction results in changes in $M_2$ that are in almost all cases smaller than the formal fitting uncertainties but systematically shift $M_2$ toward lower values. The zeropoint correction has a larger effect on systems at larger distances, where a fixed parallax shift represents a larger shift in the absolute distance. We can thus anticipate that accounting for the zeropoint will be particularly important for binaries discovered with {\it Gaia} DR4 data, which will be at larger average distances than those discovered with DR3 \citep{El-Badry2024_genmod}.

The corresponding orbital parameters after applying the source-dependent \citet{Lindegren2021} correction are reported in Table~\ref{tab:astrometric-joint-dark-l21-zpt-orbits}. As in Table~\ref{tab:astrometric-joint-dark-orbits}, the printed table shows the first eight rows, while the machine-readable version contains all 41 systems and all fitted and derived parameters.

\begin{table*}[!t]
\centering
\caption{Joint astrometry+RV solutions for reliable astrometric orbits, assuming a dark companion and applying the source-dependent parallax zeropoint correction from \citet{Lindegren2021}. The complete table is available in machine-readable form.}
\label{tab:astrometric-joint-dark-l21-zpt-orbits}
\scriptsize
\setlength{\tabcolsep}{3.2pt}
\begin{tabular}{lrrrrrrrr}
\toprule
Gaia DR3 source ID & $P_{\rm orb}$ & $e$ & $i$ & $\mathring{a}_0$ & $\varpi_{\rm corr}$ & $M_2$ & $\gamma$ & $\sigma_{\rm jit}$ \\
 & (d) &  & (deg) & (mas) & (mas) & ($M_\odot$) & (km\,s$^{-1}$) & (km\,s$^{-1}$) \\
\midrule
1522897482203494784 & $45.519_{-0.001}^{+0.001}$ & $0.050_{-0.000}^{+0.000}$ & $80.1_{-0.3}^{+0.3}$ & $2.565_{-0.005}^{+0.005}$ & $12.478_{-0.015}^{+0.015}$ & $1.32_{-0.04}^{+0.04}$ & $2.81_{-0.01}^{+0.02}$ & $0.002_{-0.001}^{+0.010}$ \\
4373465352415301632 & $185.39_{-0.01}^{+0.01}$ & $0.432_{-0.000}^{+0.000}$ & $53.8_{-0.3}^{+0.3}$ & $2.657_{-0.014}^{+0.014}$ & $2.136_{-0.016}^{+0.016}$ & $9.01_{-0.12}^{+0.13}$ & $48.90_{-0.03}^{+0.03}$ & $0.056_{-0.010}^{+0.011}$ \\
2995961897685517312 & $189.13_{-0.03}^{+0.03}$ & $0.388_{-0.001}^{+0.001}$ & $53.7_{-0.9}^{+0.9}$ & $1.233_{-0.013}^{+0.014}$ & $2.532_{-0.014}^{+0.013}$ & $1.34_{-0.04}^{+0.04}$ & $41.16_{-0.01}^{+0.01}$ & $0.003_{-0.003}^{+0.017}$ \\
220012968211559296 & $194.73_{-0.05}^{+0.05}$ & $0.235_{-0.004}^{+0.006}$ & $88.5_{-0.7}^{+0.7}$ & $1.610_{-0.019}^{+0.014}$ & $3.262_{-0.021}^{+0.021}$ & $1.22_{-0.04}^{+0.04}$ & $-4.67_{-0.10}^{+0.11}$ & $0.244_{-0.092}^{+0.263}$ \\
2080945469200565248 & $196.76_{-0.07}^{+0.07}$ & $0.002_{-0.002}^{+0.004}$ & $85.9_{-1.6}^{+1.5}$ & $0.673_{-0.005}^{+0.005}$ & $1.427_{-0.010}^{+0.010}$ & $1.43_{-0.03}^{+0.03}$ & $-14.49_{-0.09}^{+0.08}$ & $0.247_{-0.069}^{+0.127}$ \\
6481502062263141504 & $230.22_{-0.08}^{+0.07}$ & $0.307_{-0.002}^{+0.003}$ & $50.9_{-1.1}^{+1.2}$ & $0.955_{-0.012}^{+0.012}$ & $1.790_{-0.019}^{+0.018}$ & $1.30_{-0.04}^{+0.05}$ & $-5.76_{-0.02}^{+0.02}$ & $0.006_{-0.006}^{+0.049}$ \\
5820382041374661888 & $310.17_{-0.11}^{+0.11}$ & $0.534_{-0.002}^{+0.002}$ & $83.1_{-0.9}^{+0.9}$ & $0.917_{-0.009}^{+0.009}$ & $1.371_{-0.012}^{+0.012}$ & $1.38_{-0.03}^{+0.03}$ & $-42.98_{-0.10}^{+0.12}$ & $0.011_{-0.010}^{+0.053}$ \\
2919995917769953408 & $456.12_{-0.13}^{+0.12}$ & $0.811_{-0.008}^{+0.006}$ & $60.7_{-1.0}^{+1.0}$ & $2.129_{-0.042}^{+0.040}$ & $2.337_{-0.022}^{+0.023}$ & $1.26_{-0.07}^{+0.06}$ & $70.78_{-0.08}^{+0.07}$ & $0.004_{-0.003}^{+0.030}$ \\
\ldots & \ldots & \ldots & \ldots & \ldots & \ldots & \ldots & \ldots & \ldots \\
\bottomrule
\end{tabular}
\begin{minipage}{0.96\textwidth}
\vspace{2pt}\footnotesize
\textit{Note.} Values are posterior medians with 16th--84th percentile intervals. These fits fix the $G$-band flux ratio to zero and apply the source-dependent \citet{Lindegren2021} correction, $\varpi_{\rm corr}=\varpi-Z$. The machine-readable table also reports $T_p$, $\omega$, $\Omega$, $M_1$, $K_1$, and the FEROS--TRES offset.
\end{minipage}
\end{table*}

\subsection{Spectroscopic orbits}
\label{sec:spectroscopic-orbit-solutions}

\begin{figure*}[!tbp]
    \centering
    \includegraphics[width=\textwidth,height=0.93\textheight,keepaspectratio]{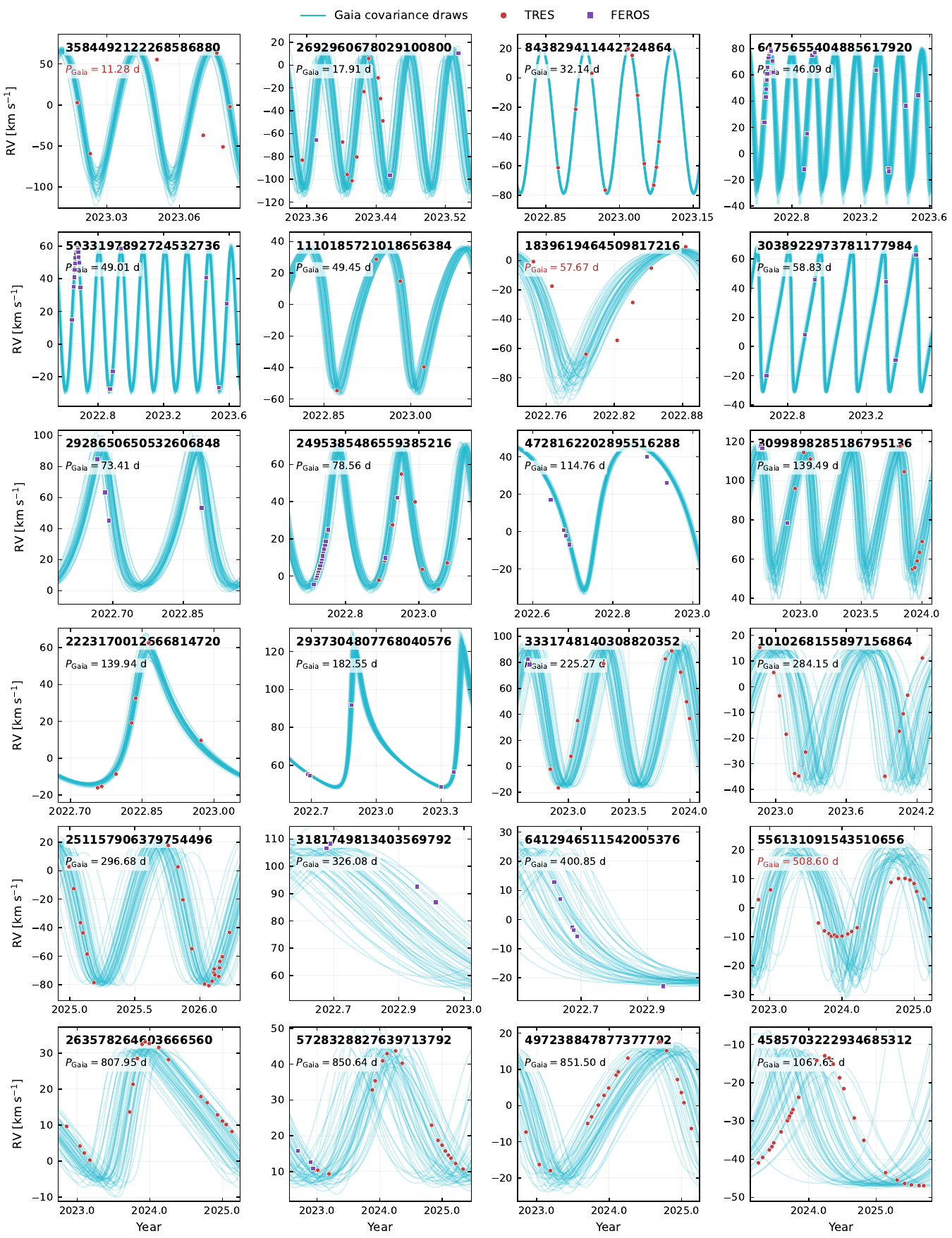}
    \caption{Observed and {\it Gaia}-predicted RVs for sources in the SB1 sample. Cyan curves show predictions for draws from the {\it Gaia} covariance matrix; red circles and purple squares show TRES and FEROS RVs. {\it Gaia} period labels are shown in red for the three sources whose SB1 solutions are inconsistent with our RVs. Fits to the follow-up RVs are shown in Figure~\ref{fig:sb1-spectroscopy-only-rv-panels}.}
    \label{fig:sb1-gaia-rv-panels}
\end{figure*}

We next analyze the spectroscopic candidates. As described in Section~\ref{sec:sb1_cands}, the sample contains 151 sources with SB1 orbits, but a majority of these were discarded without multi-epoch follow-up because they have composite SEDs, double-lined spectra, or other evidence of a luminous companion. We obtained multi-epoch RVs for 24 sources. 

Figure~\ref{fig:sb1-gaia-rv-panels} compares most of these RVs to predictions for draws from the DR3 covariance matrix for each source. A few sources have additional RVs that are not shown because they were obtained at significantly later times and do not fit within the chosen time interval. Unlike with the astrometric solutions, the SB1 solutions make a direct prediction for the RV time series of each source, with no ambiguity in inclination, flux ratio, or center-of-mass RV. As a result, even a single RV can test these solutions \citep[e.g.][]{Bashi2022}. 

21 of the 24 sources show RVs broadly consistent with predictions of the {\it Gaia} SB1 solutions. Sources 3584492122268586880, 1839619464509817216, and 556131091543510656 do not, and have RV mass functions significantly lower than predicted by the {\it Gaia} solutions. The purity of the full SB1 sample is likely lower than 21/24, because many sources with discrepant RVs and {\it Gaia} solutions were dropped early in the program (Table~\ref{tab:initial-sb1-candidates}). Considering all sources in the SB1 sample for which we obtained follow-up RVs that can meaningfully test the {\it Gaia} orbit, the fraction of sources with good solutions is 28/54 $\approx 52\%$. 

\begin{figure*}[!tbp]
    \centering
    \includegraphics[width=\textwidth,height=0.85\textheight,keepaspectratio]{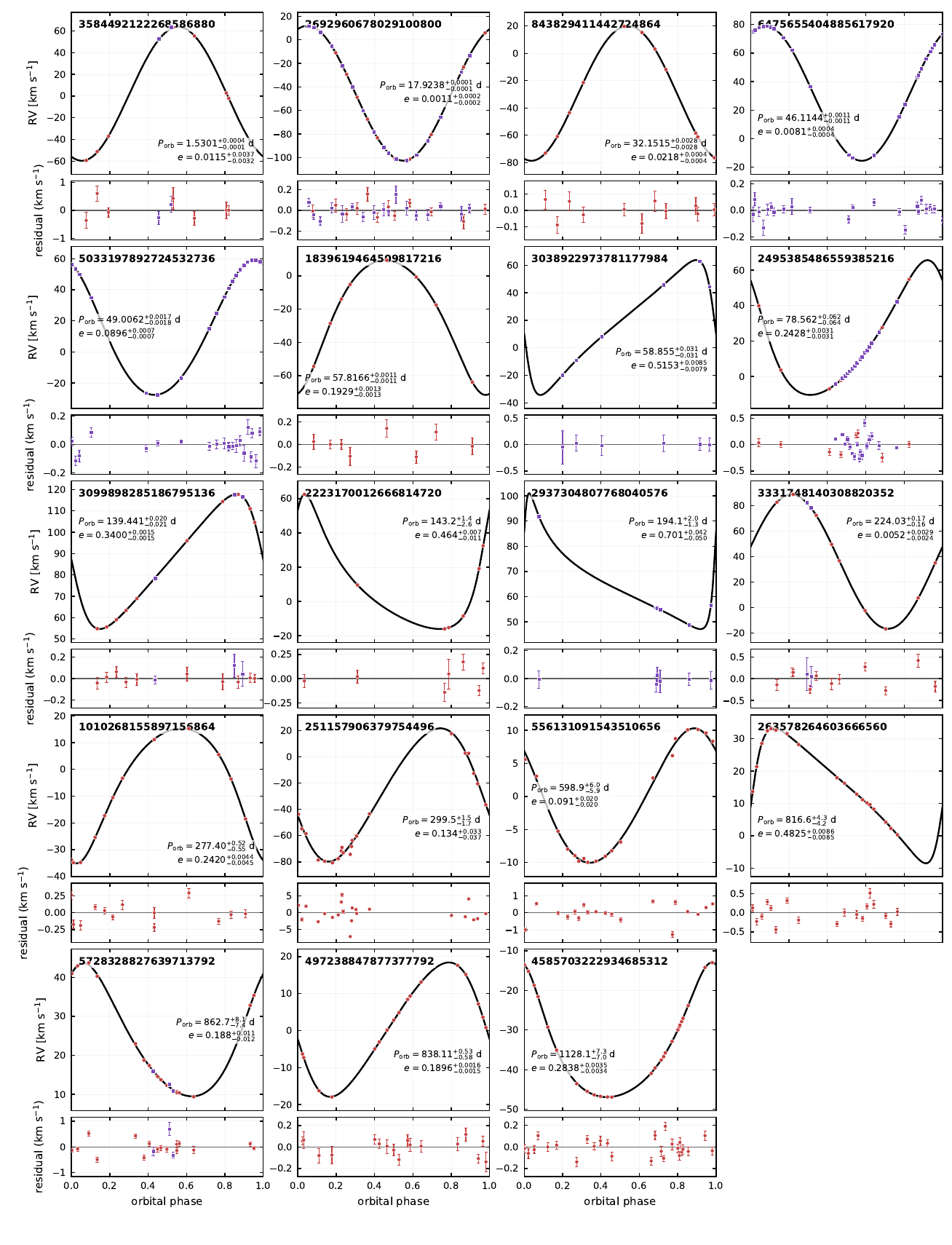}
    \caption{RV-only orbital fits for 19 candidates in the SB1 sample with sufficient RV follow-up for an independent fit. Black curves show the maximum-likelihood solution; lower panels show residuals. Red circles and purple squares show TRES and FEROS RVs. The posterior median and 16th--84th percentile interval for $P_{\rm orb}$ and $e$ from these fits are reported in each panel. 
    }
    \label{fig:sb1-spectroscopy-only-rv-panels}
\end{figure*}

19 sources in the SB1 sample have sufficient follow-up RVs and phase coverage for the orbit to be well-constrained by the follow-up RVs alone. We fit these with a Keplerian orbit, sampled with \texttt{emcee}, again including a jitter term and a TRES-FEROS offset where applicable. The results are shown in Figure~\ref{fig:sb1-spectroscopy-only-rv-panels} and reported in Table~\ref{tab:sb1-spectroscopy-only-orbits}. The fits are in most cases good, in the sense that the residuals are small relative to the observed variability amplitudes and are not structured. Several sources do, however, have typical RV residuals larger than the formal RV uncertainties, suggesting underestimated uncertainties or additional RV scatter. The source with the largest RV scatter is {\it Gaia} DR3 251157906379754496, for which we infer $\sigma_{\rm jit} = 2.5\,{\rm km\,s^{-1}}$. This source is a post-mass transfer system containing an evolved star and a companion with strong emission lines (Figure~\ref{fig:spectral-model-examples}). We suspect the measured RVs are reliable and trace intrinsic variability in the giant's photosphere. 

\begin{table*}[!t]
\centering
\caption{RV-only SB1 solutions for the systems in Figure~\ref{fig:sb1-spectroscopy-only-rv-panels}. The complete table is available in machine-readable form.}
\label{tab:sb1-spectroscopy-only-orbits}
\footnotesize
\setlength{\tabcolsep}{3.0pt}
\resizebox{\textwidth}{!}{%
\begin{tabular}{lrrrrrrrrr}
\toprule
Gaia DR3 source ID & $P_{\rm orb}$ & $T_p-2457389$ & $e$ & $\omega$ & $K_1$ & $f_m$ & $\gamma$ & $\sigma_{\rm jit}$ & $\Delta v_{\rm FEROS}$ \\
 & (d) & (d) &  & (deg) & (km\,s$^{-1}$) & ($M_\odot$) & (km\,s$^{-1}$) & (km\,s$^{-1}$) & (km\,s$^{-1}$) \\
\midrule
3584492122268586880 & $1.5301_{-0.0001}^{+0.0002}$ & $0.50_{-0.25}^{+0.22}$ & $0.011_{-0.003}^{+0.003}$ & $160.7_{-18.5}^{+19.9}$ & $61.58_{-0.22}^{+0.25}$ & $0.0370_{-0.0004}^{+0.0005}$ & $2.74_{-0.18}^{+0.18}$ & $0.003_{-0.002}^{+0.518}$ & $0.99_{-0.56}^{+0.61}$ \\
2692960678029100800 & $17.9238_{-0.0001}^{+0.0001}$ & $7.88_{-0.41}^{+0.50}$ & $0.0011_{-0.0002}^{+0.0002}$ & $341.9_{-8.3}^{+10.1}$ & $57.22_{-0.01}^{+0.01}$ & $0.3480_{-0.0003}^{+0.0003}$ & $-45.39_{-0.02}^{+0.02}$ & $0.002_{-0.001}^{+0.020}$ & $-0.18_{-0.02}^{+0.02}$ \\
843829411442724864 & $32.1515_{-0.0028}^{+0.0028}$ & $6.26_{-0.27}^{+0.28}$ & $0.0218_{-0.0004}^{+0.0004}$ & $165.9_{-1.3}^{+1.3}$ & $49.40_{-0.02}^{+0.02}$ & $0.4014_{-0.0006}^{+0.0006}$ & $-28.27_{-0.02}^{+0.02}$ & $0.002_{-0.001}^{+0.004}$ & \nodata \\
6475655404885617920 & $46.1143_{-0.0011}^{+0.0010}$ & $35.76_{-0.41}^{+0.38}$ & $0.0081_{-0.0004}^{+0.0004}$ & $331.6_{-3.3}^{+3.1}$ & $47.30_{-0.02}^{+0.02}$ & $0.5055_{-0.0006}^{+0.0006}$ & $31.44_{-0.02}^{+0.02}$ & $0.044_{-0.012}^{+0.014}$ & \nodata \\
5033197892724532736 & $49.0062_{-0.0017}^{+0.0017}$ & $37.40_{-0.09}^{+0.09}$ & $0.0896_{-0.0007}^{+0.0007}$ & $20.1_{-0.5}^{+0.4}$ & $43.39_{-0.02}^{+0.02}$ & $0.4097_{-0.0007}^{+0.0007}$ & $12.08_{-0.02}^{+0.02}$ & $0.058_{-0.013}^{+0.017}$ & \nodata \\
1839619464509817216 & $57.8166_{-0.0011}^{+0.0011}$ & $0.79_{-0.05}^{+0.05}$ & $0.193_{-0.001}^{+0.001}$ & $191.2_{-0.3}^{+0.3}$ & $40.43_{-0.05}^{+0.05}$ & $0.3740_{-0.0014}^{+0.0014}$ & $-23.51_{-0.03}^{+0.03}$ & $0.003_{-0.001}^{+0.054}$ & \nodata \\
3038922973781177984 & $58.8555_{-0.0307}^{+0.0306}$ & $19.08_{-1.37}^{+1.37}$ & $0.515_{-0.008}^{+0.009}$ & $95.3_{-1.0}^{+1.0}$ & $48.99_{-0.64}^{+0.73}$ & $0.4513_{-0.0111}^{+0.0123}$ & $17.13_{-0.31}^{+0.29}$ & $0.002_{-0.001}^{+0.004}$ & \nodata \\
2495385486559385216 & $78.5620_{-0.0645}^{+0.0625}$ & $34.40_{-2.04}^{+2.10}$ & $0.243_{-0.003}^{+0.003}$ & $45.8_{-0.6}^{+0.6}$ & $38.06_{-0.17}^{+0.17}$ & $0.4097_{-0.0047}^{+0.0048}$ & $21.12_{-0.08}^{+0.08}$ & $0.178_{-0.028}^{+0.037}$ & $0.25_{-0.15}^{+0.15}$ \\
\ldots & \ldots & \ldots & \ldots & \ldots & \ldots & \ldots & \ldots & \ldots & \ldots \\
\bottomrule
\end{tabular}
}
\begin{minipage}{0.96\textwidth}
\vspace{2pt}\footnotesize
\textit{Note.} Values are posterior medians with 16th--84th percentile intervals. These fits use only the follow-up RVs. $\Delta v_{\rm FEROS}$ is the FEROS-minus-TRES offset. The machine-readable table contains all 19 systems.
\end{minipage}
\end{table*}

\begin{figure*}[!t]
    \centering
    \includegraphics[width=0.90\textwidth]{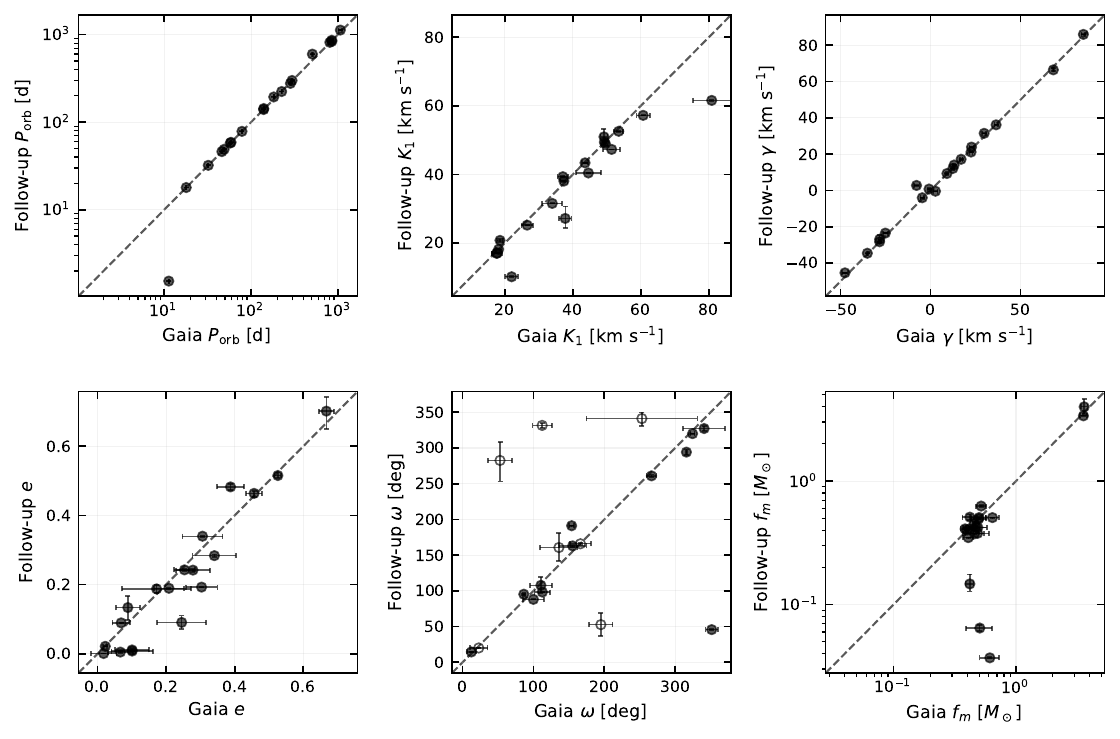}
    \caption{Best-fit SB1 solution parameters inferred from fits to our RVs (Figure~\ref{fig:sb1-spectroscopy-only-rv-panels}) compared to parameters of the DR3 SB1 solutions. Bottom right panel shows the RV mass function (Equation~\ref{eq:fm}), which is calculated from samples of $P_{\rm orb}$, $K_1$, and $e$.
 Open points in the $\omega$ panel have inferred $e<0.1$, such that $\omega$ is potentially ill-defined. One object has an inferred period completely different from the DR3 solution. The rest have periods similar to the DR3 solution, but two have significantly lower $K_1$ and thus $f_m$. Two systems have well-constrained $f_m > 3\,M_\odot$ after our follow-up (Section~\ref{sec:Be_stars}).}
    \label{fig:sb1-spectroscopy-vs-gaia-params}
\end{figure*}

Figure~\ref{fig:sb1-spectroscopy-vs-gaia-params} compares our inferred parameters for the 19 sources in Figure~\ref{fig:sb1-spectroscopy-only-rv-panels} to the {\it Gaia} parameters. Besides the three sources discussed above, most of the rest have inferred parameters consistent with the {\it Gaia} parameters: 13/16, 14/16, and 15/16 respectively have consistent $P_{\rm orb}$, $K_1$, and $e$ within $2\,\sigma$. The remaining discrepancies, while formally significant, are small in an absolute sense. We disregard sources with apparently large discrepancies in $\omega$ and near-circular orbits, since $\omega$ is ill-defined in such cases. The parameters from our follow-up in most cases have much smaller uncertainties than the {\it Gaia} parameters. For example, the median uncertainty in $K_1$ is 18 times smaller than the median {\it Gaia} uncertainty.

\subsubsection{SB1 candidates with astrometric orbits}
\label{sec:sb1_ast}
Two sources in the SB1 sample, {\it Gaia} DR3 263578264603666560 and 5728328827639713792, also have independent \texttt{Orbital} solutions. We jointly fit the astrometry and RVs of these sources following the method described in Section~\ref{sec:astrometric-joint-fits}. We applied the \citet{Lindegren2021} parallax zeropoint correction and allowed the $G$-band flux ratio, $f_G$, to vary. We infer $f_G=0.234\pm0.012$ and $0.129\pm0.010$, respectively, demonstrating that both systems have luminous companions. We suspect these systems are hierarchical triples (Section~\ref{sec:triples}).

\section{Discussion}
\label{sec:discussion}

\subsection{Nature of the SB1 sources}
\label{sec:SB1_nature}
Most of the SB1 sources have radii inferred from SED fits that are significantly larger than predicted for stars of their temperature and metallicity near the zero-age main sequence (Figure~\ref{fig:sample-summary}). Without an astrometric constraint on the flux ratio, it is in most cases not possible to rule out a luminous companion. The two sources with $f_m > 3\,M_\odot$ are discussed in Section~\ref{sec:Be_stars}. Most of the other sources with reliable orbits have $0.4 < f_m/M_\odot < 0.7$ and SED-inferred masses near $1\,M_\odot$, implying minimum companion masses of $1.3-1.7\,M_{\odot}$. Single-star companions of this mass can be rejected, since they would outshine the primaries. Companions that are themselves tight binaries \citep[e.g.][]{Czavalinga2023, Kovalev2026, Tanikawa2026}, however, would not dominate the light. They would also be difficult to identify via spectral disentangling \citep[e.g.][]{Seeburger2024} because neither component of the inner binary would move clearly in anti-phase with the primary. 

Several sources in the SB1 sample have near-circular orbits, which are not generally expected for triples but are naturally explained if the companion is a massive WD. Four of these systems were investigated by \citet{Yamaguchi2024} and found to host ultramassive WDs. We suspect that the SB1 sample contains other WD companions, and possibly also NSs, but it is not currently possible to distinguish these from systems hosting inner binaries containing two luminous stars. Most objects in our sample have sufficiently wide orbits that the combination of our RVs and astrometry from {\it Gaia} DR4 will enable direct constraints on the flux ratio. 

\subsection{Triples}
\label{sec:triples}

As noted above, 14 of the 25 eclipsing systems in the SB1 sample exhibit clear eclipses with a period much shorter than the {\it Gaia} period. Together with one source in the astrometric sample, these are secure hierarchical triples in which the {\it Gaia} solution traces the outer orbit. These sources are identified in Tables~\ref{tab:initial-astrometric-candidates} and~\ref{tab:initial-sb1-candidates}. Triples are likely much more prevalent in the SB1 sample than in the astrometric sample, because light contributions from a tertiary will significantly reduce the astrometric mass function but will not change the RV mass function. The only secure eclipsing triple we identified within the astrometric sample is 1748901959855337472, and that source was identified as a clear positive-$f_G$ outlier in our joint astrometry+RV fits (Figure~\ref{fig:fluxratio-zpt-summary}). A second astrometric candidate, 2032579979951732736, is classified only as a possible EB and is not included in this count.

\begin{figure*}[!t]
    \centering
    \includegraphics[width=\textwidth,height=0.74\textheight,keepaspectratio]{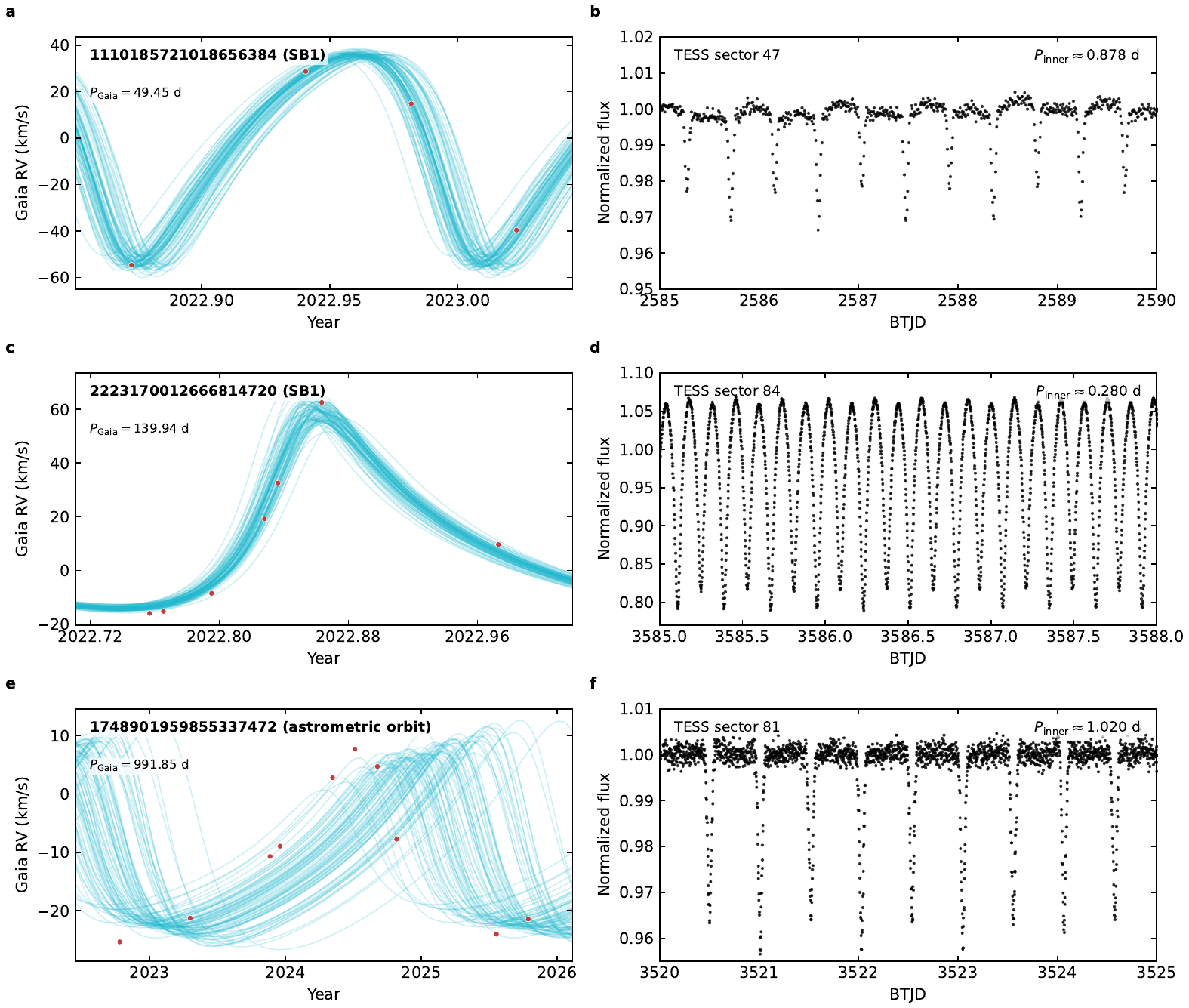}
    \caption{
    Three hierarchical triples in our sample. Top two rows show systems with {\it Gaia} SB1 orbits; bottom row shows a system with an astrometric orbit. Left panels compare our measured RVs to predictions of the {\it Gaia} solutions; right panels show TESS light curves that reveal eclipses of an inner binary. Our sample contains only one such system with an astrometric orbit, but at least 14 with spectroscopic orbits. 
    }
    \label{fig:sb1-tess-eclipses}
\end{figure*}

A majority of the eclipsing binaries have been identified in previous work (some of it published after we initiated the survey). Four of them, with DR3 source IDs 1110185721018656384, 6152018658376660992, 4219507576765009536, and 1828150428697001472, have to our knowledge not been reported previously. Figure~\ref{fig:sb1-tess-eclipses} shows RVs and TESS light curves \citep{Ricker2015} for three systems that are very likely hierarchical triples with reliable outer orbits from {\it Gaia}. None of these sources show obviously double-lined spectra (e.g. Figure~\ref{fig:spectral-model-examples}), which is unsurprising because the individual companions contribute $\lesssim 20\%$ of the light. 

Obviously, not all triples will have eclipsing inner binaries, so there are likely other as-yet-undiscovered triples within the SB1 sample. Joint fitting of astrometry and RVs after DR4 will likely provide the most straightforward opportunity to uncover them. 

\subsection{Full RV curves for Gaia BH2 and Gaia NS1}
Our program includes additional RV measurements for numerous binaries we have studied in previous work \citep[e.g.][]{ElBadry2023BH1, ElBadry2023BH2, Yamaguchi2024, Nagarajan2024, ElBadry2024NS1, El-Badry2024_ns}. Most significantly, our follow-up now includes precise RV measurements of Gaia BH2 over a full orbit. The follow-up in \citet{ElBadry2023BH2} covered only 16\% of the orbit but fortuitously covered a periastron passage, allowing the orbit to be well constrained through a joint astrometry+RV fit. Our follow-up of the system now includes 66 FEROS RVs spanning 1319 days, which is more than a full orbital period. Fitting these RVs yields a period of 1277 days, in perfect agreement with the constraints reported by \citet{ElBadry2023BH2}. Using the triple modeling framework developed by \citet{Nagarajan2024}, we find that our RVs rule out an inner BH+BH binary with $P_{\rm orb}\gtrsim 40$ d for typical inner binary orientations, phases, and mass ratios. 

We have also obtained additional RVs for Gaia NS1 \citep{ElBadry2024NS1}, with the total FEROS+TRES dataset spanning 1310 days, or almost two orbits. Our inferred parameters remain fully consistent with those reported by \citet{ElBadry2024NS1}. Inclusion of the parallax zeropoint partially alleviates the tension between the photometric and astrometric distance identified in that work and slightly reduces the best-fit mass of the NS from $1.90\,M_\odot$ to $1.88\,M_\odot$.

\subsection{NSs, massive WDs, and WD+WD binaries}
\label{sec:new-ns-candidates}

Among the 41 sources with good astrometry+RV fits, 32 have best-fit companion masses $M_2>1.25\,M_\odot$, which is the threshold \citet{El-Badry2024_ns} used to separate NS candidates from likely WD companions. The nine sources with lower inferred $M_2$ are most simply described as hosting massive WDs.

Two of the systems with $M_2>1.25\,M_\odot$ are Gaia BH1 and BH2, while the rest have inferred $M_2 < 2\,M_\odot$. After excluding three systems that we flag below as likely hosting WDs, 27 remain NS candidates. 20 of these appeared in the 21-source sample of \citet{El-Badry2024_ns}, while seven did not. Five of the seven were not ``confirmed'' in that work because their RV follow-up had not been completed: {\it Gaia} DR3 sources 1695294922548180224, 809741149368202752, 1144019690966028928, 4240540718818313984, and 1581117310088807552. A sixth source, 6054379247042197504 was not included in the AMRF sample of \citet{Shahaf2023} from which \citet{El-Badry2024_ns} selected NS candidates, because the primary is a turn-off star and is not included in the {\it Gaia} DR3 \texttt{binary\_masses} table. All of these six sources have best-fit masses at or above the maximum stable WD mass of $\approx1.39\,M_\odot$ \citep{Althaus2023GRWD}. We remind the reader that such companions could also be explained as tight, massive WD+WD inner binaries, although models also struggle to form such systems. 

The seventh new candidate, 747174436620510976, was excluded by \citet{El-Badry2024_ns} for having a problematic astrometric orbit. It marginally passes our quality cuts, but its solution should still be interpreted with caution. Conversely, source 1694708646628402048, which was considered a NS candidate by \citet{El-Badry2024_ns}, receives a slightly lower companion mass here and thus fails the $M_2 > 1.25\,M_\odot$ cut. Two systems with best-fit $M_2>1.25\,M_\odot$ have near-circular orbits, {\it Gaia} DR3~2080945469200565248 and 1522897482203494784, favoring massive WD rather than NS companions. 

\begin{figure*}
    \centering
    \includegraphics[width=\textwidth]{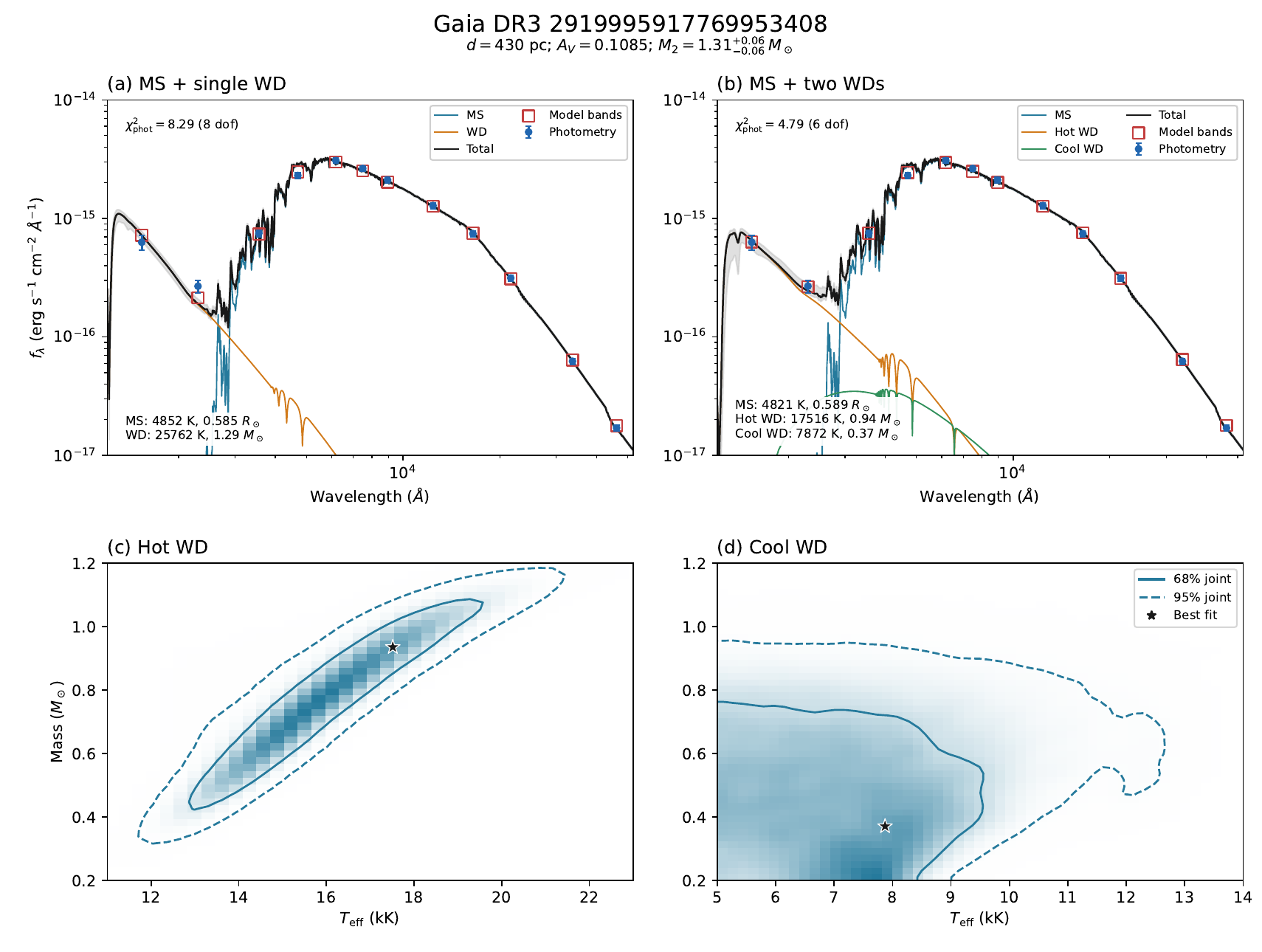}
    \caption{SED fits for Gaia DR3~2919995917769953408. The astrometric orbit implies $M_2\approx 1.3\,M_\odot$, and a strong UV excess requires a hot companion. Panel (a) shows fits in which we model the companion as a single WD. This yields a relatively poor fit: a $1.3\,M_\odot$ single WD has a small radius, requiring a temperature of $\approx 26\,$kK to match the UV SED, but the UV SED is less steep than predicted for a 26 kK WD. Panel (b) shows that the UV SED can be better fit with a sum of two WDs. Panels (c) and (d) show constraints on the temperatures of the two WDs. The presence of a second WD allows the hotter WD to have a lower mass, larger radius, and lower temperature.  }
    \label{fig:sed291999}
\end{figure*}

A particularly intriguing system is Gaia DR3~2919995917769953408, which has a good joint astrometry+RV fit, an unusually high eccentricity of $e=0.82\pm 0.01$, and an inferred companion mass of $M_2=1.31^{+0.06}_{-0.06}\,M_\odot$. These properties could all be interpreted as pointing to a NS companion. However, the source's SED (Figure~\ref{fig:sed291999}) reveals a clear UV excess due to a small and hot companion, implying that the system contains a WD. Because the orbit constrains the companion mass -- and thus, if it is a WD, its radius -- we can compare WD spectral models to the observed SED to determine whether the companion is consistent with being a single WD.

We incorporate the dynamical constraint of \(M_{2}=1.31^{+0.06}_{-0.06}\,M_\odot\) as a prior on the total companion mass, treating the companion as both a single WD and as a tight binary containing two WDs. For each likelihood call, we predict WD radii and surface gravities using the mass-radius relation from \citet{Bedard2020} and predict the WD's SED using hydrogen-atmosphere models from \citet{Koester2010}. Results are shown in Figure~\ref{fig:sed291999}. 

We find that the source's UV SED is not very well fit by any WD model when the radius is constrained to the value predicted by the dynamical mass and mass-radius relation: the UV SED prefers a larger and thus lower-mass WD than implied by the orbit. A natural possible explanation is that the companion is a tight binary containing two WDs. Allowing two WDs reduces the photometric $\chi^2$ from 8.29 to 4.79, with four and six free parameters, respectively, for 12 photometric measurements. The masses of the two WDs are not individually well constrained. We regard this result as suggestive but not yet definitive.

A scenario with two WDs in a close binary could explain both the orbit and the SED, but explaining the formation of such a tight triple with two relatively massive WDs in the inner binary would require a creative evolutionary history. Still, hierarchical triples containing compact object binaries are known to exist. PSR J0337+1715 contains an inner NS+WD binary orbited by a second WD \citep{Ransom2014}, for which \citet{Tauris2014} proposed an evolutionary history involving common-envelope evolution and multiple episodes of mass transfer. WD 0931+444, a 20-minute WD+WD binary, may have a wide M dwarf tertiary, although it might also be a chance alignment \citep{Kilic2014_wd}. These systems illustrate that complex evolutionary histories can produce compact inner binaries within surviving hierarchical systems, although the details of how such systems form are still imperfectly understood.

We hope to analyze the system in detail with UV spectroscopy in future work.  In any case, in a few Gyr the companion will fade beyond detectability in the UV, leaving the system looking in all respects like a {\it Gaia} NS candidate. Guided by this system, we remain cautious in interpreting the other NS candidates, which could also host WD+WD inner binaries. 

\subsection{Newborn Be stars}
\label{sec:Be_stars}
Two objects in the SB1 sample, with {\it Gaia} DR3 IDs 251157906379754496 and 3331748140308820352, are giants ($R \sim 20-30\,R_\odot$ and $T_{\rm eff} \approx 5000-6000$\,K) with RV mass functions above $3\,M_\odot$ and strong Balmer emission lines. Both systems' SEDs show a modest infrared excess, presumably originating in a disk.

The masses of the giants in these systems are quite uncertain given the systems' history of mass transfer, but plausible values of $1-2\,M_\odot$ imply minimum companion masses of $5-6\,M_\odot$. Neither giant is close to Roche lobe filling, so the emission lines likely track a decretion disk around the companion, not an accretion disk fed by ongoing mass transfer. Although the evolutionary state of the giants is uncertain, their properties are consistent with being recently detached donors in post-mass transfer binaries that will soon contract and subsequently be observable as stripped star + Be star binaries \citep[e.g.][]{Shenar2020, Bodensteiner2020, ElBadryQuataert2021, El-Badry2022_hd15124, MuellerHorn2025}. We thus conclude that the companions are most likely luminous stars spun-up by mass transfer, not compact objects.  

\FloatBarrier
\section{Summary and conclusions}
\label{sec:conclusions}

We have carried out a systematic search for compact object companions among astrometric and spectroscopic binary orbits published in {\it Gaia} DR3. Using a combination of analysis of archival data and many-epoch spectroscopic follow-up over a period of almost four years, the program has vetted all plausible BH candidates with DR3 orbits, a large fraction of NS candidates, and a small fraction of massive WDs. Our main results are as follows.

\begin{itemize}

\item {\it Sample selection and initial vetting}: We selected compact object binary candidates from the astrometric and spectroscopic orbit catalogs using reproducible cuts that prioritized systems with companion masses $M_2 > 1.4\,M_\odot$ (Figure~\ref{fig:sample-summary}). Many of the SB1 candidates could be immediately identified as post-interaction binaries with two-temperature spectral energy distributions; others were rejected because their light curves revealed eclipses or spectroscopy revealed multiple luminous components. Our initial sample includes 76 sources with astrometric orbits and 151 with SB1 orbits. We characterized more than 90\% of this sample through a combination of spectroscopic follow-up and analysis of archival data. 

\item {\it Spectroscopic follow-up}: We obtained multi-epoch spectra for 75 targets using the TRES and FEROS spectrographs, with the number of epochs per object ranging from 4 to 69. We estimated the spectroscopic parameters of each source using templates (Figure~\ref{fig:spectral-model-examples}) and cross-correlated these templates with the observed spectra to measure RVs. We report 1292 RVs, with a typical uncertainty of $0.05\,{\rm km\,s^{-1}}$ (Figure~\ref{fig:rv-precision}).

\item {\it Fidelity of the Gaia solutions}: We compared the measured RVs to predictions of the {\it Gaia} astrometric and spectroscopic solutions (Figures~\ref{fig:astrometric-followup-vs-gaia-params} and~\ref{fig:sb1-gaia-rv-panels}). Among 51 astrometric binaries for which we obtained many-epoch follow-up, the RVs validate the astrometric solutions for 41 sources (Figure~\ref{fig:astrometric-good-joint-rv-panels}) and are inconsistent with their predictions for 10 sources (Figure~\ref{fig:astrometric-flagged-joint-vs-sb1}). The fraction of all astrometric candidates in the sample with reliable orbits is $\sim 60\%$, since we preferentially obtained full orbits for reliable solutions. 

Most sources with inconsistent astrometry and RVs are binaries with periods comparable to the {\it Gaia} periods. The reason for the failure of the {\it Gaia} solutions is uncertain, but this suggests that the primary contaminants are not marginally resolved binaries with spurious astrometry, which would have orbital periods much longer than our observing baseline. Most spurious solutions can be eliminated with cuts on the {\it Gaia} parameters \texttt{significance} and \texttt{goodness\_of\_fit} ($F_2$). For sources with reliable astrometric solutions, joint fitting of astrometry and RVs yields much tighter orbital constraints than can be achieved with astrometry alone (Figure~\ref{fig:astrometric-followup-vs-gaia-params}).

The purity of the high-mass-function SB1 sample is somewhat lower. Of the 24 sources for which we obtained multi-epoch RVs, 21 have follow-up RVs broadly consistent with the {\it Gaia} solutions (Figure~\ref{fig:sb1-gaia-rv-panels}). When candidates rejected earlier in the program are included, only 28 of 54 systems with RVs capable of meaningfully testing the DR3 orbits have reliable solutions. All SB1 sources for which we obtained multi-epoch RVs are binaries, and in most cases the true periods are comparable to the periods reported in DR3 (Figure~\ref{fig:sb1-spectroscopy-vs-gaia-params}).

\item {\it The parallax zeropoint for astrometric orbital solutions}: Joint fitting of astrometric and RV orbits provides a unique opportunity to directly measure distances to binaries, provided that the flux ratio is known. Our sample includes 40 binaries with companions that are very likely dark, providing a unique opportunity to measure the DR3 parallax zeropoint for sources with astrometric orbital solutions (Figure~\ref{fig:fluxratio-zpt-summary}). A combined fit of all sources yields a zeropoint $Z=-0.0362\pm0.0053\,\mathrm{mas}$, under the convention $\varpi_{\rm true}=\varpi-Z$. This value is quite similar to the median zeropoint predicted by the correction from \citet{Lindegren2021} for single-star solutions, suggesting that the single-star zeropoint can and should be applied to binary solutions as well. Applying the zeropoint correction reduces inferred companion masses slightly, by a median value of $0.018\,M_\odot$ (Figure~\ref{fig:zpt-mass-comparison}). We expect the correction to be increasingly important in future {\it Gaia} data releases because the typical distances to binaries will be larger. 

\item {\it Nature of the unseen companions}: Two sources in our astrometric sample very likely host BH companions. These sources have been presented in previous work \citep{ElBadry2023BH1, ElBadry2023BH2}; the additional RVs presented here tighten constraints on their orbits. The sample also contains 27 sources with inferred $M_2 > 1.25\,M_\odot$ and eccentric orbits, which we consider NS candidates following the cuts of \citet{El-Badry2024_ns}. More than half of these sources have inferred $M_2 > 1.4\,M_\odot$. We cannot exclude the possibility that the unseen companions in these sources are tight WD+WD binaries, although such systems would be difficult to explain theoretically. Three other systems above the mass threshold are likely to host WDs rather than NSs: two have near-circular orbits, and one with $M_2 \approx 1.31\,M_\odot$ and $e=0.82$ has a strong UV excess that is best explained by a WD+WD binary (Figure~\ref{fig:sed291999}). The sample also contains dark companions with masses of $\gtrsim 1\,M_\odot$ which are likely massive single WDs. 

At least 14 sources in the SB1 sample are triples, with the SB1 solution tracing the outer orbit and light curves revealing eclipses in the inner orbit (Figure~\ref{fig:sb1-tess-eclipses}). Other triples that are not eclipsing are surely present in the sample. In most cases, this possibility cannot presently be ruled out for the SB1 candidates, but it is unlikely to apply to several SB1 candidates with near-circular orbits, which are likely to host massive WDs. Two SB1 candidates are giants with mass functions above $3\,M_\odot$, near-circular orbits, and strong emission lines. These sources are likely post-mass transfer binaries that recently detached and will soon evolve to become Be star + stripped star binaries. 

\end{itemize}

All the spectra and RVs from this program are published together with this paper. Several objects investigated here will benefit from more detailed study, and we hope that the community will use the data for such work. 

\section*{acknowledgments}
We thank the referee for a constructive report. We thank Mike Calkins, Gil Esquerdo, Perry Berlind, Sam Kim, Angela Hempel, Maren Hempel, Régis Lachaume, Alessandro Savino, and Ilya Ilyin for observing help. We are grateful to Ren{\'e} Andrae for computing {\it Gaia} XP-based metallicities using updated parallaxes from orbital solutions. This research was supported by NSF grant AST-2307232. HWR acknowledges support from the European Research Council for the ERC Advanced Grant [101054731]. This research was supported in part by the U.S. National Science Foundation (NSF) under grant PHY-1748958. This research benefited from discussions that were funded by the Gordon and Betty Moore Foundation through Grant GBMF5076.

This work has made use of data from the European Space Agency (ESA) mission {\it Gaia} (\url{https://www.cosmos.esa.int/gaia}), processed by the {\it Gaia} Data Processing and Analysis Consortium (DPAC, \url{https://www.cosmos.esa.int/web/gaia/dpac/consortium}). Funding for the DPAC has been provided by national institutions, in particular the institutions participating in the {\it Gaia} Multilateral Agreement. 

\bibliography{manuscript}
\bibliographystyle{mnras}

\clearpage
\onecolumngrid
\appendix

\section{SED fits}
\label{sec:appendix-seds}

\begin{figure}[!ht]
    \centering
    \includegraphics[width=\textwidth,height=0.82\textheight,keepaspectratio]{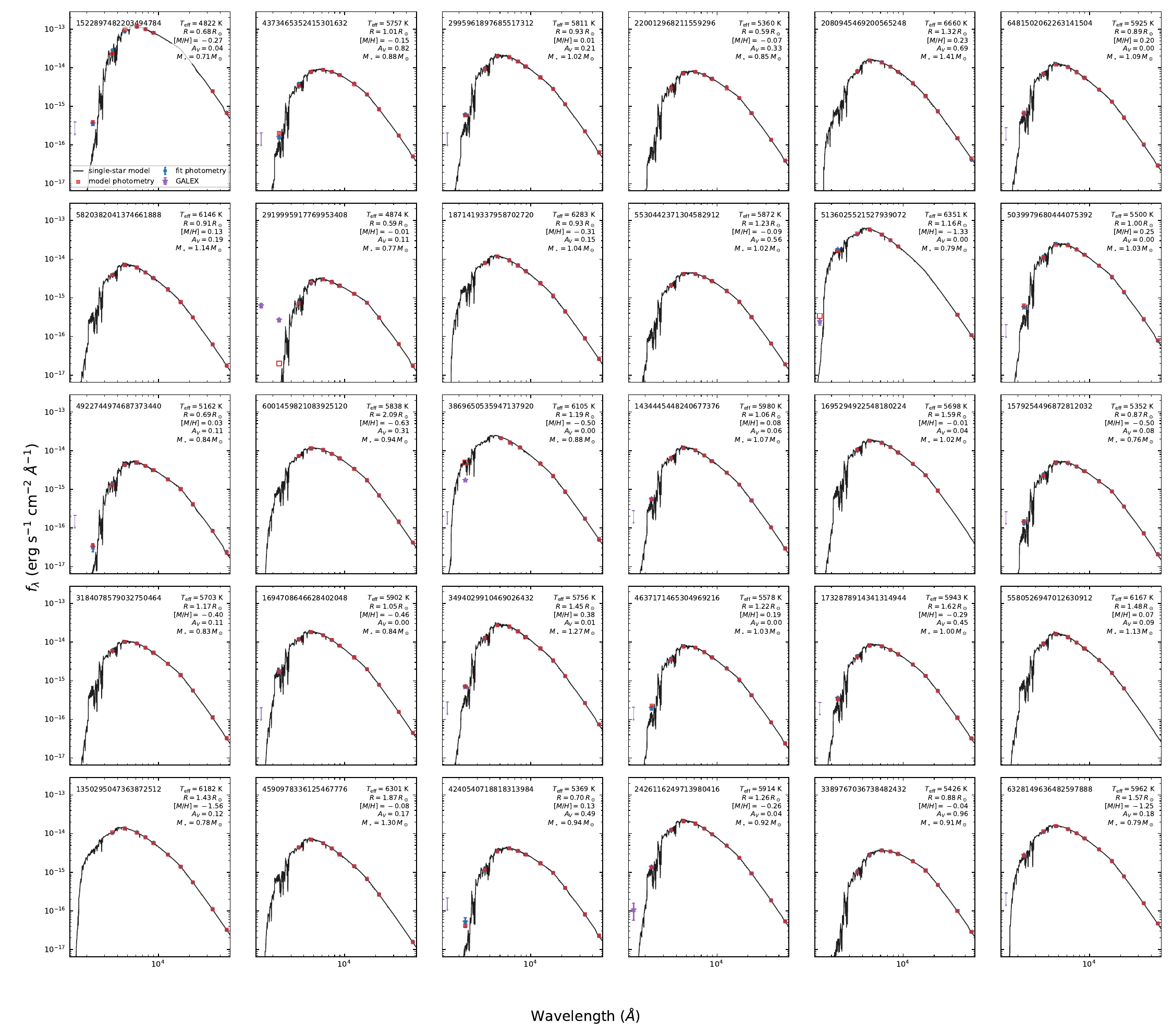}
    \caption{Single-star SED fits for astrometric candidates with RV follow-up, sorted by {\it Gaia} orbital period. Blue points are measurements included in the fit; red points are predictions of the maximum-posterior model. Purple points and arrows show GALEX detections and upper limits. Hollow red squares give the model prediction when a GALEX detection was plotted but excluded from the fit. The text in each panel reports the inferred stellar parameters used to construct the $M_1$ prior in the orbital fit. This figure is continued in Figure~\ref{fig:appendix-sed-fits}b.}
    \label{fig:appendix-sed-fits}
\end{figure}

Figures~\ref{fig:appendix-sed-fits} and~\ref{fig:appendix-sb1-sed-fits} show SED fits for all systems in the astrometric and SB1 samples with RV follow-up. 

\begin{figure}[p]
    \centering
    \includegraphics[width=\textwidth,height=0.92\textheight,keepaspectratio]{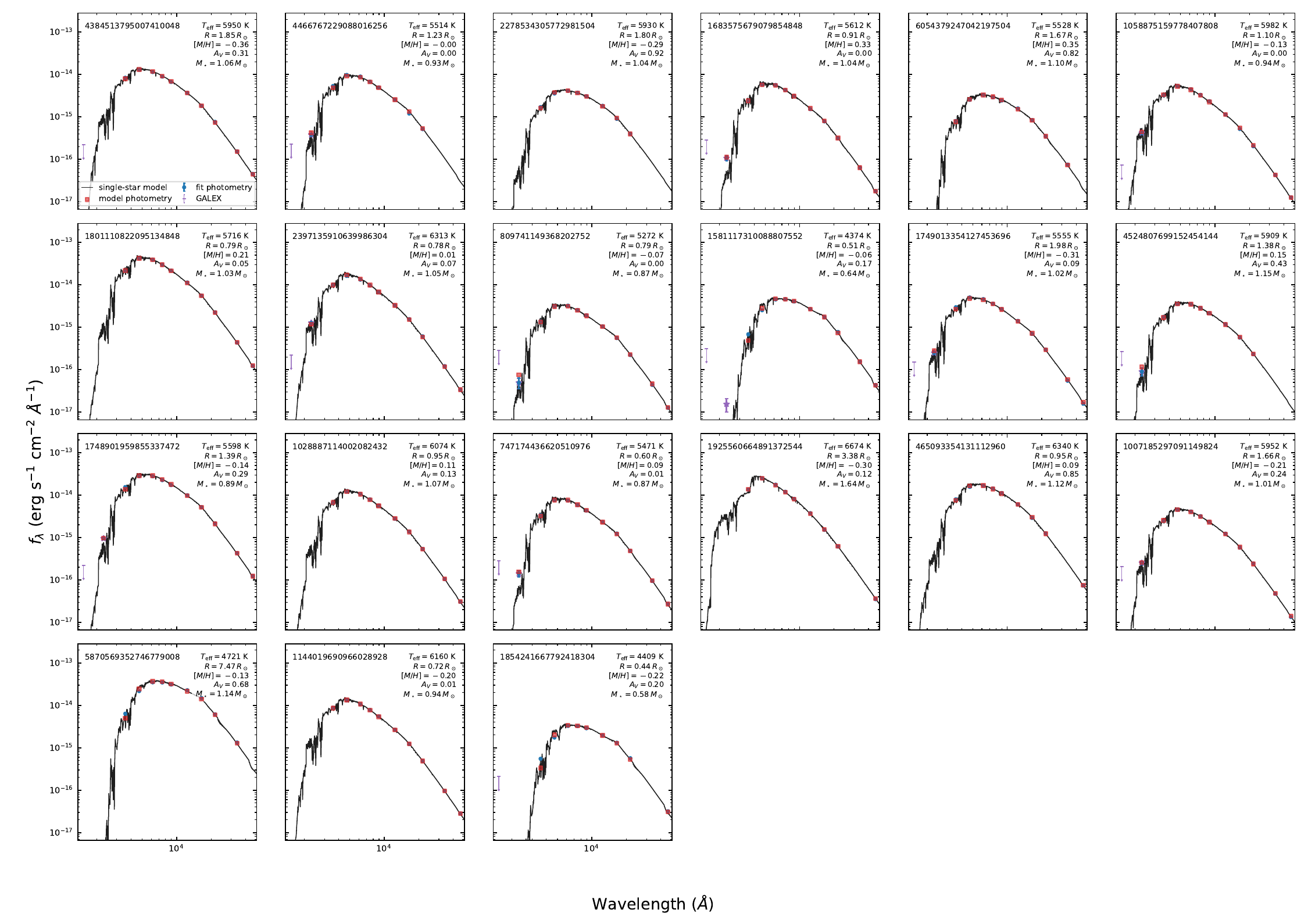}
    \vspace{0.5em}
    \noindent\footnotesize{\textbf{Figure~\ref{fig:appendix-sed-fits}b.}
    Continuation of Figure~\ref{fig:appendix-sed-fits}.}
\end{figure}

\begin{figure}[p]
    \centering
    \includegraphics[width=\textwidth,height=0.92\textheight,keepaspectratio]{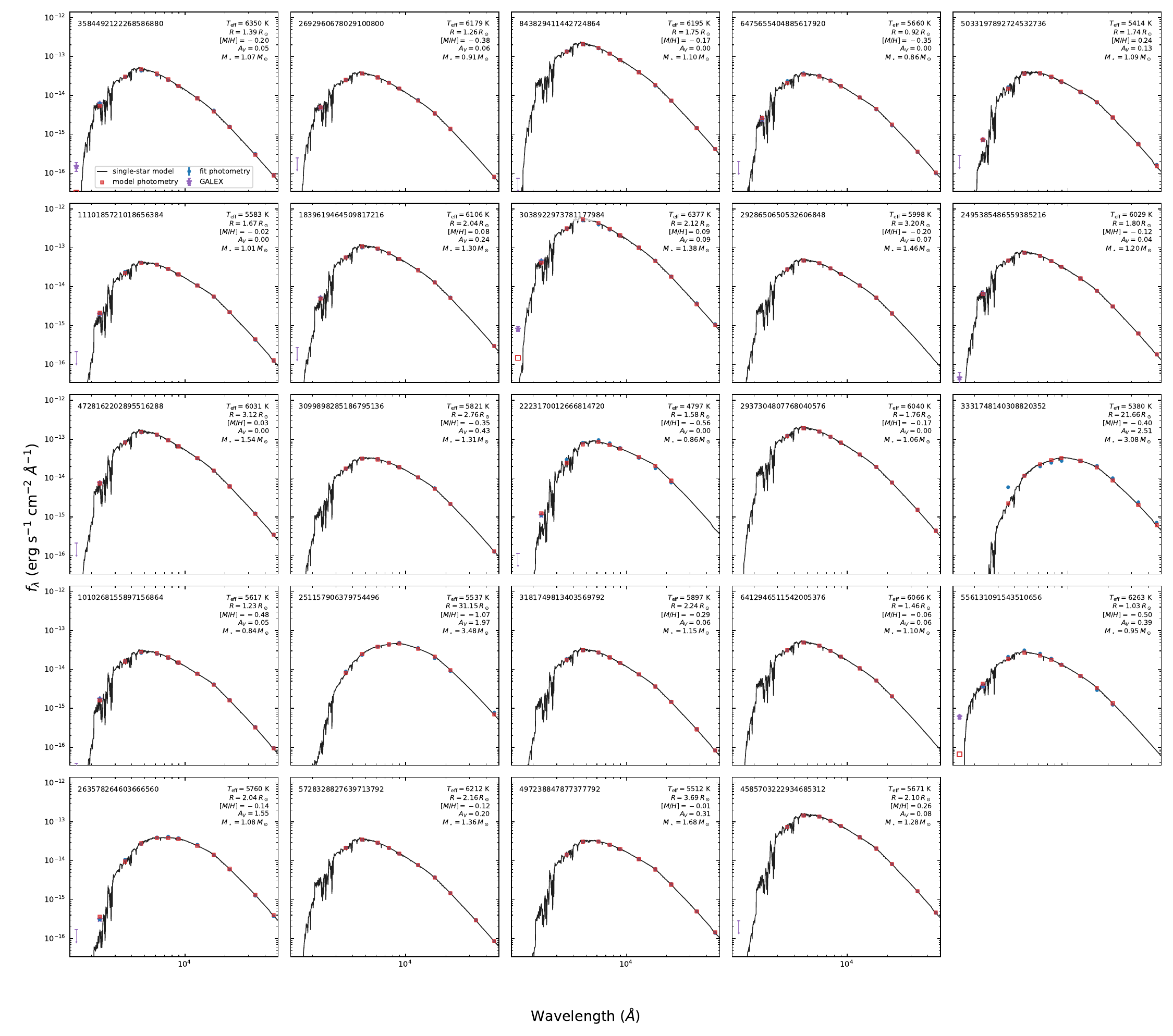}
    \caption{Single-star SED fits for SB1 candidates with RV follow-up. Symbols are as in Figure~\ref{fig:appendix-sed-fits}.} 
    \label{fig:appendix-sb1-sed-fits}
\end{figure}

\clearpage
\section{Astrometrically selected candidates}
\label{sec:appendix-astrometric}

\begingroup
\tiny
\setlength{\tabcolsep}{2pt}
\renewcommand{\arraystretch}{0.98}
\refstepcounter{table}
\label{tab:initial-astrometric-candidates}
\supertabularfirsthead{\multicolumn{11}{l}{\textbf{Table~\thetable.} Astrometrically selected Gaia DR3 compact-object candidates}\\[0.5em]
\toprule
Source ID & $P_{\rm orb}$ & $e$ & $\mathcal{A}$ & $\widetilde{M}_1$ & $\widetilde{M}_2$ & Signif. & $G$ & $E(B-V)$ & Spectroscopy & Notes\\
 & (d) & & & ($M_\odot$) & ($M_\odot$) & & (mag) & (mag) & & \\
\midrule}
\supertabularhead{\multicolumn{11}{c}{\textbf{Table~\ref{tab:initial-astrometric-candidates} (continued). Astrometrically selected Gaia DR3 compact-object candidates.}}\\[0.5em]
\toprule
Source ID & $P_{\rm orb}$ & $e$ & $\mathcal{A}$ & $\widetilde{M}_1$ & $\widetilde{M}_2$ & Signif. & $G$ & $E(B-V)$ & Spectroscopy & Notes\\
 & (d) & & & ($M_\odot$) & ($M_\odot$) & & (mag) & (mag) & & \\
\midrule}
\supertabulartail{\midrule
\multicolumn{11}{r}{\textit{Continued on next page}}\\}
\supertabularlasttail{\bottomrule}
\begin{supertabular}{@{}lrrrrrrrrp{0.12\textwidth}p{0.25\textwidth}@{}}
\texttt{3640889032890567040} & 1076.223 & 0.953 & 4.711 & 1.162 & 123.762 & 7.7 & 9.19 & 0.000 &  & BH candidate ruled out by \citet{Simon2026} \\
\underline{\texttt{4373465352415301632}} & 185.766 & 0.489 & 2.186 & 1.062 & 12.980 & 13.6 & 13.77 & 0.335 & LAMOST, MagE, TRES, FEROS & Gaia BH1 \citep{ElBadry2023BH1} \\
\texttt{6281177228434199296} & 153.947 & 0.180 & 2.071 & 1.149 & 12.218 & 24.3 & 11.26 & 0.095 & RAVE, FEROS, MagE & BH candidate ruled out by \citet{Simon2026} \\
\underline{\texttt{5870569352746779008}} & 1352.289 & 0.532 & 1.430 & 2.300 & 10.119 & 39.8 & 12.28 & 0.158 & FEROS & Gaia BH2 \citep{ElBadry2023BH2} \\
\texttt{3664684869697065984} & 1220.455 & 0.456 & 0.854 & 5.198 & 8.455 & 6.1 & 11.56 & 0.000 &  & BH candidate ruled out by \citet{ElBadry2023BH1} \\
\underline{\texttt{2278534305772981504}} & 788.822 & 0.801 & 1.128 & 1.453 & 3.920 & 5.9 & 14.59 & 0.330 & ESI, PEPSI, TRES & RVs inconsistent with orbital solution \\
\texttt{3509370326763016704} & 109.392 & 0.237 & 1.545 & 0.713 & 3.733 & 76.1 & 12.47 & 0.000 & FEROS, MagE & BH candidate ruled out by \citet{Simon2026} \\
\underline{\texttt{1925560664891372544}} & 1022.904 & 0.879 & 0.711 & 2.561 & 3.086 & 6.2 & 13.01 & 0.090 & ESI, PEPSI, TRES & RVs inconsistent with orbital solution \\
\texttt{3263804373319076480} & 510.728 & 0.278 & 1.090 & 1.160 & 2.925 & 18.1 & 12.67 & 0.080 &  LAMOST, APOGEE, TRES & BH candidate ruled out by \citet{Simon2026}  \\
\texttt{6802634430521968000} & 584.305 & 0.871 & 1.006 & 1.330 & 2.889 & 20.2 & 13.35 & 0.085 & MagE, FEROS & RVs inconsistent with orbital solution \\
\texttt{2010172929375186304} & 736.167 & 0.814 & 0.882 & 1.558 & 2.674 & 6.9 & 13.76 & 0.280 & ESI, PEPSI &  RVs inconsistent with orbital solution \\
\underline{\texttt{6328149636482597888}} & 736.018 & 0.135 & 0.958 & 1.336 & 2.656 & 89.9 & 13.34 & 0.090 & MagE, FEROS, TRES & Gaia NS1 \citep{ElBadry2024NS1} \\
\texttt{5593444799901901696} & 1038.790 & 0.442 & 0.914 & 1.447 & 2.645 & 6.5 & 14.42 & 0.040 & FEROS & RVs inconsistent with orbital solution \citep{Simon2026} \\
\texttt{6588211521163024640} & 943.333 & 0.974 & 0.995 & 1.156 & 2.459 & 10.4 & 14.19 & 0.016 & MagE, FEROS & RVs inconsistent with orbital solution  \citep{Simon2026} \\
\underline{\texttt{6001459821083925120}} & 563.833 & 0.474 & 0.825 & 1.522 & 2.336 & 19.7 & 13.60 & 0.130 & MagE, FEROS & RVs inconsistent with orbital solution  \\
\texttt{5644387063402978304} & 879.926 & 0.604 & 0.748 & 1.779 & 2.322 & 6.9 & 13.96 & 0.005 & FEROS & RVs inconsistent with orbital solution \\
\underline{\texttt{1007185297091149824}} & 1120.136 & 0.655 & 0.859 & 1.405 & 2.306 & 9.8 & 14.62 & 0.110 & ESI, PEPSI, DEIMOS, TRES & Good solution \\
\underline{\texttt{1749013354127453696}} & 932.080 & 0.508 & 0.816 & 1.457 & 2.194 & 7.9 & 14.49 & 0.092 & MagE, FEROS, PEPSI, TRES & RVs inconsistent with orbital solution  \\
\texttt{5954343888087990656} & 702.452 & 0.769 & 0.769 & 1.589 & 2.170 & 7.1 & 12.87 & 0.085 & MagE, FEROS & RVs inconsistent with orbital solution  \\
\underline{\texttt{4590978336125467776}} & 682.323 & 0.716 & 0.775 & 1.566 & 2.167 & 6.6 & 14.28 & 0.085 &  ESI, PEPSI, TRES & RVs inconsistent with orbital solution  \\
\texttt{6152333294796189568} & 1093.379 & 0.555 & 0.793 & 1.467 & 2.106 & 5.7 & 14.69 & 0.051 & 1 MagE &  RVs inconsistent with orbital solution  \\
\underline{\texttt{4384513795007410048}} & 745.201 & 0.671 & 0.795 & 1.433 & 2.063 & 16.7 & 13.50 & 0.105 & MagE, FEROS, PEPSI, TRES &  RVs inconsistent with orbital solution \\
\underline{\texttt{5580526947012630912}} & 654.326 & 0.761 & 0.790 & 1.374 & 1.961 & 12.9 & 13.36 & 0.069 & FEROS & Good solution \\
\underline{\texttt{6054379247042197504}} & 807.552 & 0.793 & 0.846 & 1.221 & 1.955 & 7.4 & 14.87 & 0.201 & MagE, FEROS & Good solution \\
\texttt{6687541573416724608} & 1125.071 & 0.781 & 0.868 & 1.128 & 1.886 & 20.1 & 12.07 & 0.029 & MagE, FEROS & RV follow-up incomplete \\
\texttt{4099356347737522432} & 718.340 & 0.750 & 0.755 & 1.411 & 1.868 & 9.4 & 13.81 & 0.330 &  & No RV follow-up \\
\underline{\texttt{3869650535947137920}} & 570.036 & 0.183 & 0.827 & 1.187 & 1.829 & 34.0 & 12.94 & 0.005 & LAMOST, TRES & RVs inconsistent with orbital solution \\
\texttt{3649963989549165440} & 892.531 & 0.357 & 0.924 & 0.972 & 1.810 & 40.2 & 14.30 & 0.050 & MagE, ESI, FEROS & NS candidate with an sdB primary \citep{Geier2023} \\
\texttt{6593763230249162112} & 679.888 & 0.613 & 0.823 & 1.180 & 1.799 & 23.3 & 13.54 & 0.010 & MagE, FEROS & RVs inconsistent with orbital solution \citep{Simon2026}  \\
\underline{\texttt{1058875159778407808}} & 835.682 & 0.417 & 0.848 & 1.117 & 1.794 & 33.0 & 14.52 & 0.005 & TRES & Good solution \\
\underline{\texttt{1350295047363872512}} & 657.242 & 0.656 & 0.757 & 1.314 & 1.747 & 34.1 & 13.52 & 0.055 & PEPSI, TRES & Good solution \\
\underline{\texttt{1732878914341314944}} & 648.097 & 0.549 & 0.738 & 1.352 & 1.727 & 6.8 & 13.94 & 0.160 & MagE, FEROS, PEPSI, TRES & Good solution \\
\underline{\texttt{809741149368202752}} & 922.389 & 0.351 & 0.964 & 0.849 & 1.706 & 13.5 & 14.91 & 0.005 & LAMOST, TRES & Good solution \\
\texttt{6092954989675820416} & 883.079 & 0.138 & 0.732 & 1.333 & 1.682 & 14.1 & 13.67 & 0.109 & MagE, FEROS  & RVs inconsistent with orbital solution \\
\texttt{3662912491312811008} & 237.156 & 0.938 & 0.854 & 1.013 & 1.647 & 7.9 & 12.61 & 0.005 & APOGEE, FEROS & RVs inconsistent with orbital solution \\
\texttt{5839182174066052224} & 331.777 & 0.897 & 0.834 & 1.050 & 1.638 & 11.0 & 12.99 & 0.130 & GALAH, MagE & RVs inconsistent with orbital solution \\
\underline{\texttt{1695294922548180224}} & 601.178 & 0.571 & 0.737 & 1.277 & 1.627 & 79.9 & 13.12 & 0.010 & ESI, PEPSI, TRES & Good solution \\
\underline{\texttt{2080945469200565248}} & 196.640 & 0.045 & 0.664 & 1.486 & 1.608 & 54.3 & 13.29 & 0.341 & TRES & Good solution \\
\underline{\texttt{2426116249713980416}} & 711.703 & 0.398 & 0.768 & 1.169 & 1.593 & 24.8 & 13.02 & 0.005 & MagE, FEROS, PEPSI, TRES & Good solution \\
\underline{\texttt{1028887114002082432}} & 994.417 & 0.657 & 0.802 & 1.062 & 1.551 & 24.5 & 13.59 & 0.065 & LAMOST, PEPSI, TRES &  Good solution  \\
\texttt{6791860144283753984} & 935.830 & 0.704 & 0.745 & 1.191 & 1.544 & 12.0 & 12.80 & 0.042 & GALAH, MagE, FEROS & RVs inconsistent with orbital solution \\
\underline{\texttt{5136025521527939072}} & 536.900 & 0.651 & 0.728 & 1.223 & 1.528 & 60.3 & 12.05 & 0.030 & FEROS, TRES & Good solution \\
\underline{\texttt{2397135910639986304}} & 916.016 & 0.564 & 0.842 & 0.960 & 1.522 & 43.8 & 13.35 & 0.000 & MagE, TRES, FEROS & Good solution \\
\underline{\texttt{1144019690966028928}} & 1401.860 & 0.381 & 0.861 & 0.923 & 1.522 & 23.6 & 13.57 & 0.028 & PEPSI, TRES & Good solution \\
\underline{\texttt{4524807699152454144}} & 947.201 & 0.483 & 0.718 & 1.244 & 1.520 & 15.9 & 14.80 & 0.147 & MagE, FEROS, TRES & Solution OK; some residual scatter \\
\underline{\texttt{5530442371304582912}} & 498.831 & 0.285 & 0.773 & 1.102 & 1.518 & 29.0 & 14.60 & 0.115 & FEROS & Good solution \\
\underline{\texttt{4637171465304969216}} & 639.151 & 0.375 & 0.767 & 1.116 & 1.516 & 55.0 & 14.01 & 0.041 & FEROS & Good solution \\
\underline{\texttt{1871419337958702720}} & 479.313 & 0.432 & 0.771 & 1.093 & 1.499 & 72.1 & 13.70 & 0.088 & LAMOST, ESI, TRES & Good solution \\
\underline{\texttt{1854241667792418304}} & 1430.334 & 0.590 & 1.065 & 0.620 & 1.499 & 55.5 & 14.87 & 0.000 & ESI, PEPSI, TRES &  Good solution \\
\underline{\texttt{3494029910469026432}} & 632.530 & 0.552 & 0.686 & 1.293 & 1.472 & 77.6 & 12.66 & 0.065 & TRES &  Good solution \\
\underline{\texttt{5820382041374661888}} & 311.140 & 0.414 & 0.783 & 1.040 & 1.460 & 39.2 & 14.19 & 0.070 & FEROS &  Good solution \\
\underline{\texttt{4922744974687373440}} & 562.381 & 0.813 & 0.919 & 0.788 & 1.455 & 15.4 & 14.48 & 0.010 & MagE, FEROS &  Good solution \\
\underline{\texttt{465093354131112960}} & 1046.370 & 0.763 & 0.756 & 1.085 & 1.440 & 24.2 & 13.09 & 0.260 & ESI, PEPSI, TRES & Good solution \\
\underline{\texttt{1748901959855337472}} & 991.847 & 0.398 & 0.717 & 1.156 & 1.410 & 6.1 & 12.55 & 0.100 & TRES & EB with $P_{\rm orb}=1.02$\,d \citep{Schanche2019} \\
\underline{\texttt{747174436620510976}} & 999.436 & 0.707 & 0.885 & 0.808 & 1.397 & 48.7 & 13.99 & 0.030 & LAMOST, TRES & RVs marginally consistent with astrometric orbit \\
\texttt{5446310318525312768} & 866.558 & 0.247 & 0.690 & 1.208 & 1.387 & 47.0 & 10.37 & 0.019 &  & No RV follow-up \\
\underline{\texttt{1581117310088807552}} & 927.314 & 0.520 & 1.000 & 0.644 & 1.383 & 175.6 & 14.51 & 0.000 & LAMOST, TRES & Good solution \\
\texttt{6037767138131854592} & 804.701 & 0.768 & 0.792 & 0.965 & 1.383 & 29.6 & 14.28 & 0.165 &  & No RV follow-up \\
\underline{\texttt{1434445448240677376}} & 572.423 & 0.301 & 0.717 & 1.114 & 1.361 & 82.5 & 13.65 & 0.055 & TRES & Good solution \\
\underline{\texttt{1801110822095134848}} & 893.583 & 0.588 & 0.820 & 0.892 & 1.354 & 92.2 & 12.19 & 0.016 & TRES & Good solution \\
\underline{\texttt{4240540718818313984}} & 691.168 & 0.618 & 0.854 & 0.834 & 1.354 & 85.2 & 14.61 & 0.175 & TRES & Good solution \\
\underline{\texttt{1694708646628402048}} & 632.034 & 0.259 & 0.730 & 1.073 & 1.348 & 114.9 & 13.20 & 0.020 & TRES & Good solution \\
\underline{\texttt{1522897482203494784}} & 45.517 & 0.046 & 0.900 & 0.752 & 1.339 & 156.6 & 11.05 & 0.000 & TRES & Good solution \\
\underline{\texttt{6481502062263141504}} & 229.684 & 0.304 & 0.746 & 1.002 & 1.301 & 48.5 & 13.58 & 0.020 & FEROS & Good solution \\
\underline{\texttt{2995961897685517312}} & 189.763 & 0.366 & 0.744 & 0.998 & 1.293 & 39.9 & 13.00 & 0.095 & FEROS, TRES & Good solution \\
\underline{\texttt{3184078579032750464}} & 614.895 & 0.153 & 0.682 & 1.102 & 1.242 & 87.4 & 13.73 & 0.065 & TRES & Good solution \\
\texttt{2032579979951732736} & 215.447 & 0.146 & 0.787 & 0.877 & 1.242 & 63.1 & 14.21 & 0.057 &  & Possible EB \citep{Chen2020, Czavalinga2023} \\
\texttt{5355633933885075328} & 574.740 & 0.111 & 0.663 & 1.139 & 1.232 & 76.8 & 13.73 & 0.100 &  & No RV follow-up \\
\underline{\texttt{4466767229088016256}} & 776.604 & 0.154 & 0.685 & 1.082 & 1.229 & 116.1 & 13.78 & 0.000 & TRES & Good solution \\
\underline{\texttt{3389767036738482432}} & 717.231 & 0.142 & 0.768 & 0.897 & 1.222 & 55.5 & 14.72 & 0.322 & FEROS, PEPSI, TRES & Good solution \\
\underline{\texttt{2919995917769953408}} & 456.774 & 0.752 & 0.852 & 0.739 & 1.196 & 23.2 & 14.97 & 0.035 & FEROS & Good solution \\
\underline{\texttt{5039979680444075392}} & 553.305 & 0.181 & 0.720 & 0.971 & 1.193 & 83.8 & 12.72 & 0.000 & FEROS & Good solution \\
\underline{\texttt{1683575679079854848}} & 795.566 & 0.621 & 0.724 & 0.939 & 1.163 & 49.6 & 14.30 & 0.000 & TRES & Good solution \\
\underline{\texttt{1579254496872812032}} & 601.462 & 0.308 & 0.745 & 0.863 & 1.120 & 64.8 & 14.50 & 0.000 & TRES & Good solution \\
\texttt{5283631903842076032} & 90.754 & 0.314 & 0.772 & 0.815 & 1.119 & 31.3 & 13.31 & 0.018 &  & No RV follow-up \\
\underline{\texttt{220012968211559296}} & 196.341 & 0.412 & 0.787 & 0.785 & 1.113 & 25.3 & 13.93 & 0.105 & TRES & Good solution \\
\midrule
\multicolumn{11}{@{}p{\textwidth}@{}}{\tiny\textit{Note.} The sample contains the astrometrically selected candidates described in Section~\ref{sec:sample}. The table is sorted by decreasing inferred companion mass. The column $\mathcal{A}$ is the dimensionless astrometric mass-ratio function defined in Equation~\ref{eq:AMRF}. The companion mass estimate $\widetilde{M}_2$ is inferred from $\mathcal{A}$ and $\widetilde{M}_1$ assuming the companion is dark. Underlined source IDs mark systems with many-epoch follow-up analyzed in this work.}\\
\end{supertabular}
\endgroup

\clearpage
\section{Initial SB1 candidates}
\label{sec:appendix}

\begingroup
\tiny
\setlength{\tabcolsep}{2pt}
\renewcommand{\arraystretch}{1.06}
\refstepcounter{table}
\label{tab:initial-sb1-candidates}
\supertabularfirsthead{\multicolumn{11}{l}{\textbf{Table~\thetable.} Initial Gaia DR3 SB1 compact-object candidates}\\[0.5em]
\toprule
Source ID & $P_{\rm orb}$ & $e$ & $f_m$ & $M_1$ & $M_{2,\min}$ & Signif. & $G$ & $E(B-V)$ & Spectroscopy & Notes\\
 & (d) & & ($M_\odot$) & ($M_\odot$) & ($M_\odot$) & & (mag) & (mag) & & \\
\midrule}
\supertabularhead{\multicolumn{11}{c}{\textbf{Table~\ref{tab:initial-sb1-candidates} (continued). Initial Gaia DR3 SB1 compact-object candidates.}}\\[0.5em]
\toprule
Source ID & $P_{\rm orb}$ & $e$ & $f_m$ & $M_1$ & $M_{2,\min}$ & Signif. & $G$ & $E(B-V)$ & Spectroscopy & Notes\\
 & (d) & & ($M_\odot$) & ($M_\odot$) & ($M_\odot$) & & (mag) & (mag) & & \\
\midrule}
\supertabulartail{\midrule
\multicolumn{11}{r}{\textit{Continued on next page}}\\}
\supertabularlasttail{\bottomrule}
\begin{supertabular}{@{}lrrrrrrrrp{0.12\textwidth}p{0.25\textwidth}@{}}
\shrinkheight{0pt}
\texttt{5824739062379517824} & 1113.338 & 0.529 & 33.039 & 6.577 & 43.723 & 12.5 & 8.91 & 0.198 & & Be star \citep{Henize1976} \\
\texttt{2174777963318889344} & 82.723 & 0.023 & 3.819 & 50.205 & 28.773 & 31.5 & 13.06 & 1.418 & FEROS & Algol-type binary \citep{Jayasinghe2023} \\
\texttt{442992311418593664} & 216.531 & 0.414 & 4.764 & 12.252 & 15.378 & 21.8 & 9.55 & 0.631 & HERMES, ESI, PEPSI & Be star + bloated stripped star \citep{MuellerHorn2025} \\
\underline{\texttt{251157906379754496}} & 296.676 & 0.089 & 3.608 & 13.531 & 13.975 & 49.2 & 12.05 & 0.979 & TRES & Post-mass-transfer binary \\
\texttt{2031113506311851904} & 35.908 & 0.057 & 3.770 & 11.652 & 13.285 & 41.6 & 13.24 & 1.125 & & Algol-type binary \citep{Jayasinghe2023} \\
\texttt{4060365702574410752} & 518.154 & 0.092 & 9.756 & 1.279 & 11.954 & 10.2 & 12.52 & 0.239 & MagE, FEROS & SB2 \citep{Simon2026} \\
\texttt{2055293686820130176} & 7.157 & 0.154 & 3.295 & 10.778 & 11.935 & 11.2 & 11.98 & 1.105 & & Two-temperature SED \\
\texttt{3326590404277028096} & 11.273 & 0.065 & 4.223 & 7.623 & 11.599 & 10.4 & 13.04 & 0.950 & & EB \citep{Hoffman2009} \\
\texttt{5857059996952633984} & 155.085 & 0.036 & 4.241 & 6.910 & 11.137 & 56.0 & 11.01 & 0.315 & RAVE & EB with $P_{\rm orb} = 1.17$\,d \citep{Mowlavi2023}\\
\texttt{1996704839648530816} & 7.546 & 0.046 & 3.741 & 7.738 & 10.921 & 38.9 & 9.80 & 0.240 & & Algol-type binary \citep{Jayasinghe2023} \\
\texttt{527155253604491392} & 149.155 & 0.040 & 3.126 & 9.286 & 10.807 & 56.4 & 13.24 & 1.095 & PEPSI & Two-temperature SED \\
\texttt{6102598776102841344} & 225.784 & 0.082 & 7.595 & 1.884 & 10.550 & 12.9 & 11.67 & 0.088 & MagE, FEROS & EB with $P_{\rm orb} = 1.18$\,d \citep{Mowlavi2023} \\
\texttt{5864217675880901760} & 32.958 & 0.007 & 2.870 & 9.099 & 10.238 & 276.2 & 9.31 & 0.289 & & Ellipsoidal variability \citep{Jayasinghe2023} \\
\texttt{4057073043204216064} & 6.565 & 0.318 & 3.055 & 7.532 & 9.670 & 11.1 & 11.89 & 0.971 & & EB \citep{Budding2004} \\
\texttt{2086448353089047808} & 69.561 & 0.128 & 6.216 & 1.836 & 9.008 & 15.9 & 11.96 & 0.068 & LAMOST MRS & EB with $P_{\rm orb} = 2.045$\,d \citep{Devor2008} \\
\texttt{2006840790676091776} & 4.073 & 0.203 & 5.460 & 2.001 & 8.378 & 28.3 & 11.18 & 0.302 & LAMOST MRS & Blended RVS spectra \citep{Holl2023} \\
\texttt{5352109964757046528} & 9.868 & 0.140 & 2.954 & 5.423 & 8.173 & 13.3 & 11.77 & 0.222 & FEROS & Multiple luminous stars \citep{Simon2026} \\
\texttt{2208323856913151360} & 84.991 & 0.063 & 2.341 & 6.960 & 8.096 & 27.7 & 12.97 & 1.165 & ESI, PEPSI & Balmer emission in ESI/PEPSI spectra \\
\texttt{2929565719083290240} & 32.473 & 0.033 & 3.079 & 4.931 & 8.024 & 49.9 & 12.88 & 0.456 & FEROS & Likely Algol-type binary; cool companion \\
\texttt{1816166713996603904} & 6.642 & 0.134 & 2.211 & 6.572 & 7.646 & 17.4 & 11.41 & 0.125 & MagE, FEROS & Likely Algol-type binary; Balmer emission \\
\texttt{1969720105580531840} & 52.246 & 0.029 & 2.284 & 6.141 & 7.530 & 72.0 & 11.87 & 0.860 & & Algol-type binary \\
\texttt{6054043964711290112} & 33.892 & 0.035 & 2.455 & 5.331 & 7.326 & 42.1 & 12.09 & 0.375 & & EB \citep{Kreiner2004} \\
\texttt{5971649891874353536} & 49.976 & 0.013 & 2.336 & 4.789 & 6.791 & 56.1 & 10.19 & 0.349 & & EB \citep{Otero2005} \\
\texttt{2021374066702077312} & 44.413 & 0.007 & 2.387 & 4.538 & 6.710 & 249.1 & 10.04 & 0.367 & & Two-temperature SED \\
\texttt{5229989171064887936} & 23.518 & 0.332 & 2.672 & 3.896 & 6.689 & 13.0 & 12.20 & 0.166 & & Two-temperature SED \\
\texttt{5866096676853138432} & 76.373 & 0.083 & 1.974 & 5.500 & 6.618 & 24.0 & 11.42 & 0.376 & & Two-temperature SED \\
\texttt{2204345647749753216} & 41.268 & 0.145 & 1.955 & 4.349 & 5.899 & 28.4 & 12.72 & 0.442 & & Two-temperature SED + ellipsoidal \citep{Jayasinghe2023} \\
\texttt{2143064916926398848} & 54.130 & 0.004 & 1.553 & 5.175 & 5.676 & 331.7 & 7.97 & 0.025 & & Two-temperature SED + ellipsoidal \citep{Jayasinghe2023} \\
\texttt{4116115997318757120} & 41.298 & 0.219 & 1.929 & 3.779 & 5.494 & 80.1 & 10.61 & 0.255 & & Two-temperature SED \\
\texttt{5352456452682656256} & 59.850 & 0.021 & 1.629 & 4.547 & 5.465 & 126.0 & 12.03 & 0.204 & & Two-temperature SED \\
\texttt{6000420920026118656} & 15.318 & 0.006 & 2.055 & 3.221 & 5.306 & 127.6 & 10.73 & 0.138 & GALAH & Algol-type binary (\citealt{ElBadryRix2022}) \\
\texttt{1850548988047789696} & 15.302 & 0.051 & 1.833 & 3.677 & 5.277 & 29.4 & 12.03 & 0.145 & & Algol-type binary (\citealt{ElBadryRix2022}) \\
\texttt{2079141028874426624} & 21.582 & 0.006 & 1.529 & 4.380 & 5.195 & 68.7 & 12.10 & 0.267 & & Two-temperature SED + ellipsoidal \citep{Jayasinghe2023} \\
\texttt{448452383082046208} & 23.494 & 0.072 & 1.470 & 4.513 & 5.164 & 34.0 & 12.48 & 0.906 & & Algol-type binary (\citealt{ElBadryRix2022}) \\
\texttt{2950633087724402688} & 24.204 & 0.044 & 1.575 & 4.055 & 5.086 & 21.6 & 12.26 & 0.320 & & Two-temperature SED \\
\texttt{5336217383170465792} & 31.832 & 0.046 & 1.370 & 4.358 & 4.895 & 24.7 & 12.54 & 0.278 & & Two-temperature SED \\
\texttt{4514813786980451840} & 22.020 & 0.136 & 1.555 & 3.688 & 4.834 & 24.6 & 12.57 & 0.285 & & Algol-type binary (\citealt{ElBadryRix2022}) \\
\texttt{5611909581558507648} & 21.963 & 0.034 & 1.393 & 4.011 & 4.745 & 157.0 & 9.81 & 0.065 & & Two-temperature SED \\
\texttt{6734611563148165632} & 14.344 & 0.008 & 1.691 & 3.118 & 4.689 & 106.4 & 10.94 & 0.112 & & Algol-type binary (\citealt{ElBadryRix2022}) \\
\texttt{358522257594654592} & 11.018 & 0.093 & 1.705 & 3.067 & 4.675 & 19.4 & 12.39 & 0.180 & & EB \citep{Mowlavi2023} \\
\texttt{2925811436628649984} & 714.215 & 0.574 & 1.190 & 4.524 & 4.642 & 13.8 & 12.08 & 0.205 & & EB \citep{Mowlavi2023} \\
\texttt{2197954362764248192} & 17.510 & 0.043 & 1.467 & 3.555 & 4.606 & 20.4 & 12.69 & 0.445 & & Algol-type binary (\citealt{ElBadryRix2022}) \\
\texttt{5694373091078326784} & 12.885 & 0.018 & 1.656 & 3.031 & 4.576 & 30.8 & 11.91 & 0.015 & & Algol-type binary (\citealt{ElBadryRix2022}) \\
\texttt{1868420557434760960} & 24.995 & 0.083 & 1.348 & 3.654 & 4.461 & 14.8 & 12.59 & 0.180 & & Two-temperature SED + ellipsoidal \\
\texttt{548272473920331136} & 13.371 & 0.047 & 1.607 & 2.930 & 4.434 & 32.1 & 12.58 & 0.260 & & Algol-type binary (\citealt{ElBadryRix2022}) \\
\texttt{5243109471519822720} & 14.914 & 0.023 & 1.561 & 2.888 & 4.335 & 170.1 & 10.45 & 0.106 & GALAH & Algol-type binary (\citealt{ElBadryRix2022}) \\
\texttt{209386703724361088} & 22.025 & 0.013 & 1.219 & 3.689 & 4.252 & 189.4 & 10.27 & 0.230 & & Two-temperature SED \\
\texttt{6077083852882264320} & 897.558 & 0.034 & 1.260 & 3.550 & 4.246 & 71.5 & 10.81 & 0.083 & & Two-temperature SED \\
\texttt{948585824160038912} & 8.202 & 0.077 & 1.468 & 2.894 & 4.194 & 21.8 & 11.96 & 0.070 & & Algol-type binary (\citealt{ElBadryRix2022}) \\
\texttt{5722942457613931264} & 38.100 & 0.025 & 1.112 & 3.923 & 4.179 & 115.0 & 9.94 & 0.045 & RAVE & Two-temperature SED \\
\texttt{5644846693617714560} & 17.117 & 0.075 & 1.087 & 4.009 & 4.175 & 39.6 & 10.68 & 0.025 & & EB \citep{Mowlavi2023} \\
\texttt{2907796252142974720} & 1018.612 & 0.005 & 1.179 & 3.618 & 4.139 & 269.3 & 8.84 & 0.046 & & Two-temperature SED \\
\texttt{2933630927108779776} & 14.718 & 0.021 & 1.223 & 3.431 & 4.113 & 69.0 & 11.07 & 0.202 & & Algol-type binary (\citealt{ElBadryRix2022}) \\
\texttt{182029303939094784} & 107.155 & 0.263 & 1.551 & 2.455 & 4.022 & 11.6 & 12.36 & 0.248 & LAMOST & LAMOST RVs inconsistent with orbit \\
\texttt{5536105058044762240} & 12.177 & 0.086 & 1.177 & 3.345 & 3.984 & 24.0 & 12.15 & 0.187 & & Algol-type binary (\citealt{ElBadryRix2022}) \\
\texttt{5877599080364216576} & 16.673 & 0.014 & 1.245 & 3.141 & 3.983 & 67.9 & 10.77 & 0.207 & & Two-temperature SED \\
\texttt{2219809419798508544} & 10.865 & 0.075 & 1.284 & 2.918 & 3.914 & 38.4 & 12.48 & 0.370 & & Algol-type binary (\citealt{ElBadryRix2022}) \\
\texttt{3372818683473351936} & 12.208 & 0.059 & 1.085 & 3.389 & 3.842 & 13.5 & 11.61 & 0.085 & LAMOST MRS & Two-temperature SED \\
\texttt{2966694650501747328} & 10.398 & 0.019 & 1.155 & 2.977 & 3.733 & 79.5 & 11.20 & 0.045 & & Algol-type binary (\citealt{ElBadryRix2022}) \\
\texttt{3998734261904478848} & 691.876 & 0.213 & 2.116 & 1.135 & 3.641 & 12.5 & 12.49 & 0.000 & LAMOST & LAMOST RVs inconsistent with orbit \\
\texttt{5353377151908876032} & 76.852 & 0.261 & 0.898 & 3.577 & 3.584 & 13.9 & 12.18 & 0.183 & FEROS & RVs consistent with orbit; evolved star \\
\texttt{5610703902634458112} & 1003.250 & 0.021 & 0.964 & 3.116 & 3.472 & 48.1 & 11.01 & 0.085 & & Two-temperature SED \\
\texttt{209136938490844928} & 14.292 & 0.043 & 0.869 & 3.201 & 3.336 & 30.4 & 12.40 & 0.336 & & Two-temperature SED + ellipsoidal \citep{Jayasinghe2023} \\
\texttt{3107879743172451456} & 27.305 & 0.027 & 1.079 & 2.502 & 3.318 & 21.3 & 11.90 & 0.091 & PEPSI, FEROS & SB2 \\
\texttt{4295254307232999296} & 10.833 & 0.046 & 0.907 & 2.966 & 3.285 & 29.6 & 11.55 & 0.190 & & Two-temperature SED + ellipsoidal \citep{Jayasinghe2023}\\
\texttt{4510976285245111296} & 7.655 & 0.021 & 0.895 & 2.943 & 3.251 & 18.8 & 12.48 & 0.240 & & Two-temperature SED \\
\texttt{4513319417560405376} & 10.378 & 0.177 & 0.801 & 3.089 & 3.145 & 15.7 & 12.68 & 0.620 & & Two-temperature SED \\
\texttt{2034914999055297664} & 60.718 & 0.071 & 0.917 & 2.592 & 3.096 & 11.1 & 11.38 & 0.090 & ESI, PEPSI, TRES & EB with $P_{\rm orb} = 4.96\,{\rm d}$ \\
\texttt{5363143465161744512} & 951.923 & 0.348 & 0.871 & 2.664 & 3.054 & 31.2 & 11.53 & 0.094 & GALAH & Two-temperature SED \\
\texttt{1842500631652332544} & 4.580 & 0.105 & 0.810 & 2.407 & 2.801 & 22.0 & 11.80 & 0.128 & & EB \citep{Mowlavi2023} \\
\texttt{395989834300077184} & 755.929 & 0.237 & 0.861 & 2.113 & 2.718 & 41.8 & 11.82 & 0.160 & ESI, PEPSI & EB \citep{Mowlavi2023} \\
\texttt{4239611700214875904} & 56.124 & 0.404 & 1.110 & 1.510 & 2.700 & 23.6 & 11.91 & 0.175 & FEROS & RVs inconsistent with orbit \\
\texttt{6057458699887995904} & 2.380 & 0.450 & 0.727 & 2.050 & 2.451 & 13.3 & 12.42 & 0.182 & FEROS & RVs unreliable \\
\texttt{6609971199873698944} & 18.342 & 0.194 & 0.777 & 1.469 & 2.179 & 36.5 & 11.52 & 0.000 & MagE, FEROS & RVs inconsistent with orbit \\
\underline{\texttt{2928650650532606848}} & 73.414 & 0.310 & 0.575 & 1.988 & 2.141 & 19.2 & 12.16 & 0.040 & FEROS & RVs consistent with orbit; EB with $P_{\rm orb} = 0.61$\,d \citep{Mowlavi2023} \\
\texttt{4039283116865495424} & 79.549 & 0.128 & 0.656 & 1.660 & 2.101 & 20.9 & 11.99 & 0.074 & FEROS & RVs inconsistent with orbit \\
\underline{\texttt{4728162202895516288}} & 114.763 & 0.380 & 0.550 & 1.975 & 2.086 & 73.6 & 10.90 & 0.014 & FEROS & RVs consistent with orbit; SB2 \\
\texttt{4341334461557059328} & 353.710 & 0.380 & 1.087 & 0.794 & 2.077 & 21.7 & 13.51 & 0.265 & PEPSI, MagE & SB2 \\
\texttt{1198284197571176704} & 584.032 & 0.277 & 0.530 & 1.991 & 2.056 & 12.0 & 12.28 & 0.055 & PEPSI, MagE, FEROS & RVs inconsistent with orbit \\
\texttt{5698084561292928384} & 126.311 & 0.328 & 0.528 & 1.938 & 2.024 & 143.2 & 10.46 & 0.025 & PEPSI, FEROS & SB2 \\
\texttt{1821800302351693312} & 653.991 & 0.250 & 0.566 & 1.723 & 1.981 & 15.2 & 12.66 & 0.174 & PEPSI, MagE, FEROS & SB2 \\
\underline{\texttt{497238847877377792}} & 851.501 & 0.208 & 0.497 & 1.956 & 1.972 & 21.0 & 12.43 & 0.130 & TRES & Short-timescale variability \\
\texttt{5355234746758153728} & 211.036 & 0.102 & 0.710 & 1.264 & 1.938 & 10.1 & 12.32 & 0.052 & FEROS & RVs inconsistent with orbit \\
\texttt{2220072241734878720} & 331.191 & 0.179 & 0.501 & 1.856 & 1.930 & 21.4 & 11.75 & 0.170 & PEPSI & SB2 \\
\texttt{3216989916989330048} & 43.229 & 0.341 & 0.618 & 1.453 & 1.912 & 12.9 & 12.68 & 0.275 & ESI, FEROS & RVs inconsistent with orbit \\
\texttt{5836463597206648320} & 68.791 & 0.308 & 0.526 & 1.731 & 1.911 & 13.4 & 12.28 & 0.238 & MagE, FEROS & SB2 \\
\texttt{4278307358382586880} & 11.963 & 0.174 & 0.708 & 1.182 & 1.879 & 29.5 & 12.53 & 0.221 & LAMOST & RVs inconsistent with orbit \\
\underline{\texttt{3099898285186795136}} & 139.495 & 0.306 & 0.488 & 1.800 & 1.875 & 11.6 & 12.43 & 0.198 & PEPSI, FEROS, DEIMOS, TRES & Short-timescale variability \\
\texttt{3079530794349424128} & 726.969 & 0.182 & 0.463 & 1.848 & 1.849 & 37.0 & 11.35 & 0.040 & FEROS, DEIMOS & SB2 \\
\texttt{3102189152022476288} & 6.085 & 0.153 & 0.674 & 1.196 & 1.838 & 53.6 & 11.96 & 0.030 & ESI, PEPSI, FEROS & SB2; RVs inconsistent with orbit \\
\underline{\texttt{3181749813403569792}} & 326.080 & 0.128 & 0.525 & 1.596 & 1.834 & 19.9 & 12.58 & 0.040 & ESI, PEPSI, FEROS & RVs consistent with orbit \\
\texttt{20478992478455936} & 587.294 & 0.067 & 0.503 & 1.635 & 1.817 & 31.5 & 10.31 & 0.220 & LAMOST & RVs inconsistent with orbit \\
\texttt{4219507576765009536} & 474.996 & 0.400 & 0.540 & 1.514 & 1.815 & 13.0 & 10.98 & 0.035 & PEPSI, MagE, FEROS & EB with $P_{\rm orb} = 5.5$\,d \\
\underline{\texttt{3584492122268586880}} & 11.276 & 0.101 & 0.607 & 1.305 & 1.804 & 14.7 & 12.31 & 0.000 & TRES, LAMOST & Period significantly different from Gaia \\
\texttt{459778005683388928} & 16.104 & 0.129 & 0.454 & 1.776 & 1.797 & 21.8 & 12.81 & 0.570 & LAMOST & Two-temperature SED \\
\texttt{5975773133493068160} & 339.119 & 0.167 & 0.543 & 1.446 & 1.781 & 11.8 & 12.36 & 0.339 & MagE, FEROS & Large $v \sin i$; RVs unreliable \\
\underline{\texttt{5728328827639713792}} & 850.636 & 0.172 & 0.471 & 1.638 & 1.758 & 11.5 & 12.51 & 0.070 & FEROS, DEIMOS, TRES & RV orbit measured in this work \\
\texttt{6551893071549223040} & 79.395 & 0.093 & 0.842 & 0.782 & 1.757 & 36.3 & 11.89 & 0.003 & MagE, FEROS & SB2 \\
\texttt{2589496668215074944} & 2.974 & 0.127 & 0.504 & 1.509 & 1.749 & 11.9 & 11.69 & 0.110 & FEROS & EB \citep{Hoffman2009} \\
\underline{\texttt{3038922973781177984}} & 58.827 & 0.526 & 0.456 & 1.641 & 1.731 & 103.8 & 9.64 & 0.020 & FEROS, GALAH & RVs consistent with orbit \\
\underline{\texttt{4585703222934685312}} & 1067.654 & 0.341 & 0.502 & 1.478 & 1.728 & 12.7 & 10.83 & 0.034 & ESI, PEPSI, LAMOST MRS, TRES & RV orbit measured in this work \\
\texttt{4973506334166805376} & 57.523 & 0.282 & 0.472 & 1.569 & 1.724 & 98.4 & 11.73 & 0.010 & RAVE & EB with $P_{\rm orb} = 0.57$\,d \citep{Mowlavi2023} \\
\texttt{1773718654551084544} & 126.468 & 0.138 & 0.480 & 1.518 & 1.710 & 15.4 & 12.48 & 0.130 & LAMOST & RVs inconsistent with orbit \\
\underline{\texttt{1839619464509817216}} & 57.671 & 0.304 & 0.458 & 1.540 & 1.681 & 12.3 & 11.25 & 0.060 & ESI, PEPSI, TRES & Ellipsoidal variability with $P = 4.69$\,d \\
\texttt{4181886412034683648} & 43.720 & 0.038 & 0.568 & 1.191 & 1.668 & 19.1 & 12.48 & 0.125 & MagE, FEROS & RVs inconsistent with orbit \\
\texttt{5978015823680404224} & 569.309 & 0.407 & 0.493 & 1.392 & 1.664 & 14.4 & 12.49 & 0.335 & MagE, FEROS & RVs inconsistent with orbit \\
\texttt{3880367303984519936} & 12.538 & 0.069 & 0.530 & 1.257 & 1.648 & 17.2 & 12.44 & 0.000 & FEROS & RVs inconsistent with orbit \\
\texttt{947097089712447744} & 521.943 & 0.079 & 0.563 & 1.145 & 1.631 & 11.9 & 11.98 & 0.045 & ESI, PEPSI, LAMOST & RVs inconsistent with orbit \\
\underline{\texttt{6475655404885617920}} & 46.095 & 0.101 & 0.639 & 0.967 & 1.625 & 20.5 & 12.37 & 0.019 & MagE, FEROS, GALAH & Ultramassive WD \citep{Yamaguchi2024} \\
\texttt{6490750737462184064} & 1211.791 & 0.072 & 0.429 & 1.497 & 1.605 & 19.9 & 12.59 & 0.013 & MagE, FEROS & RVs inconsistent with orbit \\
\texttt{5218112727494813952} & 139.790 & 0.395 & 0.434 & 1.445 & 1.586 & 13.5 & 12.47 & 0.116 & FEROS & RVs consistent with orbit\\
\underline{\texttt{263578264603666560}} & 807.952 & 0.388 & 0.422 & 1.484 & 1.583 & 21.3 & 12.08 & 0.483 & ESI, PEPSI, LAMOST, TRES & RVs consistent with orbit \\
\texttt{5532246425068569216} & 459.232 & 0.258 & 0.478 & 1.283 & 1.575 & 16.8 & 12.66 & 0.048 & FEROS & RVs consistent with orbit \\
\texttt{4918705781283781888} & 603.254 & 0.423 & 0.458 & 1.342 & 1.573 & 23.2 & 12.07 & 0.010 & MagE, FEROS & RVs consistent with orbit \\
\texttt{6443896148956045568} & 1007.887 & 0.084 & 0.463 & 1.302 & 1.559 & 10.0 & 11.84 & 0.042 & MagE, FEROS & RVs inconsistent with orbit \\
\texttt{3431326755205579264} & 120.869 & 0.193 & 0.391 & 1.547 & 1.556 & 53.6 & 10.73 & 0.047 & ESI, PEPSI, TRES, LAMOST & RV orbit measured in this work \\
\underline{\texttt{2937304807768040576}} & 182.553 & 0.667 & 0.421 & 1.436 & 1.556 & 20.5 & 10.67 & 0.011 & FEROS & RV orbit measured in this work \\
\texttt{1926668079258767744} & 490.043 & 0.248 & 0.449 & 1.315 & 1.541 & 12.5 & 12.67 & 0.135 & ESI, PEPSI & RVs inconsistent with orbit \\
\texttt{5639517669994071424} & 577.090 & 0.218 & 0.534 & 1.061 & 1.530 & 10.1 & 12.66 & 0.035 & FEROS & RVs inconsistent with orbit \\
\underline{\texttt{843829411442724864}} & 32.143 & 0.023 & 0.398 & 1.466 & 1.528 & 229.3 & 10.61 & 0.000 & LAMOST MRS, TRES & Ultramassive WD \citep{Yamaguchi2024} \\
\texttt{6217167780337157888} & 68.528 & 0.062 & 0.491 & 1.155 & 1.519 & 23.5 & 12.12 & 0.105 & & Photometric variability with $P = 3.4$\,d \\
\underline{\texttt{556131091543510656}} & 508.599 & 0.246 & 0.512 & 1.094 & 1.516 & 11.5 & 12.63 & 0.075 & ESI, PEPSI, TRES & RV orbit measured in this work \\
\texttt{6831004854174666112} & 222.298 & 0.422 & 0.388 & 1.481 & 1.516 & 20.0 & 12.41 & 0.035 & MagE, FEROS, DEIMOS & SB2 \\
\underline{\texttt{2223170012666814720}} & 139.943 & 0.456 & 0.520 & 1.031 & 1.489 & 23.9 & 11.31 & 0.003 & ESI, PEPSI, TRES & RV orbit measured in this work; EB with $P_{\rm orb} = 0.28$\,d \citep{Mowlavi2023} \\
\underline{\texttt{2495385486559385216}} & 78.563 & 0.254 & 0.384 & 1.441 & 1.488 & 36.2 & 11.67 & 0.010 & PEPSI, TRES, FEROS, RAVE & RV orbit measured in this work \\
\underline{\texttt{1110185721018656384}} & 49.452 & 0.283 & 0.413 & 1.329 & 1.483 & 40.2 & 12.22 & 0.055 & PEPSI, DEIMOS, TRES & RVs consistent with orbit; EB with $P_{\rm orb} = 0.9$\,d \\
\texttt{2230168502956738048} & 1076.703 & 0.539 & 0.388 & 1.412 & 1.482 & 10.8 & 12.05 & 0.220 & & EB with $P_{\rm orb} = 1.32$\,d \citep{Mowlavi2023} \\
\underline{\texttt{1010268155897156864}} & 284.150 & 0.278 & 0.489 & 1.078 & 1.470 & 15.7 & 12.67 & 0.020 & TRES, DEIMOS & RV orbit measured in this work\\
\texttt{1553851414586014720} & 380.200 & 0.161 & 0.398 & 1.352 & 1.468 & 25.0 & 11.71 & 0.010 & TRES, APOGEE & RVs consistent with orbit \\
\texttt{5402382213356987520} & 332.629 & 0.318 & 0.410 & 1.308 & 1.467 & 24.4 & 12.12 & 0.050 & MagE, FEROS & RVs consistent with orbit \\
\texttt{5943754392898655872} & 904.890 & 0.416 & 0.377 & 1.417 & 1.462 & 25.2 & 11.13 & 0.226 & MagE, FEROS & SB2 \\
\texttt{6152018658376660992} & 430.476 & 0.284 & 0.522 & 0.983 & 1.461 & 18.6 & 11.87 & 0.039 & MagE, FEROS, GALAH & RVs consistent with orbit; EB with $P_{\rm orb} = 0.8$\,d \\
\underline{\texttt{5033197892724532736}} & 49.010 & 0.069 & 0.418 & 1.248 & 1.448 & 45.7 & 12.27 & 0.035 & MagE, FEROS & Ultramassive WD \citep{Yamaguchi2024} \\
\underline{\texttt{2692960678029100800}} & 17.912 & 0.018 & 0.414 & 1.254 & 1.445 & 31.2 & 12.47 & 0.045 & PEPSI, MagE, FEROS, TRES & Ultramassive WD \citep{Yamaguchi2024} \\
\texttt{1919455592216723072} & 114.578 & 0.109 & 0.443 & 1.152 & 1.438 & 18.4 & 12.61 & 0.190 & ESI, PEPSI & SB2 \\
\underline{\texttt{6412946511542005376}} & 400.846 & 0.444 & 0.378 & 1.301 & 1.403 & 18.6 & 12.12 & 0.014 & MagE, FEROS, GALAH & RVs consistent with orbit\\
\texttt{649099683051543552} & 74.005 & 0.160 & 0.378 & 1.298 & 1.401 & 15.4 & 12.62 & 0.025 & FEROS, LAMOST & RVs inconsistent with orbit \\
\midrule
\multicolumn{11}{@{}l}{\textit{Sources that do not pass the main-sequence primary cut with $f_m > 3\,M_\odot$}}\\
\midrule
\texttt{206292746724589824} & 347.002 & 0.028 & 4.517 & \nodata & \nodata & 134.8 & 11.57 & 0.885 & & Balmer emission \\
\texttt{512307478642441984} & 145.578 & 0.027 & 3.135 & \nodata & \nodata & 229.1 & 11.07 & 0.978 & & EB with $P_{\rm orb} = 145.6$\,d \\
\texttt{878555832642451968} & 837.099 & 0.095 & 4.576 & \nodata & \nodata & 24.3 & 11.88 & 0.030 & LAMOST MRS & RVs inconsistent with orbit\\
\texttt{1828150428697001472} & 333.688 & 0.013 & 3.529 & \nodata & \nodata & 276.6 & 9.07 & 0.822 & & EB with $P_{\rm orb} = 2.8$\,d \\
\texttt{3112097229257687680} & 260.978 & 0.010 & 3.321 & \nodata & \nodata & 97.6 & 11.31 & 0.412 & & EB with $P_{\rm orb} = 261$\,d \\
\underline{\texttt{3331748140308820352}} & 225.268 & 0.067 & 3.547 & \nodata & \nodata & 41.5 & 12.72 & 0.655 & ESI, PEPSI, FEROS, TRES & RV orbit measured in this work \\
\texttt{4661290764764683776} & 204.930 & 0.095 & 13.638 & \nodata & \nodata & 29.4 & 11.77 & 0.000 & APOGEE & Balmer emission \\
\texttt{5259215388421037696} & 62.707 & 0.184 & 11.506 & \nodata & \nodata & 42.1 & 12.21 & 0.291 & & SB2 \citep{GaiaCollaboration2023} \\
\texttt{5307567722242799232} & 207.182 & 0.088 & 6.145 & \nodata & \nodata & 11.8 & 12.90 & 0.263 & & Two-temperature SED + ellipsoidal \\
\texttt{5863544023161862144} & 10.606 & 0.096 & 7.781 & \nodata & \nodata & 60.7 & 9.98 & 0.537 & & EB \citep{Alfonso-Garzon2012} \\
\texttt{5869320651099982464} & 63.924 & 0.014 & 3.085 & \nodata & \nodata & 257.2 & 9.88 & 0.533 & & Two-temperature SED + ellipsoidal \\
\texttt{5941011867264508288} & 12.522 & 0.068 & 5.102 & \nodata & \nodata & 14.2 & 12.30 & 0.523 & & Two-temperature SED \\
\texttt{5941689303870322688} & 559.387 & 0.266 & 4.519 & \nodata & \nodata & 10.1 & 13.29 & 0.414 & & Two-temperature SED \\
\texttt{5963629779180627968} & 11.369 & 0.030 & 3.184 & \nodata & \nodata & 25.3 & 11.82 & 0.409 & & Two-temperature SED + ellipsoidal \\
\midrule
\multicolumn{11}{@{}p{\textwidth}@{}}{\tiny\textit{Note.} The sample contains the 151 sources satisfying the initial SB1-candidate cuts described in Section~\ref{sec:sample}. It is sorted by decreasing minimum companion mass. Underlined source IDs mark systems in the follow-up RV sample analyzed in this work. The spectroscopy column lists sources of archival spectra or follow-up RVs obtained through this program. \nodata\ values of $M_1$ and $M_{2,\min}$ indicate sources that do not satisfy the main-sequence cut used to estimate these masses; these enter the sample through the high-mass-function branch, $f_m>3\,M_\odot$.}\\
\end{supertabular}
\endgroup

\end{document}